\documentclass[aps,prb,superscriptaddress,twocolumn,floatfix,10pt]{revtex4-2}

\usepackage{times}

\usepackage[colorlinks=true,citecolor=blue,linkcolor=blue,urlcolor=blue]{hyperref}

\usepackage[utf8]{inputenc}
\usepackage{graphicx}

\usepackage{amsmath}
\usepackage{amssymb}
\usepackage{wasysym} % for \permil % load after amssymb!
\usepackage{braket}

\usepackage[capitalise]{cleveref}
\crefname{equation}{Eq.\!}{Eqs.\!}
\crefname{figure}{Fig.\!}{Figs.\!}
\crefname{chapter}{Chap.\!}{Chaps.\!}
\crefname{section}{Sec.\!}{Secs.\!}
\crefname{app}{App.\!}{Apps.\!}

\usepackage[shortlabels]{enumitem}

\newcommand{\mb}[1]{{\mathbb{#1}}}
\newcommand{\mc}[1]{{\mathcal{#1}}}
\newcommand{\tl}[1]{{\tilde{#1}}}
\newcommand{\bs}[1]{{\boldsymbol{#1}}}

\newcommand{\bH}{\mb{H}}
\newcommand{\bS}{\mb{S}}
\newcommand{\bSb}{{\overline{\mb{S}}}}
\newcommand{\bX}{\mb{X}}
\newcommand{\bY}{\mb{Y}}

\newcommand{\bM}{\mb{M}}
\newcommand{\bN}{\mb{N}}

\newcommand{\bO}{\mb{O}}
\newcommand{\bU}{\mb{U}}
\newcommand{\bV}{\mb{V}}
\newcommand{\bG}{\mb{G}}
\newcommand{\bR}{\mb{R}}
\newcommand{\bF}{\mb{F}}

\newcommand{\tH}{\tl{H}}

\newcommand{\G}{\varGamma}

\newcommand{\p}{{r}}

\newcommand{\lvec}{{\bs{\ell}}}

\usepackage[english]{babel}

\makeatletter
\let\ORIbbl@fixname\bbl@fixname
\def\bbl@fixname#1{%
  \@ifundefined{languagealias@\expandafter\string#1}
    {\ORIbbl@fixname#1}
    {\edef\languagename{\@nameuse{languagealias@#1}}}%
}
\newcommand{\definelanguagealias}[2]{%
  \@namedef{languagealias@#1}{#2}%
}
\makeatother

\definelanguagealias{en}{english}
\definelanguagealias{En}{english}
\begin{document}

\title{Flat bands by latent symmetry}

\author{C.\,V. Morfonios}
\affiliation{Zentrum f\"ur Optische Quantentechnologien, Fachbereich Physik, Universit\"at Hamburg, 22761 Hamburg, Germany}

\author{M. R\"ontgen}
\affiliation{Zentrum f\"ur Optische Quantentechnologien, Fachbereich Physik, Universit\"at Hamburg, 22761 Hamburg, Germany}

\author{M. Pyzh}
\affiliation{Zentrum f\"ur Optische Quantentechnologien, Fachbereich Physik, Universit\"at Hamburg, 22761 Hamburg, Germany}

\author{P. Schmelcher}
\affiliation{Zentrum f\"ur Optische Quantentechnologien, Fachbereich Physik, Universit\"at Hamburg, 22761 Hamburg, Germany}
\affiliation{The Hamburg Centre for Ultrafast Imaging, Universit\"at Hamburg, 22761 Hamburg, Germany}

% \date{\today}

\begin{abstract}
Flat energy bands of model lattice Hamiltonians provide a key ingredient in designing dispersionless wave excitations and have become a versatile platform to study various aspects of interacting many-body systems.
Their essential merit lies in hosting compactly localized eigenstates which originate from destructive interference induced by the lattice geometry, in turn often based on symmetry principles.
We here show that flat bands can be generated from a hidden symmetry of the lattice unit cell, revealed as a permutation symmetry upon reduction of the cell over two sites governed by an effective dimer Hamiltonian.
This so-called latent symmetry is intimately connected to a symmetry between possible walks of a particle along the cell sites, starting and ending on each of the effective dimer sites.
The summed amplitudes of any eigenstate with odd parity on the effective dimer sites vanish on special site subsets called walk multiplets.
We exploit this to construct flat bands by using a latently symmetric unit cell coupled into a lattice via walk multiplet interconnections.
We demonstrate that the resulting flat bands are tunable by different parametrizations of the lattice Hamiltonian matrix elements which preserve the latent symmetry.
The developed framework may offer fruitful perspectives to analyze and design flat band structures.

\end{abstract}

\maketitle

\section{Introduction}

Wave excitations in a lattice system are governed by the form of its energy band structure and the corresponding eigenstates.
Since the dawn of quantum mechanics, substantial efforts have been made to understand the response properties of crystals in terms of their energy bands.
With the technological advances of the past decades, however, also artificial lattice systems have been realized with ever increasing accuracy.
This has enabled an unprecedented engineering of bands with targeted properties.
A most intriguing case is that of ``flat'' bands with vanishing curvature, which have become a subject of intense research for designed lattice setups \cite{Leykam2018_APX_3_1473052_ArtificialFlatBandSystems}.
Those range among various spatial scales and different technological platforms, such as photonic waveguide or resonator arrays  \cite{Leykam2018_AP_3_070901_PerspectivePhotonicFlatbands,Mukherjee2015_PRL_114_245504_ObservationLocalizedFlatBandState,Vicencio2015_PRL_114_245503_ObservationLocalizedStatesLieb}, optical lattices for trapped atoms \cite{Taie2015_SA_1_e1500854_CoherentDrivingFreezingBosonic,Apaja2010_PRA_82_041402_FlatBandsDiracCones}, superconducting wire networks \cite{Abilio1999_PRL_83_5102_MagneticFieldInducedLocalization}, nanostructured electronic lattices \cite{Drost2017_NP_13_668_TopologicalStatesEngineeredAtomic}, optomechanical setups \cite{Wan2017_SR_7_15188_HybridInterferenceInducedFlat}, or electric circuit networks \cite{Helbig2019_PRB_99_161114_BandStructureEngineeringReconstruction}.

The remarkable features induced by flat bands essentially originate from the vanishing group velocity---or, equivalently, diverging effective mass---of the eigenstates residing in them.
This allows for dispersionless wave excitations over the whole crystal-momentum range of the flat band \cite{Mukherjee2015_PRL_114_245504_ObservationLocalizedFlatBandState}, which may be exploited for their robust storage and transfer \cite{Rontgen2019_PRL_123_080504_QuantumNetworkTransferStorage}.
In turn, transport properties of flat band states can be manipulated by weak perturbations which set a dominant energy scale for them \cite{Leykam2018_APX_3_1473052_ArtificialFlatBandSystems}.
In particular, flat bands have been used, e.\,g., to model certain types of superfluidity \cite{Peotta2015_NC_6_8944_SuperfluidityTopologicallyNontrivialFlat,Julku2016_PRL_117_045303_GeometricOriginSuperfluidityLiebLattice,Kopnin2011_PRB_83_220503_HightemperatureSurfaceSuperconductivityTopological,Iglovikov2014_PRB_90_094506_SuperconductingTransitionsFlatbandSystems,Kobayashi2016_PRB_94_214501_SuperconductivityRepulsivelyInteractingFermions,Tovmasyan2016_PRB_94_245149_EffectiveTheoryEmergentSU2,Liang2017_PRB_95_024515_BandGeometryBerryCurvature} or topological phases of matter \cite{Tang2011_PRL_106_236802_HighTemperatureFractionalQuantumHall,Sun2011_PRL_106_236803_NearlyFlatbandsNontrivialTopology,Neupert2011_PRL_106_236804_FractionalQuantumHallStates,Pal2018_PRB_98_245116_NontrivialTopologicalFlatBands,Bhattacharya2019_PRB_100_235145_FlatBandsNontrivialTopological}.
Flat bands have also been explored very recently to generate many-body localization  \cite{Danieli2020_PRB_102_041116_ManybodyFlatbandLocalization,Kuno2020_NJP_22_013032_FlatbandManybodyLocalizationErgodicity,Orito2021_PRB_103_L060301_NonthermalizedDynamicsFlatbandManybody} and ``caging'' \cite{Danieli2020_ACP_NonlinearCagingAllBandsFlatLattices,Danieli2020_ACP_QuantumCagingInteractingManyBody} in the presence of interactions, or to control superradiance via synthetic gauge fields \cite{He2021_PRL_126_103601_FlatBandLocalizationCreutzSuperradiance}.

Flat bands of discrete lattice Hamiltonians rely on the occurrence of eigenstates which are strictly localized on a subset of sites, with vanishing amplitude in the remainder of the lattice \cite{Rhim2019_PRB_99_045107_ClassificationFlatBandsAccording}.
Such ``compact localized states'' (CLSs) can be classified according to the number of unit cells they occupy \cite{Maimaiti2019_PRB_99_125129_Universald1FlatBand}.
Notably, they do not violate the translational invariance of the lattice since they can, due to their macroscopic degeneracy at the flat band energy, be linearly combined into extended Bloch states.
A CLS originates from the destructive interference of its amplitudes on the neighboring lattice sites coupled to the site subset the CLS occupies.
This mechanism may result directly from the geometric symmetry of the lattice unit cell under a site permutation operation \cite{Flach2014_E_105_30001_DetanglingFlatBandsFano,Rontgen2018_PRB_97_035161_CompactLocalizedStatesFlat}. 
It may also be caused by a bipartite (or chiral) symmetry of a lattice composed of sublattices \cite{Ramachandran2017_PRB_96_161104_ChiralFlatBandsExistence}, or induced ``accidentally'' by tuning the Bloch Hamiltonian matrix elements into the CLS condition.
Various schemes for generating flat bands from CLSs have been proposed, based e.\,g. on local permutation symmetries \cite{MoralesInostroza2016_PRA_94_043831_SimpleMethodConstructFlatband,Rontgen2018_PRB_97_035161_CompactLocalizedStatesFlat}, ``origami'' rules \cite{Dias2015_SR_5_16852_OrigamiRulesConstructionLocalized}, local basis transformations \cite{Flach2014_E_105_30001_DetanglingFlatBandsFano}, solving inverse eigenvalue problems \cite{Maimaiti2019_PRB_99_125129_Universald1FlatBand,Maimaiti2017_PRB_95_115135_CompactLocalizedStatesFlatband}, and, as shown very recently, using the properties of Gram matrices \cite{Xu2020_PRA_102_053305_BuildingFlatbandLatticeModels} or combining lattice deformations with site additions \cite{Lee2019_PRB_100_045150_HiddenMechanismEmbeddingFlat}.
Despite the great value of such approaches, the question remains whether flat bands may be systematically invoked by symmetry principles beyond the existing paradigms.

In the present work, we propose a scheme to create flat bands which is based on a type of hidden symmetry in the unit cell Hamiltonian of a lattice.
This so-called \emph{latent symmetry}, introduced recently in graph theory \cite{Smith2019_PASMaiA_514_855_HiddenSymmetriesRealTheoretical}, is revealed as a permutation symmetry once reducing the unit cell Hamiltonian over a particular subset of sites to an effective subsystem Hamiltonian.
Very recently, latent symmetries were proposed as a novel possibility to explain seemingly accidental spectral degeneracies of generic Hamiltonian matrices \cite{Rontgen2021_PRL_126_180601_LatentSymmetryInducedDegeneracies}.
Reduction over a \emph{pair} of latently exchange-symmetric sites---as we will focus on here---results in an effective two-site symmetric dimer, and the symmetry-induced parity of this dimer's eigenstates is inherited in the original unit cell;
that is, any of its eigenstates is locally even or odd on the latently symmetric sites.
Latent symmetry of two sites can be intuitively interpreted as a collective symmetry of so-called \emph{walks} \cite{Kempton2020_LAaiA_594_226_CharacterizingCospectralVerticesIsospectral} (i.\,e., sequential hoppings) along the coupled sites of the unit cell, starting and ending at each of those two sites.
Equivalently, the latent symmetry is simply expressed in terms of powers of the Hamiltonian.
We here combine latent symmetry with the occurrence of special subsets of sites called \emph{walk multiplets}.
On each such site subset, the amplitudes of any non-degenerate eigenvector with odd parity on the latently symmetric sites sum to zero.
As we show, periodic lattices generated by interconnection of walk multiplets between latently symmetric unit cells host flat bands with corresponding CLSs which occupy single unit cells.
Importantly, the underlying latent symmetry persists upon the simultaneous variation of certain parameters in the lattice Hamiltonian, making the generated flat bands systematically tunable.
With our results applicable to arbitrary dimensions, we demonstrate the principle for one- and two-dimensional lattices with simple prototype cells possessing latent symmetries.

After introducing the concepts of latent symmetry and walk equivalence in \cref{sec:latSymmetry}, we show how to combine them to generate flat band lattices in \cref{sec:flatBands}, illustrating the principle with prototype examples.
We discuss possible extensions in \cref{sec:discussion}, while \cref{sec:conclusion} concludes this work.

%%%%% fig: graphA_reduction %%%%%%%%%%%%%%%%%%%%%%%%%%%%%%%%%%%%%%%%%%%%%%%%%%
\begin{figure}[t!]
\center
\includegraphics[width=.9\columnwidth]{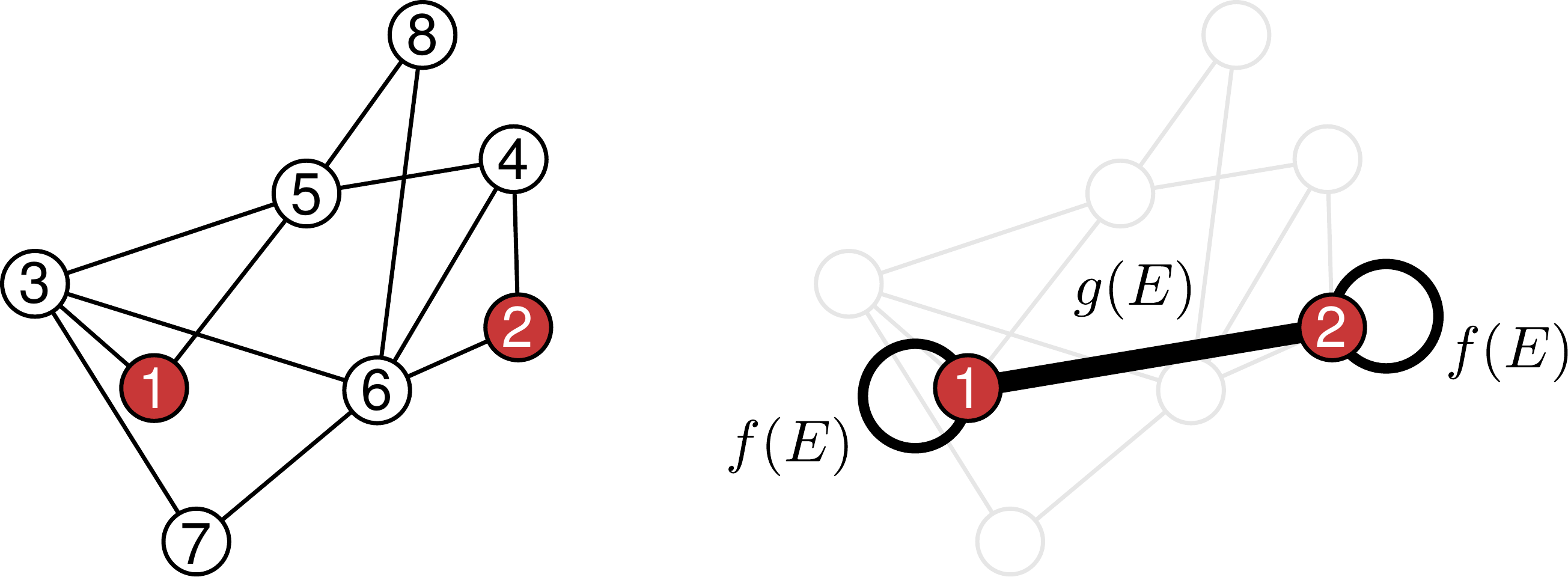}
\caption{
\emph{Left:} 
An unweighted graph, representing a Hamiltonian $H$ with unit hopping (thin edge lines) and zero onsite elements, with two cospectral vertices $\mathbb{S} = \{u,v\} = \{1,2\}$ forming a latently symmetric site pair.
\emph{Right:}
$H$ is reduced over $\bS$ to the effective Hamiltonian $\tH_\bS$ describing a symmetric two-site dimer with onsite elements $f$ (visualized as loop edges in the graph) and hopping $g$ depending functionally on the eigenvalue $E$;
in this example 
$f(E) = (8 - 4 E - 10 E^2 + 2 E^3 + 2 E^4)/d(E)$ and $g(E) = (-6 + 6 E^2 + 2 E^3)/d(E)$, where $d(E) = 7 E - 2 E^2 - 8 E^3 + E^5$.
}
\label{fig:graphA_reduction}
\end{figure}
%%%%%%%%%%%%%%%%%%%%%%%%%%%%%%%%%%%%%%%%%%%%%%%%%%%%%%%%%%%%%%%%%%%%%%%%%%%%%%

\section{Latent symmetry, cospectrality, and walk multiplets}
\label{sec:latSymmetry}

Consider the eigenvalue problem $H \ket{\varphi} = E \ket{\varphi}$ for a real symmetric $N \times N$ Hamiltonian matrix $H$ represented in the orthonormal basis of single orbitals $\ket{n}$ on $N$ coupled sites, $n \in \bH \equiv \{1,\dots,N\}$. 
To introduce the notion of latent symmetry, let us partition the system into a selected subset $\bS \subset \bH$ with $N_\bS = |\bS|$ sites  and its complement $\bSb = \bH \setminus \bS$.
The reduced $N_\bS \times N_\bS$ Hamiltonian $\tH_\bS$ effectively describing subsystem $\bS$ under the influence of the rest of the system $\bSb$ is then given by \cite{Priyadarshy1996_JCP_104_9473_BridgemediatedElectronicInteractionsDifferences,Jin2011_PRA_83_062118_PartitioningTechniqueDiscreteQuantum,Grosso2013_SolidStatePhysics}
\begin{equation} \label{eq:effectiveHam}
 \tH_\bS(E) = H_\bS + \G [E - H_\bSb]^{-1} \G^\top  \equiv H_\bS + \varSigma_\bSb(E),
\end{equation}
where the diagonal blocks $H_\bX = H_{\bX\bX}$ of $H$ are the Hamiltonians of the isolated subsystems $\bX = \bS,\bSb$ and $\G = H_{\bS\bSb}$ is the coupling from $\bSb$ to $\bS$.
This is essentially Feshbach’s projection operator method \cite{Feshbach1962_AoP_19_287_UnifiedTheoryNuclearReactions} applied to the present discrete model, while the term $\varSigma_\bSb(E)$ can be recognized as the ``self-energy'' \cite{Grosso2013_SolidStatePhysics,Datta1995_ElectronicTransportMesoscopicSystems} of $\bS$ induced by its coupling to $\bSb$.
It amounts to ``renormalized'' matrix elements in the resulting Hamiltonian $\tH_\bS(E)$, in analogy to decimation procedures in real-space renormalization group theory \cite{Grosso2013_SolidStatePhysics}, which has been applied to study, e.\,g., localization in disordered and quasiperiodic tight-binding structures \cite{Pal2013_E_102_17004_CompleteAbsenceLocalizationFamily,Pal2014_PELSaN_60_188_AbsolutelyContinuousEnergyBands,Pal2014_PLA_378_2782_EngineeringBandsExtendedElectronic}.
The reduced eigenvalue problem now has a smaller dimension, but is nonlinear due to the $E$-dependence of $\tH_\bS$.
The spectrum $\sigma(\tH_\bS)$ of $\tH_\bS$, given by $\det(E - \tH_\bS(E))=0$, coincides with that of $H$ after removing $E$-values which happen to be eigenvalues of $H_\bSb$ and for which $\tH_\bS$ is not defined; symbolically, $\sigma(\tH_\bS) = \sigma(H) - \sigma(H_\bSb)$ (note that $\sigma$ is a multiset in the presence of repeated eigenvalues).
Most importantly, any eigenvector $\ket{\tl{\varphi}}$ of $\tH_\bS$ equals the \emph{restriction} of that of $H$, with the same eigenenergy, to the subsystem $\bS$ \cite{Duarte2015_LAaiA_474_110_EigenvectorsIsospectralGraphTransformations}: $\braket{s|\tl{\varphi}} = \braket{s|\varphi}$ for $s \in \bS$.

A \emph{latent symmetry} is a permutation symmetry $\varPi_\bS$ of the reduced Hamiltonian $\tH_\bS$ such that any extended permutation $\varPi_\bS \oplus \varPi_\bSb$  (including the identity $\varPi_\bSb = I_\bSb$) is \emph{not} a symmetry of the original Hamiltonian $H$.
Throughout this work, $\bS$ will consist of two sites $u$ and $v$, and by `latent symmetry' we will always mean symmetry under transposition (i.\,e., exchange) of $u$ and $v$.
Then $\tH_\bS$ effectively behaves like a two-site dimer with $E$-dependent onsite potentials and coupling; see \cref{fig:graphA_reduction}.
If $u$ and $v$ are latently symmetric in $H$, this effective dimer is symmetric under exchange of $u$ and $v$.
If non-degenerate, its eigenstates $\ket{\tl{\varphi}}$ accordingly have definite parity, $\braket{u|\tl{\varphi}} = \pm \braket{v|\tl{\varphi}}$, and the same holds for the corresponding eigenstates, with the same eigenenergies, of $H$: $\braket{u|\varphi} = \pm \braket{v|\varphi}$.
Such a parity of amplitudes on $u$ and $v$ in the original, extended system $H$, is usually traced back to an involutory site permutation symmetry.
Remarkably, the global parity in $\tH_\bS$ is here inherited to the eigenstates $\ket{\varphi}$ as a \emph{local parity} in $H$, where there is \emph{no permutation symmetry producing it}.

%%%%% fig: graphA_multiplets %%%%%%%%%%%%%%%%%%%%%%%%%%%%%%%%%%%%%%%%%%%%%%%%%
\begin{figure}[t!]
\center
\includegraphics[width=.8\columnwidth]{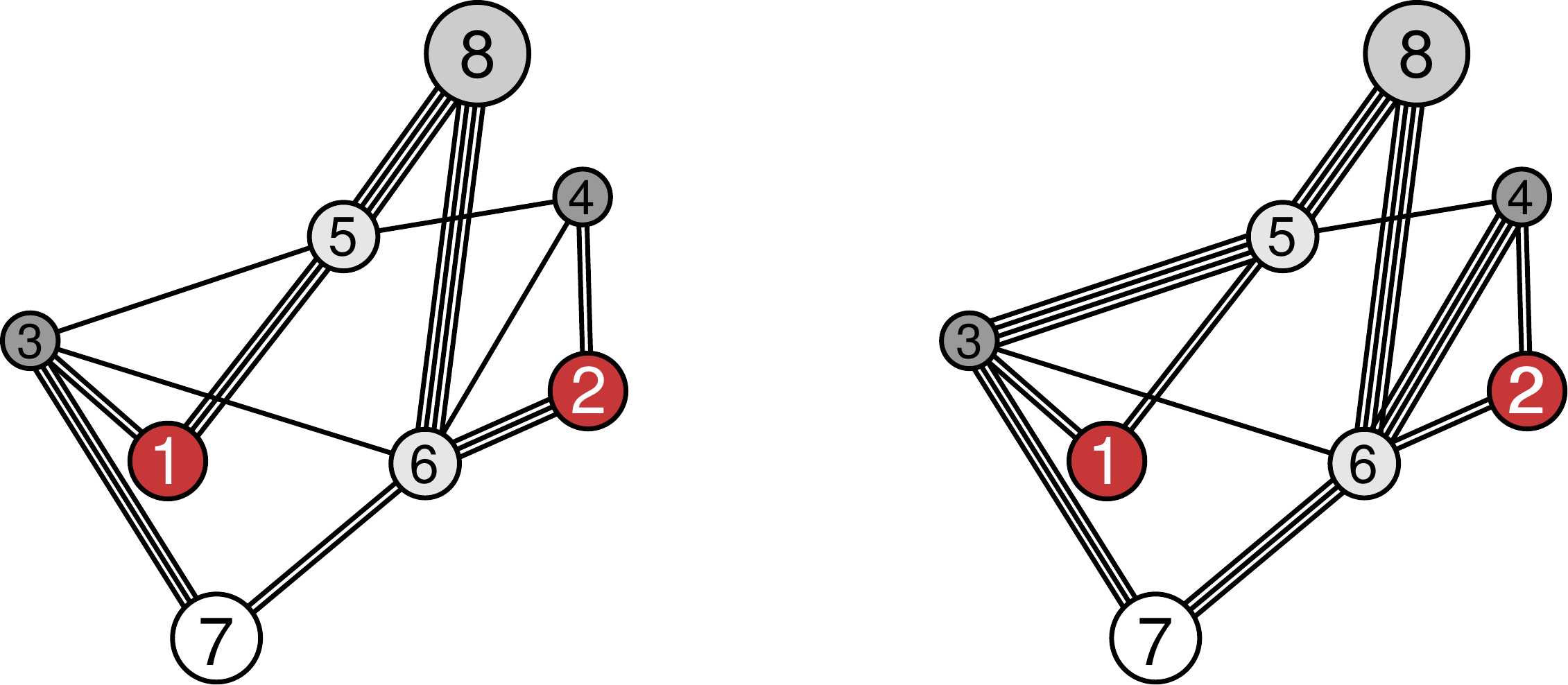}
\caption{
The same graph as in \cref{fig:graphA_reduction}, but now parametrically weighted in the two different ways which preserve the cospectrality of $\bS = \{u,v\} = \{1,2\}$.
Different edges and loops (hopping and onsite elements in $H$) are visualized by different line numbers and vertex sizes, respectively.
Subsets $\mathbb{M}_\mu$ ($\mu = 1,2,\dots, 5$) of sites with same shading (also with same vertex size) are ``walk multiplets'' with respect to $\bS$ (with $u,v$ being ``walk equivalent'' relative to any $\bM_\mu$), fulfilling \cref{eq:walkequivalent}.
Here, the doublets are $\{1,2\},\{3,4\},\{5,6\}$, and the singlets $\{7\}$, $\{8\}$.
}
\label{fig:graphA_multiplets}
\end{figure}
%%%%%%%%%%%%%%%%%%%%%%%%%%%%%%%%%%%%%%%%%%%%%%%%%%%%%%%%%%%%%%%%%%%%%%%%%%%%%%

Latent symmetries were introduced very recently \cite{Smith2019_PASMaiA_514_855_HiddenSymmetriesRealTheoretical,Bunimovich2019_AMNS_4_231_FindingHiddenStructuresHierarchies} in the context of isospectral graph reductions \cite{Bunimovich2014_IsospectralTransformationsNewApproach}.
There, $H$ is the (weighted) adjacency matrix of a connected graph with vertex set $\bH$ and edges with weights $H_{mn} = \braket{m|H|n}$ between vertices.
For brevity, we will refer to the graph itself simply as $H$. 
The \emph{isospectral reduction} of the graph over a subset $\bS$ of its vertices is exactly the graph with adjacency matrix given in \cref{eq:effectiveHam}.
An example graph is shown in \cref{fig:graphA_reduction}, containing the two latently symmetric vertices $\bS = \{u,v\} = \{1,2\}$.

A crucial fact, promoting the treatment of latent symmetry with the tools of graph theory, is the following \cite{Kempton2020_LAaiA_594_226_CharacterizingCospectralVerticesIsospectral}:
Two latently symmetric vertices $u,v$ of a graph are \emph{cospectral}, meaning that the spectra of the ``vertex-deleted'' graphs $H_{\overline{u}} = H - u$ and $H_{\overline{v}} = H - v$ (where vertices $u$ and $v$, as well as edges incident to them, have been deleted, respectively) coincide, $\sigma(H_{\overline{u}}) = \sigma(H_{\overline{v}})$.
Alternatively, and of more use for our purposes here, cospectral vertices are defined by the property that their corresponding diagonal entries in any non-negative power $\p$ of $H$ coincide \cite{Godsil2017_AM_StronglyCospectralVertices},
\begin{equation} \label{eq:cospectrality}
 [H^\p]_{uu} = [H^\p]_{vv} \quad \forall~ \p \in \mb{N}.
\end{equation}

In general, for an unweighted graph ($H_{mn} \in \{0,1\}$) the element $[H^\p]_{mn}$ gives the total number of all possible \emph{walks} of length $\p$ from vertex $m$ to $n$ 
\cite{Estrada2015_FirstCourseNetworkTheory,Tsuji2019_JCP_150_204123_InfluenceLongrangeInteractionsQuantum}, that is, sequences 
\begin{equation} \label{eq:walk}
 \alpha = (a_1=m,b_1)(a_2,b_2)\cdots (a_\p,b_\p=n)
\end{equation}
of $\p$ possibly repeated edges $(a_i,b_i)$ with $a_{i+1} = b_i$.
For example, the walk with steps $1 \to 3 \to 5 \to 8 \to 6 \to 2$ along the sites of \cref{fig:graphA_reduction} is denoted as the sequence of edges $(1,3)(3 , 5)(5 , 8)(8 , 6)(6 , 2)$.
Note that also \emph{loops}, with $b_i = a_i$, may be included in a walk, representing onsite potentials $H_{a_i a_i}$.
\cref{eq:cospectrality} concerns the special case of closed walks ($n=m$) starting and ending at each cospectral vertex $u$ or $v$.
For instance, \cref{eq:cospectrality} can easily be verified in the unweighted graph of \cref{fig:graphA_reduction} for the first few powers $\p$, by counting all closed walks of length $\p$ starting at $u=1$ or $v=2$.
The closed walks of length $\p = 3$, for example, are (in simplified step notation) $1 \to 3 \to 5 \to 1$ and $2 \to 4 \to 6 \to 2$, plus the same in opposite directions, in accordance with $[H^3]_{11} = [H^3]_{22} = 2$.

For a weighted graph, a weight $w(\alpha)$ is assigned to each walk, equal to the product of edge weights along it, $w(\alpha) = \prod_{i=1}^\p w(a_i,b_i) = \prod_{i=1}^\p H_{a_i b_i}$.
The above interpretation of matrix powers in terms of walks is then generalized to a sum over walk weights \cite{Brualdi2008_CombinatorialApproachMatrixTheory}, % Theorem 3.1.2
\begin{equation} \label{eq:powers}
 [H^\p]_{mn} = \sum_{\alpha \in \mc{A}^{(\p)}_{mn}} w(\alpha), \quad \p \in \bN
\end{equation}
where $\mc{A}^{(\p)}_{mn}$ is the set of all walks of length $\p$ from $m$ to $n$.

Fortunately, there is no need to evaluate \cref{eq:cospectrality} beyond $k = N-1$ since, by the Cayley-Hamilton theorem, any higher powers $H^{k \geqslant N}$ can be expressed as lower order polynomials in $H$.
As a consequence, if \cref{eq:cospectrality} holds for $\p = 0,..,N-1$, it automatically holds for all $\p$.
This enables the use of the $N \times N$ \emph{walk matrix} \cite{Godsil2012_AC_16_733_ControllableSubsetsGraphs,Liu2019_AM_UnlockingWalkMatrixGraph} $W_\bM$ of a subset $\bM \subseteq \bH$ to encode walks ending in $\bM$, constructed by the action of $H^\p$ on the indicator vector $\ket{e_\bM}$ of $\bM$ (with $\braket{m|e_\bM} = 1$ for $m \in \bM$ and $0$ otherwise):
\begin{equation} \label{eq:walkmatrix}
 W_\bM = [~\ket{e_\bM}, H \ket{e_\bM}, \dots, H^{N-1} \ket{e_\bM}~],
\end{equation}
also known as the Krylov matrix of $H$ generated by $\ket{e_\bM}$ \cite{Meyer2000_MatrixAnalysisAppliedLinear,Xu2011_LAaiA_434_185_FunctionsMatrixKrylovMatrices}.
The $\p$-th column of $W_\bM$ is given by $[W_\bM]_{*\p} = H^{\p-1} \ket{e_\bM}$ ($*$ denoting all indices) and its element
\begin{equation}
 [W_\bM]_{s\p} = \sum_{m \in \bM} [H^{\p-1}]_{sm}
\end{equation}
yields the sum over weighted walks---in the sense of \cref{eq:powers}---of length ${\p-1}$ from vertex $s$ to all vertices in $\bM$.

We call two vertices $u,v$ \emph{walk equivalent} \cite{Morfonios2021_LAaiA_624_53_CospectralityPreservingGraphModifications} relative to $\bM$ if their summed walks to $\bM$ are equal for any walk length $\p$, that is, if the corresponding rows of $W_\bM$ are equal, 
\begin{equation} \label{eq:walkequivalent}
 [W_\bM]_{u*} = [W_\bM]_{v*}.
\end{equation}
Conversely, we say that $\bM$ then constitutes a \emph{walk multiplet} with respect to $\{u,v\}$; 
specifically, a walk $M$-let of size $M = |\bM|$ (singlet for $M=1$, doublet for $M=2$, etc.)
\footnote{
We note that the occurrence of a walk equivalent pair relative to some $\bM$ necessarily renders $W_\bM$ non-invertible by reducing its rank.
Incidentally, this lifts the so-called ``controllability'' \cite{Godsil2012_AC_16_733_ControllableSubsetsGraphs,Farrugia2014_LAaiA_454_138_ControllabilityUndirectedGraphs,Aguilar2015_ITAC_60_1611_GraphControllabilityClassesLaplacian} of $\bM$ relative to $H$, and as a consequence allows for local permutation symmetries in $H$ mapping $\bM$ to itself \cite{Godsil2012_AC_16_733_ControllableSubsetsGraphs}. %[Lemma 1.1].
}.

Examples of walk singlets ($M = 1$) and doublets ($M = 2$) are shown in \cref{fig:graphA_multiplets}.
There, the weights of the graph in \cref{fig:graphA_reduction} have also been \emph{parametrized} in two different ways such that the cospectral pair $\{1,2\}$, and each shown walk multiplet relative to it, are preserved \footnote{
We find such parametrizations numerically by starting with the unweighted graph and then (i) setting one weight to a random value (representing an independent parameter) and scanning through the graph for edges which can be set to that same value while preserving cospectrality, (ii) repeating successively for unaltered edges until all edges are parametrized.
}. 
More specifically, the cospectrality of $\{1,2\}$ and walk multiplets relative to it remain intact for any arbitrary value---a parameter of $H$---of edge weights (including loops) which are equal.
For instance, in the right parametrization the equal weights $H_{13} = H_{15} = H_{24} = H_{26}$ can be varied together arbitrarily while retaining the cospectrality and walk multiplets of $\{1,2\}$.

As we will see further below, this cospectrality- and multiplet-preserving parametrization will allow for a flexible tuning of flat bands.
In the next section, we will start by showing how the combination of walk equivalence with cospectrality for vertex pairs may be used to generate CLSs and corresponding flat band lattices.

Before continuing, let us note that, if two vertices $u$ and $v$ are related by a permutation symmetry of the graph, i.\,e., there is some permutation matrix $\varPi$ commuting with $H$ for which $\varPi \ket{u} = \ket{v}$, then \cref{eq:cospectrality} is automatically fulfilled.
In this work we focus on cospectral pairs $u,v$ for which \cref{eq:cospectrality} is \emph{not} induced by permutation symmetry, but which correspond to \emph{latent} symmetry as defined above.
As will be discussed later on (see \cref{sec:occurrenceOfLatSym}), constructing such graphs is not a trivial task.
We have here resorted to numerical validation of \cref{eq:cospectrality} for a fixed small graph size ($N = 8$), with a latently symmetric example provided in \cref{fig:graphA_reduction}.
It should thus be clear that the graphs we utilize as representative examples in this work are special cases whose structure supports latent symmetry.
Modifying them \emph{arbitrarily} (e.\,g. by adding or deleting edges) would in general invalidate \cref{eq:cospectrality} and thus break the latent symmetry between the selected vertices $u,v$ in each case.
Nevertheless, as shown in Ref.\,\cite{Morfonios2021_LAaiA_624_53_CospectralityPreservingGraphModifications}, there \emph{exist} systematic graph modifications which do preserve the cospectrality of a given vertex pair.
Those modifications are outlined below (in \cref{sec:modifications}) and will constitute the key ingredient in generating flat bands by combining latent symmetry with walk multiplets.

\section{Flat bands induced by walk equivalent cospectral sites} 
\label{sec:flatBands}

Let us now consider a graph $H$ with cospectral vertices $u,v$ which are walk equivalent relative to a multiplet $\bM$, like in \cref{fig:graphA_multiplets} (with $\bM$ chosen as one of the multiplets $\bM_\mu$).
Due to cospectrality, any non-degenerate eigenvector $\ket{\varphi_\nu}$ has (or, if degenerate, can be chosen to have) local parity on $\{u,v\}$ \cite{Eisenberg2019_DM_342_2821_PrettyGoodQuantumState},
\begin{equation} \label{eq:localParity}
 \braket{u|\varphi_\nu^\pm} = \pm \braket{v|\varphi_\nu^\pm}
\end{equation}
with $+$ ($-$) denoting even (odd) parity.
This local parity on the cospectral pair $\{u,v\}$ is equivalent to a symmetry $Q$ of $H$ with $Q^2 = I$ which exchanges $u$ and $v$, that is, $Q \ket{e_u} = \ket{e_v}$, while acting as a general orthogonal transformation on the complement $\bH \setminus \{u,v\}$ \cite{Godsil2017_AM_StronglyCospectralVertices,Rontgen2021_PRL_126_180601_LatentSymmetryInducedDegeneracies}, as described in detail in \cref{app:Qmatrix}.

Now, by inserting the spectral decomposition $H = \sum_\nu E_\nu \ket{\varphi_\nu}\bra{\varphi_\nu}$ into \cref{eq:walkequivalent} and using \cref{eq:localParity}, one can show \cite{Morfonios2021_LAaiA_624_53_CospectralityPreservingGraphModifications} that the amplitude sum of any odd $\{u,v\}$-parity eigenstate over any walk multiplet relative to $\{u,v\}$ vanishes, that is,
\begin{equation} \label{eq:zeroMultipletAmpSum}
 \sum_{m \in \bM} \braket{m|\varphi_\nu^-} = \braket{e_\bM|\varphi_\nu^-} = 0,
\end{equation}
where, in the case of degenerate $\ket{\varphi_\nu^-}$, it has been chosen to be the only $\{u,v\}$-odd eigenstate to its eigenvalue $E_\nu$, given by the projection of the vector $\ket{u} - \ket{v}$ onto that degenerate subspace \cite{Morfonios2021_LAaiA_624_53_CospectralityPreservingGraphModifications}.
In particular, $\braket{m|\varphi_\nu^-}$ vanishes on any walk \emph{singlet} $\bM = \{m\}$.
We note that walk singlets are fixed (that is, each mapped onto itself) under the action of $Q$, as shown in \cref{app:Qmatrix}.

The generation of flat bands from a latently symmetric $H$ will ultimately consist in converting it into a Bloch Hamiltonian by interconnecting any of its walk multiplets \emph{within the same graph $H$ itself} via edges with corresponding complex weights.
To develop and demonstrate the principle step-by-step in the following subsections, we will first provide the necessary graph modification rules in \cref{sec:modifications}; 
apply them to construct a periodic 1D lattice, or directly its Bloch Hamiltonian, hosting CLSs in \cref{sec:multipletInterconnection}; 
demonstrate how the corresponding flat bands can be parametrically tuned in \cref{sec:parametricInvariance}; and combine the above in a 2D example in \cref{sec:singletAugmentation}.

%%%%% fig: cospectralGraphModifications %%%%%%%%%%%%%%%%%%%%%%%%%%%%%%%%%%%%%%
\begin{figure}[t!]
\center
\includegraphics[width=.8\columnwidth]{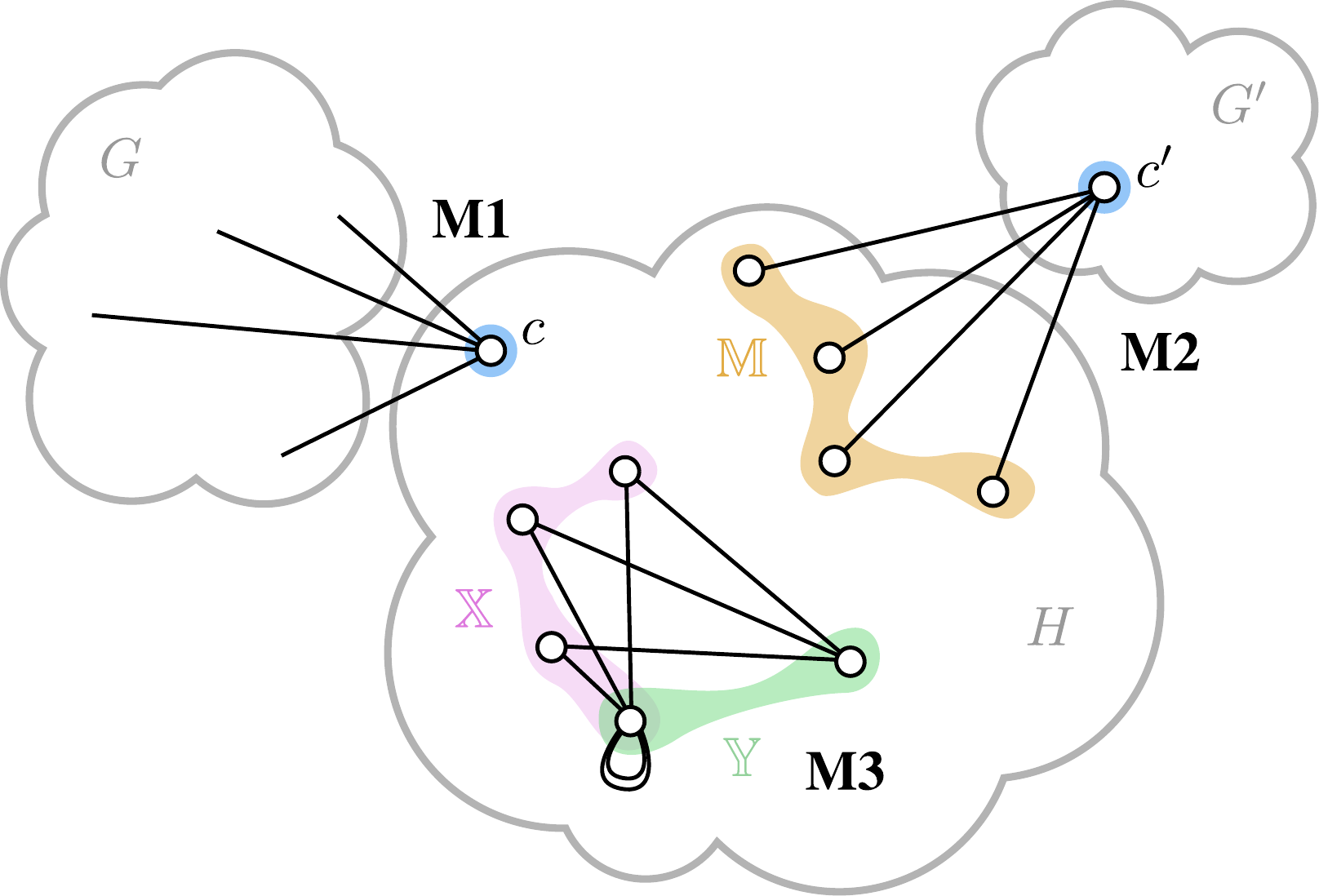}
\caption{
Schematically depicted graph modifications addressed in \cref{sec:modifications} and \cref{app:modifications}:
\ref{mod:singletExtension} connection of an arbitrary graph $G$ exclusively to a walk singlet $c$ of $H$, 
\ref{mod:multipletExtension} connection of a walk multiplet $\bM$ of $H$ to a single vertex $c'$ of an arbitrary graph $G'$,
\ref{mod:multipletInterconnection} interconnection of two overlapping walk multiplets $\bX$ and $\bY$ of $H$.
}
\label{fig:cospectralGraphModifications}
\end{figure}
%%%%%%%%%%%%%%%%%%%%%%%%%%%%%%%%%%%%%%%%%%%%%%%%%%%%%%%%%%%%%%%%%%%%%%%%%%%%%%

\subsection{Graph modifications preserving walk multiplets}
\label{sec:modifications}

As shown in Ref.\,\cite{Morfonios2021_LAaiA_624_53_CospectralityPreservingGraphModifications}, certain modifications can be performed on a graph $H$ such that the cospectrality of vertex pair $\{u,v\}$ together with the walk multiplets relative to it remain intact.
For clarity, we here focus on their simplest form (see \cref{sec:discussion} below for related generalizations).

The cospectrality of $\{u,v\}$, as well as any walk multiplet $\bM$ of $H$, are preserved in the new graph $H'$ obtained by performing the following modifications:
\begin{enumerate}[(M1)]
 \item \label{mod:singletExtension} Connection of an arbitrary graph exclusively to any walk singlet $c$ of $H$ via edges of arbitrary weights, whereby all vertices of the added graph become walk singlets in $H'$;
 \item \label{mod:multipletExtension} Connection of all vertices of any walk multiplet of $H$ to a single vertex $c'$ of an arbitrary graph via edges of uniform weight, whereby all vertices of the added graph become walk singlets in $H'$;
 \item \label{mod:multipletInterconnection} Interconnection of any two walk multiplets of $H$ via edges of uniform weight between all vertices of one multiplet and each vertex of the other (added to any already existing edge weights),
\end{enumerate}
where any walk multiplet is implied relative to $\{u,v\}$.
In \cref{app:modifications} we provide brief proofs of the above properties in their general form.
A generic schematic of the modifications is given in \cref{fig:cospectralGraphModifications}.
Note that if the two multiplets in \ref{mod:multipletInterconnection} overlap (that is, have common vertices), then the vertices in the overlap are interconnected by double (additional) edges, like the double loop in \cref{fig:cospectralGraphModifications}; see \cref{app:modifications}.

In the following, we will employ the above modifications \ref{mod:singletExtension}--\ref{mod:multipletInterconnection} for the construction of flat band lattices, illustrated in concrete examples.

\subsection{Flat bands via walk multiplet interconnections}
\label{sec:multipletInterconnection}

In the following principle for constructing flat band lattices, an original latently symmetric Hamiltonian $H$ featuring walk multiplets will be used as a unit cell of a lattice with Hamiltonian $H^\sharp$.
The unit cells are interconnected using the modifications described above in \cref{sec:modifications}.
In this way, the latent symmetry in any copy of $H$ is inherited by the whole lattice in the sense that it remains present after the interconnection. 
We stress that, in order to induce flat bands, the walk multiplets used in the those interconnections are relative to a given cospectral site pair $\{u,v\}$, as described above, and not to any arbitrary site pair.

%%%%% fig: graphA_cellInterconnection %%%%%%%%%%%%%%%%%%%%%%%%%%%%%%%%%%%%%%
\begin{figure}[t!]
\center
\includegraphics[width=\columnwidth]{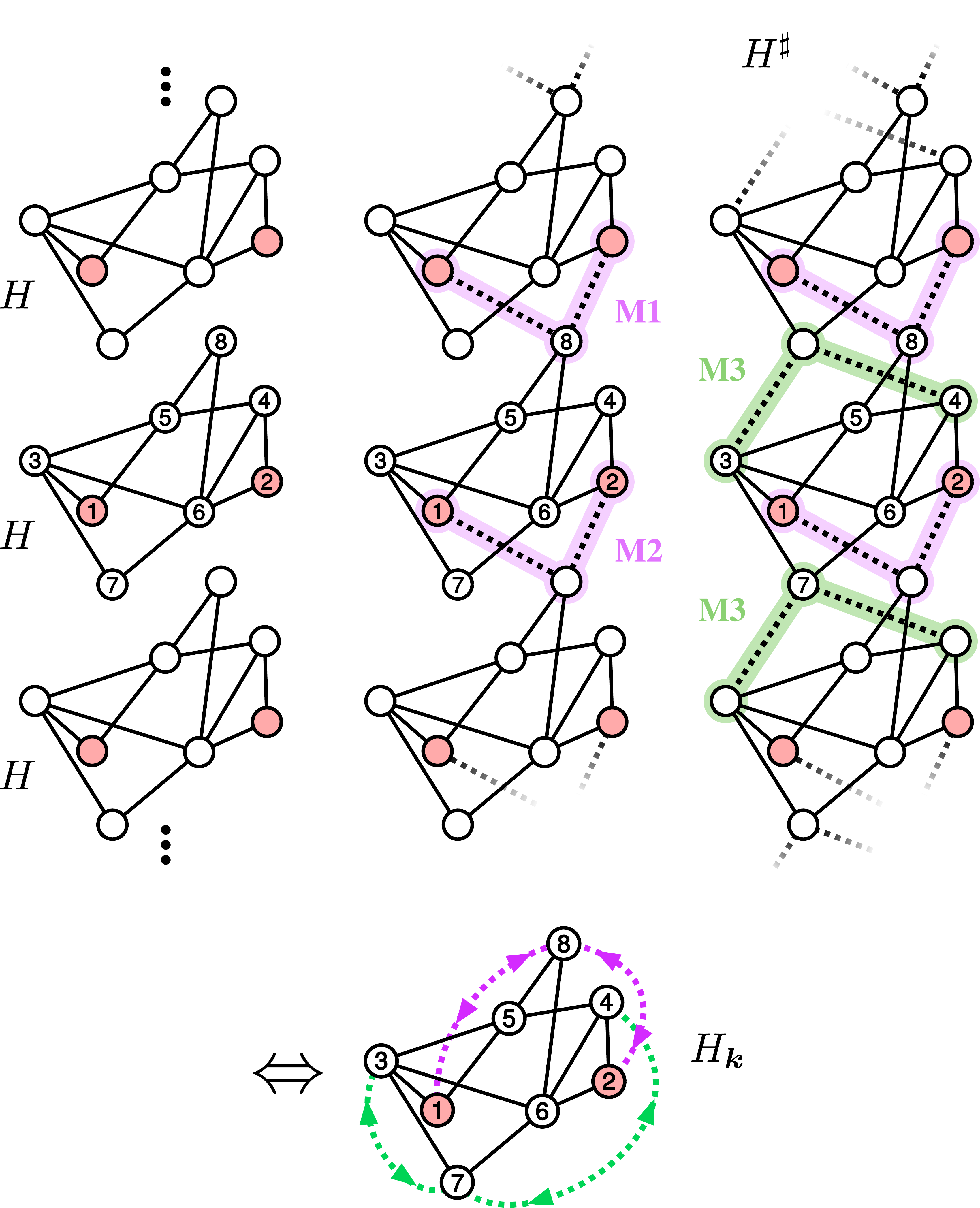}
\caption{
\textit{Top:}
Construction of a lattice Hamiltonian $H^\sharp$ using the Hamiltonian $H$ of \cref{fig:graphA_reduction} as an isolated unit cell, via modifications \ref{mod:singletExtension}--\ref{mod:multipletInterconnection} of \cref{sec:modifications} with respect to its cospectral site pair $\{1,2\}$:
For each of the copies of $H$ \textit{(left)}, setting the labeled graph as reference,  first we connect the walk singlet $\{8\}$ to sites $1,2$ of the cell above using \ref{mod:singletExtension} and the walk doublet $\{1,2\}$ to site $8$ of the cell below using \ref{mod:multipletExtension} \textit{(middle)}, and then connect the walk multiplets $\{3,4\}$ and $\{7\}$ to walk singlets of the resulting graph using \ref{mod:multipletInterconnection} \textit{(right)}, with inter-cell connections $h$ indicated by dotted lines.
\textit{Bottom:}
The Bloch Hamiltonian $H_\bs{k}$ corresponding to the lattice Hamiltonian $H^\sharp$ can be constructed from $H$ via \ref{mod:multipletInterconnection} by interconnecting the singlet $\{8\}$ ($\{7\}$) to the doublet $\{1,2\}$ ($\{3,4\}$), though with complex Hermitian couplings $h e^{\pm ikL}$ indicated by purple (green) double-arrowed dotted lines, with $k = |\bs{k}|$ and lattice constant $L$; see \cref{sec:multipletInterconnection}.
}
\label{fig:graphA_cellInterconnection}
\end{figure}
%%%%%%%%%%%%%%%%%%%%%%%%%%%%%%%%%%%%%%%%%%%%%%%%%%%%%%%%%%%%%%%%%%%%%%%%%%%%%%

In \cref{fig:graphA_cellInterconnection}\,(top), we illustrate the modifications \ref{mod:singletExtension}--\ref{mod:multipletInterconnection} as applied to our example Hamiltonian $H$ of \cref{fig:graphA_reduction} to create a periodic lattice $H^\sharp$ with $H$ as a unit cell.
First, for a given reference cell $H$ (the cell with labeled sites in \cref{fig:graphA_cellInterconnection}; simply ``cell'' will mean ``unit cell'' from here on), we connect the walk singlet $\{8\}$ to sites $1,2$ in the cell above and the doublet $\{1,2\}$ to site $8$ in the cell below.
Thus, the cospectrality of the pair $\{1,2\}$ and relative multiplets are preserved by simultaneous application of \ref{mod:singletExtension} (connecting a singlet to the graph above) and \ref{mod:multipletExtension} (connecting a multiplet to the graph below).
Note that, after this interconnection, all sites in the remainder of the lattice (outside the reference cell) are singlets relative to $\{1,2\}$ in the reference cell.
Second, in the resulting graph, we apply \ref{mod:multipletInterconnection} by interconnecting the doublet $\{3,4\}$ with the site $7$ of the cell above (a singlet relative to $\{1,2\}$ in the reference cell) and the singlet $\{7\}$ to sites $3,4$ of the cell below (both singlets relative to $\{1,2\}$ in the reference cell).

Note that the same interconnections as for the reference cell to adjacent cells can be performed simultaneously for all periodically arranged copies of $H$, without affecting the cospectrality of $\{u,v\} = \{1,2\}$ in the reference cell.
Thus, since the reference cell is chosen arbitrarily, each unit cell in $H^\sharp$ inherits the cospectral pair $\{u,v\}$ (in local labeling for that cell) and its relative walk multiplets from the isolated graph $H$.
We also underline that the distinction between the different inter-cell connections in \cref{fig:graphA_cellInterconnection} by the labels M1, M2, M3 refers only to the way the lattice is constructed by sequential application of those graph modification rules.
In the final lattice, those physical connections are qualitatively equivalent; 
in fact, the connections labeled M1 and M2 constitute the same inter-cell coupling, translated by one unit cell.

The inter-cell connection scheme previously outlined can be more compactly expressed directly at the level of the Bloch Hamiltonian $H_\bs{k}$ of the lattice.
$H_\bs{k}$ is generally obtained by Fourier transformation of the lattice Hamiltonian elements \cite{Vanderbilt2018_BerryPhasesElectronicStructure} %[Eq.(2.75)]
as 
\begin{equation} \label{eq:blochHamiltonian}
 [H_\bs{k}]_{mn} = \sum_\lvec e^{i\bs{k}\cdot \lvec} \braket{m|H^\sharp|n_\lvec}, 
\end{equation}
where $\ket{n_\lvec}$ is the orbital $\ket{n}$ in the cell at position $\lvec$, with $\ket{n} \equiv \ket{n_{\lvec = \bs{0}}}$ for the reference unit cell at $\lvec = \bs{0}$. 
The eigenvalues $E_\nu({\bs{k}})$ of $H_\bs{k}$ constitute the band structure of the lattice.
Interconnections between \emph{different} cells in the lattice graph $H^\sharp$ (e.\,g. with some coupling $\braket{m|H^\sharp|n_\lvec} = h \in \bR$ between sites $m,n_{\lvec \neq \bs{0}}$) are equivalent to the corresponding interconnections in the \emph{single} cell graph $H$, though additionally weighted with conjugate Bloch phases (i.\,e. coupling $[H_\bs{k}]_{mn} = [H_\bs{k}]_{nm}^* = h\,e^{i\bs{k}\cdot \lvec}$ between sites $m,n$).
This is shown in \cref{fig:graphA_cellInterconnection}\,(bottom) for the example lattice.
The resulting Bloch graph $H_\bs{k}$ is directed, with complex conjugate edge weights in opposite directions between any vertex pair being interconnected.
In fact, $H_\bs{k}$ can be seen as resulting from the cospectrality-preserving modification \ref{mod:multipletInterconnection} (interconnection of two walk multiplets) on $H$, though with additional uniform prefactors $e^{\pm i\bs{k}\cdot \lvec}$ in either direction of the connection (see \cref{fig:graphA_cellInterconnection}).
In \cref{app:hermiticity} we explicate that site pair cospectrality and corresponding latent symmetry are preserved under walk multiplet interconnections \ref{mod:multipletInterconnection} with complex Hermitian coupling weights.

%%%%% fig: graphA_bandsLattice %%%%%%%%%%%%%%%%%%%%%%%%%%%%%%%%%%%%%%%%%%%%%%%%%%%%%%
\begin{figure}[t!]
\center
\includegraphics[width=.9\columnwidth]{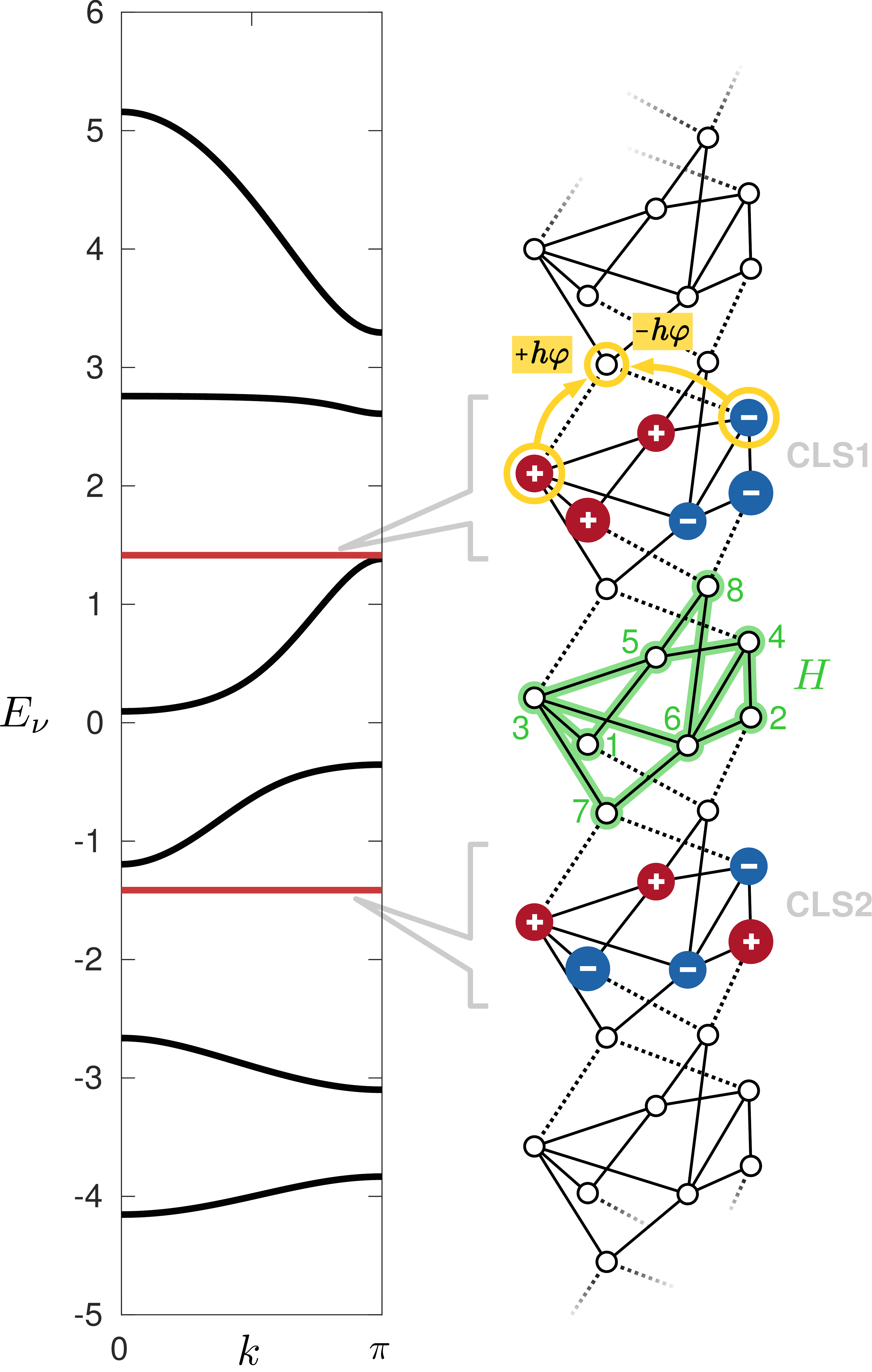}
\caption{
Band structure \textit{(left)} of the lattice $H^\sharp$ in \cref{fig:graphA_cellInterconnection}, with unit intra-cell couplings and inter-cell couplings $h = 2$.
It features two flat bands at $E = \pm\sqrt{2}$ (red lines) corresponding to the two latent-symmetry induced CLSs ``CLS1'' and ``CLS2'' depicted \textit{(right)}.
The amplitudes of each shown CLS are real with indicated relative sign $+$ (red) and $-$ (blue), with magnitudes proportional to the areas of the corresponding signed (red/blue) circles, and zero on all other sites. 
The isolated unit cell Hamiltonian $H$ and site labeling are highlighted (green) in the middle.
An example of destructive interference of one of CLS1's amplitudes on the connected site of an adjacent cell is indicated by (orange) arrows, with $\pm \varphi$ denoting the amplitude on the walk doublet $\{3,4\}$.
Energies are in units of the uniform intra-cell site couplings, and lengths in units of the lattice constant $L = 1$ (quasimomentum $k$ in units of $1/L$).
}
\label{fig:graphA_bandsLattice}
\end{figure}
%%%%%%%%%%%%%%%%%%%%%%%%%%%%%%%%%%%%%%%%%%%%%%%%%%%%%%%%%%%%%%%%%%%%%%%%%%%%%%

Let us now explain how the multiplet interconnection described above can induce CLSs and corresponding flat bands for the resulting lattice.
Specifically, any $\{u,v\}$-odd eigenstate $\ket{\varphi_{\nu}^-}$ of the initially isolated unit cell $H$ constitutes a CLS in the lattice $H^\sharp$ constructed via multiplet interconnection between unit cells.
Indeed, consider the infinite-length column vector $\ket{\varphi_{\nu;\lvec}^-}$ defined to have the components of $\ket{\varphi_{\nu}^-}$ on the cell at $\lvec$, padded with zeros on all other cells $\lvec' \neq \lvec$, that is: $\braket{n_{\lvec'}|\varphi_{\nu;\lvec}^-} = \braket{n | \varphi_{\nu}^-}\,\delta_{\lvec\lvec'}$.
In other words, $\ket{\varphi_{\nu;\lvec}^-}$ is a CLS occupying the cell at $\lvec$.

Notice now that $\ket{\varphi_{\nu;\lvec}^-}$ is an eigenstate of the lattice Hamiltonian $H^\sharp$ to the eigenenergy $E_\nu$ (the eigenenergy of $\ket{\varphi_{\nu}^-}$ in the isolated cell $H$).
To see this, let us write $H^\sharp$ in the form
\begin{equation} \label{eq:latticeHam}
 H^\sharp = \bigoplus_{\lvec} H + \sum_{\lvec \neq \lvec', \bX, \bY} \left( h^{\lvec\lvec'}_{\bX\bY}\ket{e_{\bX;\lvec}}\bra{e_{\bY;\lvec'}} + \text{h.c.} \right),
\end{equation}
where $H$ is repeated on the block-diagonal and $\ket{e_{\bX;\lvec}}\bra{e_{\bY;\lvec'}}$ contains the off-diagonal block coupling multiplet $\bY$ in cell $\lvec'$ to multiplet $\bX$ in cell $\lvec$ with uniform coupling strength $h^{\lvec\lvec'}_{\bX\bY}$, and zeros otherwise ($\ket{e_{\bX;\lvec}}$ being the infinite column with components $\ket{e_{\bX}}$ on cell $\lvec$ and zeros otherwise)---see e.\,g. colored couplings in \cref{fig:graphA_cellInterconnection}\,(top right).
Now, acting with $H^\sharp$ on $\ket{\varphi_{\nu;\lvec}^-}$ directly yields
\begin{equation}
   H^\sharp \ket{\varphi_{\nu;\lvec}^-} = E_\nu \ket{\varphi_{\nu;\lvec}^-},
\end{equation}
since $H \ket{\varphi_{\nu}^-} = E_\nu \ket{\varphi_{\nu}^-}$ for block $\lvec$ (corresponding to the only cell occupied  by $\ket{\varphi_{\nu;\lvec}^-}$), while $\braket{e_{\bX;\lvec}|\varphi_{\nu;\lvec}^-} = \braket{e_{\bX}|\varphi_{\nu}^-} = 0$ by \cref{eq:zeroMultipletAmpSum}.

As an example, the lattice constructed in \cref{fig:graphA_cellInterconnection} features two different CLS types, which are illustrated in \cref{fig:graphA_bandsLattice}\,(right panel) in two different unit cells of the lattice.
The orange arrows indicate an example of how the CLS amplitudes cancel out (interfere destructively) on a neighboring cell site upon action of $H^\sharp$ due to the multiplet condition, \cref{eq:zeroMultipletAmpSum}.

Analogously to the above, $\ket{\varphi_{\nu}^-}$ is an eigenvector of the Bloch Hamiltonian $H_\bs{k}$ constructed from $H$ via multiplet interconnections. 
More specifically, $H_\bs{k}$ can be written as
\begin{equation} \label{eq:blochHamiltonianMultiplets}
 H_\bs{k} = H + \sum_{\lvec, \bX, \bY} \left( h^{\lvec}_{\bX\bY}e^{i \bs{k}\cdot\lvec}\ket{e_{\bX}}\bra{e_{\bY}} + \text{h.c.} \right),
\end{equation}
summing over all interconnected multiplet pairs $\bX,\bY$ with $\bY$ in cell $\lvec$ and $\bX$ in the reference cell $\lvec \equiv \bs{0}$ connected with uniform coupling weight $h^{\lvec}_{\bX\bY}$.
Acting with $H_\bs{k}$ on $\ket{\varphi_{\nu}^-}$ immediately yields
\begin{equation} \label{eq:blochEVP}
 H_\bs{k} \ket{\varphi_{\nu}^-} = E_\nu \ket{\varphi_{\nu}^-},
\end{equation}
again due to \cref{eq:zeroMultipletAmpSum}.
This holds \emph{for any} $\bs{k}$, so $\ket{\varphi_{\nu}^-}$ corresponds to a flat band at the $\bs{k}$-independent eigenenergy $E_\nu$ in the band structure of the lattice.

In \cref{fig:graphA_bandsLattice}, the band structure of the lattice constructed in \cref{fig:graphA_cellInterconnection} is shown \footnote{For clarity, we note that the bands were computed by standard numerical matrix diagonalization of the Bloch Hamiltonian $H_\bs{k}$ (in varying $\bs{k}$) and not from the corresponding nonlinear eigenvalue problem $\tH_{\bS;\bs{k}}(E)\ket{\varphi}=E\ket{\varphi}$ of the reduced $H_\bs{k}$ by finding the roots of $\det(E - \tH_{\bS;\bs{k}}(E))=0$ (which, as mentioned in \cref{sec:latSymmetry}, would generally yield a subset of the full eigenvalue spectrum).}.
As we see, there are two flat bands at $E = \pm \sqrt{2}$, corresponding to the two CLSs ``CLS1'' and ``CLS2'' depicted on the right, with odd parity on the cospectral sites $\{1,2\}$.

We would like to underline here that the constructed flat bands are \emph{independent} of the inter-cell coupling strength used in the walk multiplet interconnections.
Indeed, as evidenced by \cref{eq:zeroMultipletAmpSum}, the hopping elements $h^{\lvec}_{\bX\bY}$ connecting the lattice cells do not enter the eigenvalue problem in \cref{eq:blochEVP}.
The corresponding flat band energy $E_\nu$ is therefore unaffected by the value of the $h^{\lvec}_{\bX\bY}$, which however generally do affect the rest of the energy spectrum.
Thus, the inter-cell coupling strengths used in the walk multiplet interconnections can be flexibly tuned to modify the dispersive part of the band structure around the constructed flat bands.

The above construction of CLSs and flat bands from latent symmetry and walk multiplets can be seen as a generalization of the construction from local permutation symmetries $\varPi$ which are involutory ($\varPi^2 = I$) and leave certain sites of the unit cell fixed.
If $n$ is such a fixed site, i.\,e. $\varPi \ket{n} = \ket{n}$, then any eigenstate $\ket{\varphi}$ with odd parity under $\varPi$ has $\braket{n|\varphi} = \braket{n|\varPi^2\varphi} = - \braket{n|\varphi} = 0$, that is, has a node (vanishing amplitude) on the fixed site.
Interconnecting unit cells into a lattice by coupling such $\varPi$-fixed sites from cell to cell, any $\varPi$-odd eigenstate of the isolated unit cell yields a CLS and thus a corresponding flat band for the lattice.
This scenario constitutes a special case of the construction described in the present work (based on walk equivalent cospectral sites), where (a) the cospectral sites are related by a common permutation symmetry exchanging those sites and (b) the unit cells are interconnected via walk singlets relative to the pair of exchanged sites \footnote{Note that two sites $u,v$ related by an involutory permutation symmetry $\varPi$ are automatically cospectral, since $\braket{u|H^\p|u} = \braket{u|H^\p\varPi^2|u} = \braket{u|\varPi H^\p\varPi|u} = \braket{v|H^\p|v}$ $\forall\,\p$ (see \cref{eq:cospectrality}), and any site $c$ fixed by $\varPi$ is a walk singlet relative to $\{u,v\}$, since $\braket{u|H^\p|c} = \braket{u|H^\p\varPi|c} = \braket{u|\varPi H^\p|c} = \braket{v|H^\p|c}$ $\forall\,\p$.}.
We note that such a local exchange symmetry is, in turn, a special case of general local permutation symmetries inducing CLSs, as addressed in Ref.\,\cite{Rontgen2018_PRB_97_035161_CompactLocalizedStatesFlat} in terms of so-called \emph{equitable partitions} of graphs.
Relating that approach to latent symmetries involving site subsets $\bS$ of more than two sites is an interesting direction of further research.

We stress that in the present case (\cref{fig:graphA_bandsLattice}), the zeros of the CLSs within the unit cell (on sites $7,8$) are not induced by any permutation symmetry of the cell fixing those nodal sites, but rather by latent symmetry and walk equivalence (of the cospectral sites relative to walk singlets), as described above.

%%%%% fig: graphA_doubletInterconnection %%%%%%%%%%%%%%%%%%%%%%%%%%%%%%%%
\begin{figure}[t!]
\center
\includegraphics[width=.8\columnwidth]{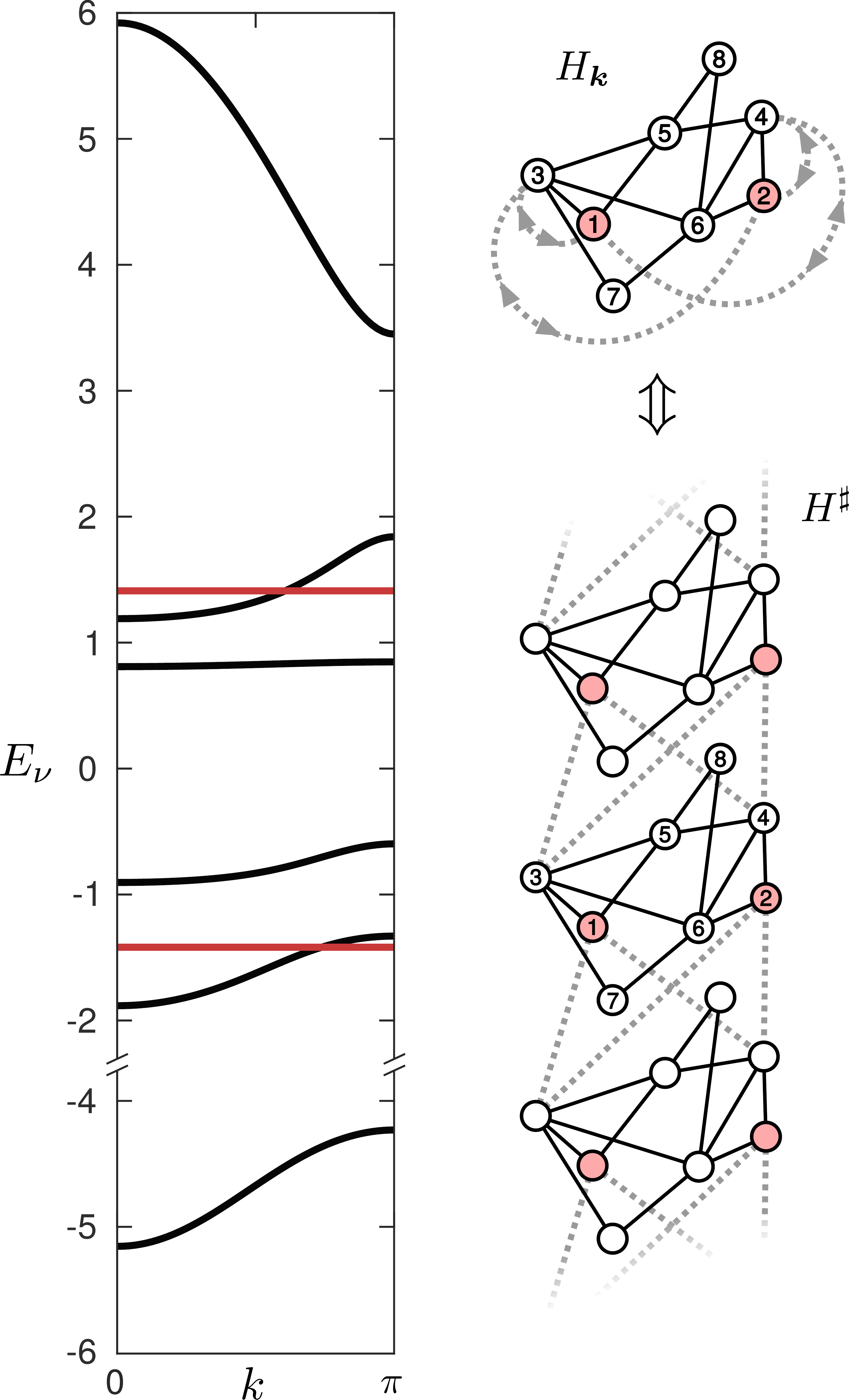}
\caption{
Band structure \textit{(left)} of a lattice constructed from $H$ in \cref{fig:graphA_reduction} as a unit cell by interconnecting its walk doublets $\{1,2\}$ and $\{3,4\}$ as indicated by dotted edges in $H_\bs{k}$ and $H^\sharp$ \textit{(right)}, with unit intra-cell couplings and inter-cell couplings $h = 2$.
The two flat bands at $E = \pm\sqrt{2}$ (red lines) correspond to the same CLSs as in \cref{fig:graphA_bandsLattice}.
Note that the band around $E \approx 0.8$ is dispersive and only looks rather flat at the used plotting scale of the $E$-axis.
}
\label{fig:graphA_doubletInterconnection}
\end{figure}
%%%%%%%%%%%%%%%%%%%%%%%%%%%%%%%%%%%%%%%%%%%%%%%%%%%%%%%%%%%%%%%%%%%%%%%%%%%%%%

For simplicity, we have applied walk singlet-to-doublet inter-cell connections in the above example (\cref{fig:graphA_cellInterconnection,fig:graphA_bandsLattice}).
It is clear from the above, however, that the procedure to generate flat bands applies naturally for any walk multiplet interconnection as inter-cell coupling; see \cref{eq:blochHamiltonianMultiplets,eq:blochEVP}.

As an example, in \cref{fig:graphA_doubletInterconnection} we start with the same graph $H$ (as in \cref{fig:graphA_cellInterconnection}) but now interconnect the walk doublets $\{1,2\}$ and $\{3,4\}$ in $H_\bs{k}$, i.\,e. each site $1,2$ to both $3,4$ with complex Hermitian couplings (including Bloch phases), and corresponding real inter-cell couplings in $H^\sharp$, as explained above.
This lattice maintains the same CLSs and flat bands as before (\cref{fig:graphA_bandsLattice}), though generally with modified dispersive bands.

In general, any Hermitian walk multiplet interconnection \ref{mod:multipletInterconnection} with complex Bloch phases, applied to a unit cell $H$, is mapped to an inter-cell connection in the lattice Hamiltonian $H^\sharp$ preserving the latent symmetry in each cell.
This allows for great flexibility in generating flat bands with a given latently symmetric prototype cell.

To summarize, the proposed flat band construction principle consists in 
\begin{enumerate}[(i)]
 \item starting with a Hamiltonian $H$ in the form of a graph having two latently exchange-symmetric, cospectral vertices $\{u,v\}$,
 \item identifying walk multiplets of $H$ relative to $\{u,v\}$, and
 \item using $H$ as the unit cell of a lattice constructed by periodically interconnecting any walk multiplet of each cell to any walk multiplet of other cells (which can be neighboring cells but also more remote ones).
\end{enumerate}
The resulting lattice $H^\sharp$ then features a flat band for each eigenstate $\ket{\varphi_{\nu}^-}$ of $H$ with odd parity on $\{u,v\}$, which becomes a macroscopically degenerate CLS in $H^\sharp$ occupying one unit cell.

%%%%% fig: graphA_bandsEvolution %%%%%%%%%%%%%%%%%%%%%%%%%%%%%%%%%%%%%%%%%%%%%
\begin{figure}[t!]
\center
\includegraphics[width=.9\columnwidth]{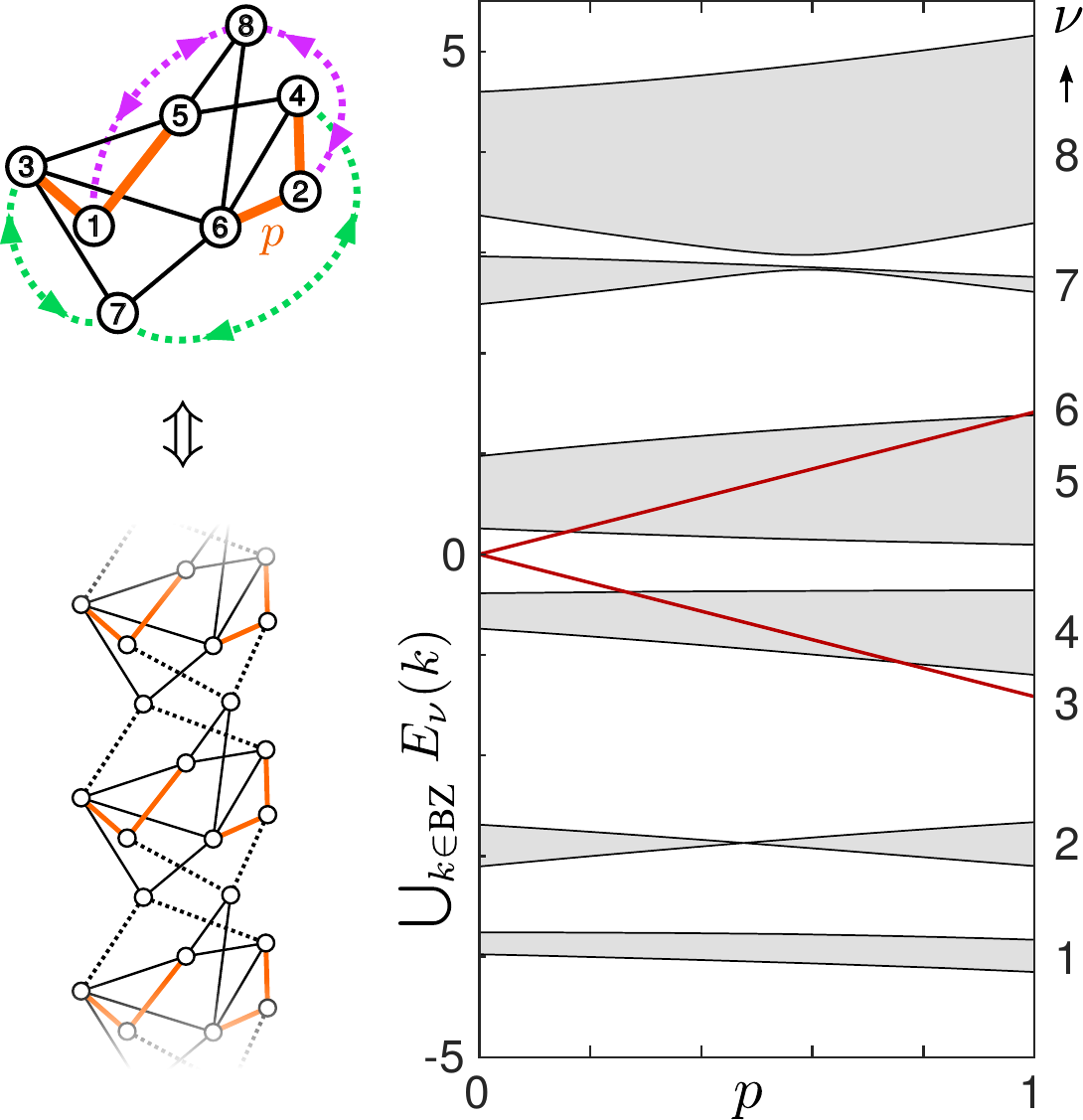}
\caption{
Band edges (black lines) and band projections $\bigcup_{k \in \textrm{BZ}} \, E_\nu(k)$ (gray shades) of the bands $E_\nu(k)$ over the Brillouin zone (BZ) for the lattice in \cref{fig:graphA_bandsLattice} with varying coupling parameter $H_{13} = H_{15} = H_{24} = H_{26} = p$ indicated (orange lines) in the schematic on the left; $p = 1$ corresponds to the band structure in \cref{fig:graphA_bandsLattice}.
The two flat bands induced by latent symmetry of sites $1,2$ occur at any $p$, with energies varying in $p$ (thick red lines).
}
\label{fig:graphA_bandsEvolution}
\end{figure}
%%%%%%%%%%%%%%%%%%%%%%%%%%%%%%%%%%%%%%%%%%%%%%%%%%%%%%%%%%%%%%%%%%%%%%%%%%%%%%

\subsection{Parametric invariance of latent symmetry flat bands}
\label{sec:parametricInvariance}

It is important to notice that the generation of CLSs and resulting flat bands from latent symmetry and walk multiplets of a graph $H$ is not restricted to a fixed set of edge weight values $H_{mn}$.
Indeed, there is a certain freedom in changing $H$'s elements \emph{parametrically} while still inducing flat bands from the same latent symmetry and walk multiplets.
Specifically, this parametrization means that there exist groups of the elements $H_{mn}$ which can be set to a common arbitrary real value per group, without breaking the given latent symmetry and selected walk multiplets.
For example, the weight parametrizations shown in \cref{fig:graphA_multiplets} preserve the cospectrality of $\{u,v\}$ as well as the multiplets interconnected to form the lattice in \cref{fig:graphA_cellInterconnection}.
Thus, when varying the weight parameters (that is, the common value of each group of elements $H_{mn}$), flat bands are still induced for the constructed lattice.
Their energy positions, however, generally depend on the weight parameters, which allows for tuning the flat bands relative to the rest of the band structure.

We demonstrate this parametric invariance of the flat bands for our navigating example graph in \cref{fig:graphA_bandsEvolution}, where the band edges for the lattice in \cref{fig:graphA_bandsLattice} are plotted for a continuous variation of selected couplings in the unit cell.
Specifically, using the cospectrality- and multiplet-preserving edge weight parametrization of \cref{fig:graphA_multiplets}\,(right), a selected subset of couplings is set to a common varying value $p$ (see \cref{fig:graphA_bandsEvolution} caption).
As we see, while the dispersive band widths vary with $p$, the flat bands constructed by latent symmetry for $p = 1$ remain flat for any $p$ (see red lines, whose vertical cross sections at any $p$ are single points at the corresponding $E_\nu$).
This is in contrast to flat bands that may appear ``accidentally'' when varying $p$, as seen e.\,g. for the second lowest band which becomes flat at a single point around $p \approx 0.4765$.

Further, in this example the upper (lower) flat band energy increases (decreases) linearly with $p$ across the dispersive bands and the gaps between them. 
This demonstrates the possibility to tune the flat band positions relative to dispersive bands without invoking any apparent symmetry of the unit cell.

For clarity, let us here underline the qualitative difference of inter-cell and intra-cell variations regarding their influence on the constructed flat bands.
The inter-cell couplings used in walk multiplet interconnections in the unit cell (the $h^{\lvec}_{\bX\bY}$ in \cref{eq:blochHamiltonianMultiplets} for each interconnected multiplet pair $\bX, \bY$) can be varied at will leaving the flat bands intact in energy.
In contrast, the intra-cell couplings must first be parametrized into groups of common values, as described above, whose variation then retains the occurrence of the flat bands but may generally alter their energy position.
Combined, those inter- and intra-cell coupling variations constitute a flexible way to design the overall band structure featuring flat bands induced by latent symmetry.

%%%%% fig: graphB_bandsLattice %%%%%%%%%%%%%%%%%%%%%%%%%%%%%%%%%%%%%%%%%%%%%%%%%%%%%%
\begin{figure}[t!]
\center
\includegraphics[width=\columnwidth]{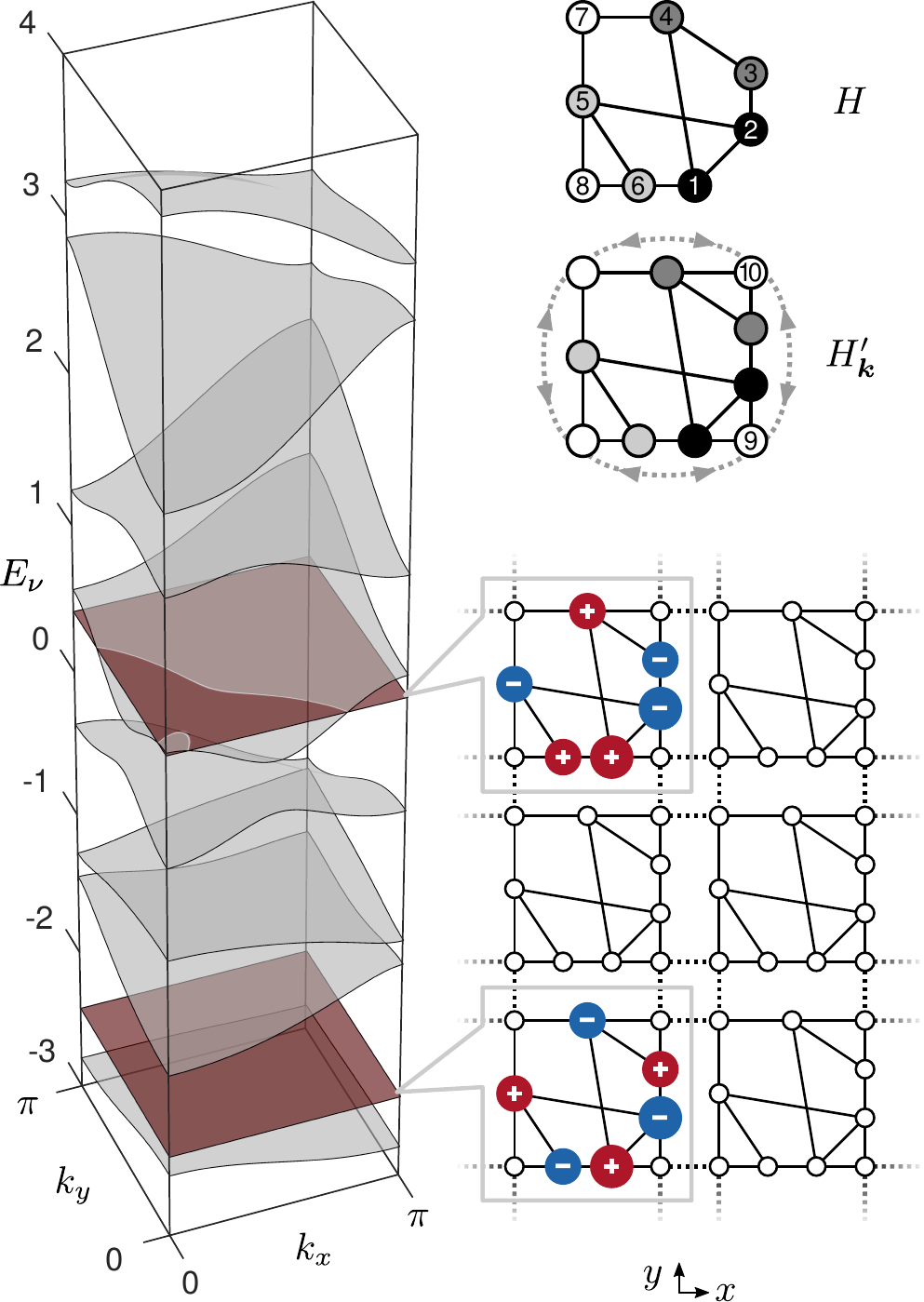}
\caption{
Band structure \textit{(left)} of a 2D lattice \textit{(bottom right)} constructed by repetition of an unweighted $8$-vertex graph $H$ (with unit intra-cell couplings) augmented by two vertices $9$ and $10$  connected to the graph's cospectral pair $\{1,2\}$ and the walk doublet $\{3,4\}$, respectively \textit{(top right)}, with dotted double-arrowed edges indicating complex couplings $h e^{\pm i k_{x(y)}L}$ in $\pm x$($y$)-direction in the Bloch Hamiltonian $H'_\bs{k}$ (see text). 
Inter-cell edges with unit weight $h = 1$ (dotted lines) connect the walk singlets $\{7,8,9,10\}$ of the graph in $x$- and $y$-direction, preserving its two CLSs (depicted in the lattice; colormap as in \cref{fig:graphA_bandsLattice}) which correspond to two flat bands at $E = \pm\sqrt{2}-1$.
}
\label{fig:graphB_bandsLattice}
\end{figure}
%%%%%%%%%%%%%%%%%%%%%%%%%%%%%%%%%%%%%%%%%%%%%%%%%%%%%%%%%%%%%%%%%%%%%%%%%%%%%%

\subsection{Flat bands via walk singlet augmentation}
\label{sec:singletAugmentation}

Another variation of using the graph modifications in \cref{sec:modifications} for flat band construction is to first modify a latently symmetric graph $H$ itself, before interconnecting it into a lattice.
In particular, using \ref{mod:multipletExtension} we can augment $H$ by connecting new vertices to walk multiplets relative to a cospectral pair $\{u,v\}$.
In the resulting graph $H'$, each such new vertex $c'$ will be a walk singlet, which will in turn have vanishing amplitude in any non-degenerate eigenvector with odd parity on $\{u,v\}$; see \cref{eq:zeroMultipletAmpSum}.
This `singlet augmentation' may be used, e\,g., to bring a given unit cell into a more preferable shape for connection into a lattice.

We demonstrate this procedure by constructing a 2D flat band lattice in \cref{fig:graphB_bandsLattice}.
The original $8$-vertex graph $H$ (upper right of figure) has four doublets relative to the cospectral pair $\{u,v\} = \{1,2\}$, with one of them further
consisting of the two singlets $\{7\}$, $\{8\}$.
The graph has two eigenvectors with odd $\{u,v\}$-parity which vanish on those singlets.
Note that, like the graph in \cref{fig:graphA_multiplets}, also this one can be parametrized in its edge weights while keeping its latent symmetry and corresponding compact eigenvectors, as we will see below.
For simplicity, we first keep its unweighted version.
We now connect two new vertices $9$ and $10$ to two doublets using modification \ref{mod:multipletExtension}, which thus yields two more singlets on which the previous compact eigenvectors also vanish.
Then, we connect the new graph $H'$ into a 2D lattice---similarly to the procedure in \cref{sec:multipletInterconnection}---via its four corner singlet vertices, as shown, described by the corresponding Bloch Hamiltonian $H'_\bs{k}$.
The resulting band structure $E_\nu(\bs{k})$ features two flat bands at $E = \pm \sqrt{2} - 1$, with the corresponding CLSs depicted in two unit cells of the lattice.

%%%%% fig: graphB_bandsEvolution %%%%%%%%%%%%%%%%%%%%%%%%%%%%%%%%%%%%%%%%%%%%%
\begin{figure}[t!]
\center
\includegraphics[width=\columnwidth]{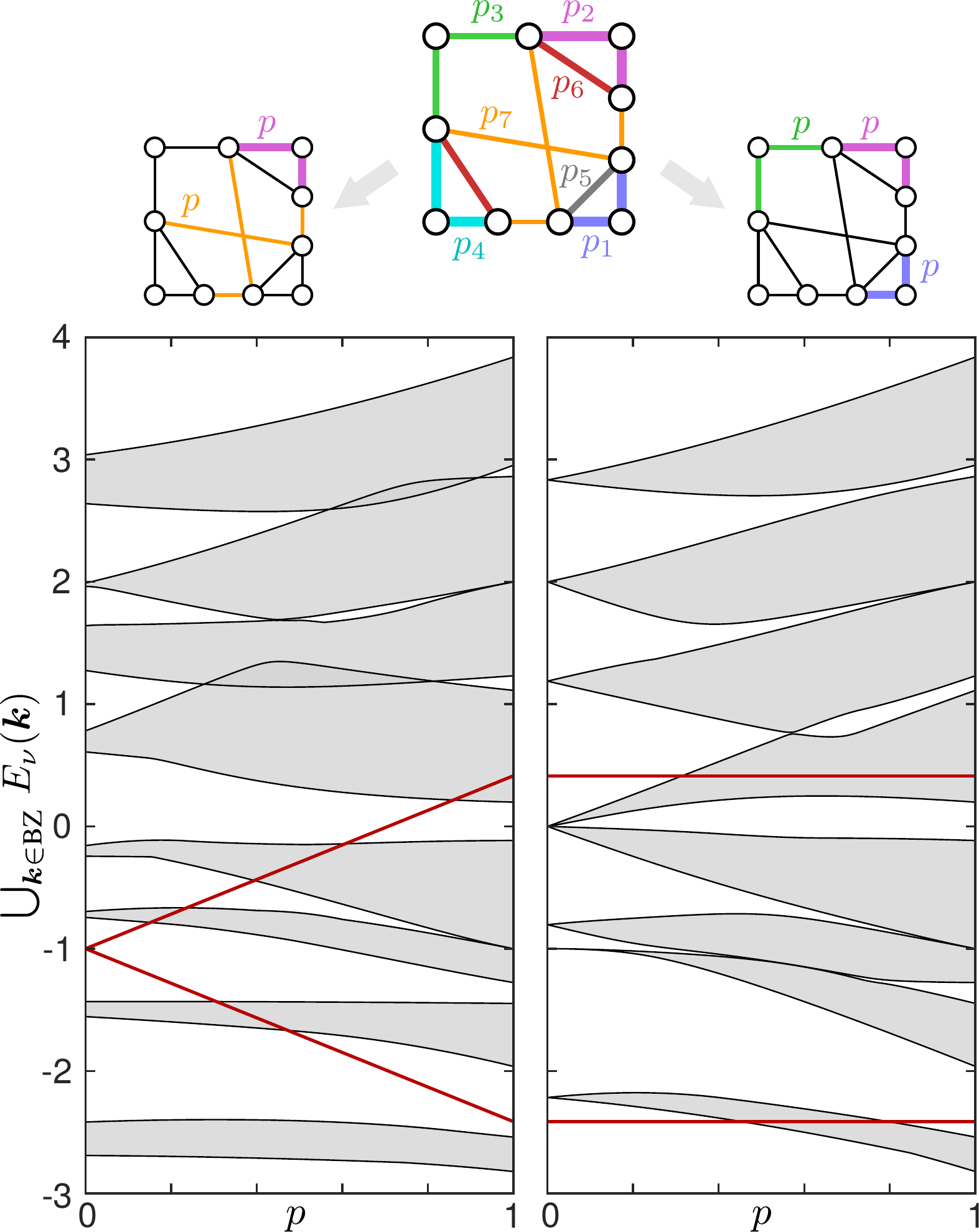}
\caption{
Band edges and projections as in \cref{fig:graphA_bandsEvolution} but for the 2D bands of the lattice in \cref{fig:graphB_bandsLattice} for varying parameter $p$ in two different cases of the edge weight parametrization as shown at the top; $p = 1$ corresponds to the band structure in \cref{fig:graphB_bandsLattice}.
}
\label{fig:graphB_bandsEvolution}
\end{figure}
%%%%%%%%%%%%%%%%%%%%%%%%%%%%%%%%%%%%%%%%%%%%%%%%%%%%%%%%%%%%%%%%%%%%%%%%%%%%%%

We emphasize that the CLSs are induced by the latent symmetry of the site pair $\{1,2\}$, and not by a permutation symmetry of the lattice cell.
Specifically, the cell is indeed reflection symmetric about one diagonal (the line passing through sites $7$ and $9$), and the CLSs are odd under this reflection with nodes on this diagonal, as expected (recall discussion on permutation symmetry $\varPi$ in \cref{sec:multipletInterconnection}).
This symmetry does not explain, however, the other two CLS nodes at sites $8$ and $10$.
Each of those are instead fixed under the latent symmetry operation $Q$ (see \cref{app:Qmatrix}) induced by the cospectrality between $u,v$.
In fact, a general weight parametrization preserving the walk multiplet structure violates the cell's reflection symmetry, though retains the compactness of the CLSs, that is, their nodes on the singlet sites, and the corresponding flat bands.

The latter is demonstrated in \cref{fig:graphB_bandsEvolution}, where a cospectrality- and multiplet-preserving parametrization of the edge weights by real parameters $p_{i=1,2,\dots,7}$ is considered (top panel).
Parametrization of the onsite elements, or loops, is also possible but not shown for simplicity.
The band edge evolution for two parametrical variations is plotted.
In the first case (left plot), we set $p_2 = p_7 = p$ (other intracell hoppings to unity) and vary $p$, whereby the flat bands (red lines) are preserved with linearly varying energy.
In the second case (right plot), we set $p_1 = p_2 = p_3 = p$ (other intracell hoppings again equal unity), whose variation modifies the dispersive bands but leaves the flat band energies fixed.
We thus see that such parametrizations of the unit cell Hamiltonian preserving its latent symmetry, together with the chosen inter-cell couplings (whose variation, not shown here, evidently also preserves the latent symmetry), can be used to tune the induced flat bands flexibly in relation to the surrounding band structure.

Finally, we note that in this example we interconnected the unit cells via their corner walk singlets for simplicity. 
One could instead, or additionally, interconnect larger multiplets between the cells, still preserving the same CLSs and concomitant flat bands---though generally changing the dispersive bands.
For example, the walk doublet $\{3,4\}$ (see \cref{fig:graphB_bandsLattice}) of the cell at each $\lvec$ could be connected diagonally in the lattice to the doublet $\{5,6\}$ of the cell at $\lvec + L\hat{\bs{x}} + L\hat{\bs{y}}$.

\section{Discussion} 
\label{sec:discussion}

Having demonstrated how latent symmetry, in combination with walk multiplets, may be employed to induce flat bands, let us now discuss some aspects and extensions of the presented framework.

\subsection{Number and spatial extension of CLSs}

In each of the above examples, \cref{fig:graphA_bandsLattice,fig:graphB_bandsLattice}, there were two CLSs per unit cell associated with a cospectral site pair $\{u,v\}$ in $H$.
The number of such CLSs depends on the structure of the graph used as a cell.
Specifically, the number of eigenstates of $H$ with odd $\{u,v\}$-parity is given by the dimension of the Krylov subspace generated by the vector $\ket{-} \equiv \ket{u}-\ket{v}$ \cite{Eisenberg2019_DM_342_2821_PrettyGoodQuantumState}, that is, the rank of the corresponding Krylov matrix $[\ket{-},H\ket{-},H^2\ket{-},\dots,H^{N-1}\ket{-}]$.
Also, there may be more than one cospectral pair in the graph $H$, each of which may induce different CLSs in a corresponding multiplet-interconnected lattice.
Of course, such latently symmetric cospectral pairs may further coexist with cospectral pairs corresponding to permutation symmetries swapping only two vertices $u,v$.
Clearly, for such pairs \emph{all} other sites in $H$ are walk singlets, with corresponding CLSs confined to $\{u,v\}$ in each lattice cell.
In the examples shown here, we have chosen cell graphs having only latent symmetries for clarity.

Note, further, that in the above flat band construction scheme (see \cref{sec:multipletInterconnection}) we have explicitly considered the original graph $H$ (or some augmented one, see \cref{sec:singletAugmentation}) as the  unit cell of the generated lattice $H^\sharp$.
The induced CLSs then occupy $U = 1$ unit cell each, using the number of occupied unit cells $U$ as a flat band classifier \cite{Maimaiti2019_PRB_99_125129_Universald1FlatBand} (recently generalized accordingly for lattice dimensions $d > 1$ \cite{Maimaiti2021_PRB_103_165116_FlatbandGeneratorTwoDimensions}).
One could in principle, however, start with a supercell $H'$ of a target lattice $H^\sharp$, consisting of $U > 1$ interconnected copies of $H$, and look for new cospectral pairs $\{u',v'\}$ which are not cospectral in $H$. 
Then, CLSs induced by $\{u',v'\}$-odd eigenstates of $H'$ will generally occupy $U > 1$ primitive unit cells within the supercell.
The key challenge here would be to design inter-$H$ connections which coincide with walk multiplet interconnections between supercells $H'$.
We leave this endeavour for future work. 

\subsection{Generalizations of walk multiplets}
\label{sec:generalizedMultiplets}

The concept of walk multiplets can be generalized \cite{Morfonios2021_LAaiA_624_53_CospectralityPreservingGraphModifications} by replacing the indicator vector of $\bM$ in \cref{eq:walkmatrix} with a nonuniform version $\ket{e^\gamma_\bM}$, with a tuple $\gamma$ of generally different amplitudes 
\begin{equation} \label{eq:weightedIndicator}
 \gamma_m = \braket{m|e^\gamma_\bM}
\end{equation}
for $m \in \bM$ and $0$ otherwise.
$\gamma_m = 1$, up to a global factor, corresponds to the uniform walk multiplets considered so far.
If a new vertex $c'$ is connected to $\bM$ via those weights $\gamma_m$, then the associated cospectrality is preserved and $c$ becomes a walk singlet if \cref{eq:walkequivalent} is fulfilled---now with the walk matrix generated by $\ket{e^\gamma_\bM}$.
In other words, the modification \ref{mod:multipletExtension} of \cref{sec:modifications} is generalized to such nonuniform walk multiplets, as is, similarly, the multiplet-interconnection \ref{mod:multipletInterconnection}; see \cref{app:modifications}.
A particular case is that of \emph{overlapping} uniform multiplets $\bM_\mu$ ($\mu = 1,2,\dots$), whose union yields a nonuniform multiplet with indicator vector $\ket{e^\gamma_{\bM}} = \sum_\mu \ket{e_{\bM_\mu}}$, where $\ket{e_{\bM_\mu}}$ is the usual indicator vector of multiplet $\bM_\mu$.
An example is schematically shown as overlapping multiplets $\bX$ and $\bY$ in \cref{fig:cospectralGraphModifications}.

Another variation is to consider walk \emph{anti-equivalence} by replacing \cref{eq:walkequivalent} with $[W_\bM]_{u*} = - [W_\bM]_{v*}$.
In this case, $\bM$ is a walk \emph{anti-multiplet} relative to $\{u,v\}$ and the role of parity is swapped:
Now the eigenvectors $\ket{\varphi_\nu^+}$ with \emph{even} parity on $\{u,v\}$ become CLSs, with vanishing amplitudes on anti-singlets \cite{Morfonios2021_LAaiA_624_53_CospectralityPreservingGraphModifications}.  
These generalizations of the concept of walk multiplets offer an even larger flexibility in generating flat band lattices from graphs with latent symmetries.

\subsection{Occurrence and construction of latently symmetric graphs}
\label{sec:occurrenceOfLatSym}

In all of the above, we have assumed that the original graph $H$ is latently exchange-symmetric, that is, features some pair of vertices $u$ and $v$ which are cospectral but not exchange-symmetric in $H$.
We also assumed the given graph to feature some walk multiplets relative to $\{u,v\}$.
The aim was to show how these properties, when given, can be used instead of common symmetries---that is, permutation operations commuting with $H$---to induce CLSs and corresponding flat bands for periodic lattice structures.

The systematic construction of latently symmetric graphs is far from trivial.
To date, and to the best of our knowledge, there is indeed no general procedure for constructing undirected, latently symmetric graphs; it is rather a subject of ongoing research.
One approach is based on ``unpacking'' the isospectrally reduced form of a graph \cite{Kempton2020_LAaiA_594_226_CharacterizingCospectralVerticesIsospectral}, by applying partial fraction decomposition to its functional dependence on the eigenvalue $E$, and then accordingly constructing a generally directed graph with complex weights.
Another recent, semi-empirical approach \cite{Rontgen2020_PRA_101_042304_DesigningPrettyGoodState}, starts from a graph with trivially cospectral vertices---that is, induced by some permutation symmetry---which is then modified by adding vertices and edges such that the permutation symmetry is broken while the cospectrality is not. 

In fact, the defining property of vertex cospectrality, \cref{eq:cospectrality} evaluated up to $\p = N-1$, makes it straightforward to resort to numerical iteration for verifying it.
In this spirit, Ref.\,\cite{Rontgen2020_PRA_101_042304_DesigningPrettyGoodState} reports on the occurrence of latently symmetric graphs out of all possible unweighted graphs of given small size.
Specifically, we have created a database of all unweighted graphs (adjacency matrices) of size up to $N=11$ which have at least one cospectral vertex pair and \emph{no} permutation symmetry.
For $N \leqslant 7$, there is no such graph.
For $N = 8,9,10,11$, there are $78, 2\,247, 78\,489, 3\,714\,397$ such graphs, respectively.
Although this is, in each case, a small portion ($\approx 7.0, 8.6, 10.0, 4.7$ $\permil$) of all possible graphs, the analysis shows that there is a substantial number of latently symmetric unweighted graphs even for such small sizes.
This means that latent symmetry would in principle not be hard to design in a targeted setup, consulting e.\,g. the above database.

For larger graphs ($N \gg 1$), there is numerical evidence that the occurrence of latent symmetries is correlated to that of common permutation symmetries, in the sense that their percentage has been found to follow the same trend when varying a structural parameter for a class of randomly generated graphs, as stated in Ref.\,\cite{Bunimovich2019_AMNS_4_231_FindingHiddenStructuresHierarchies}.
The exact reason for this behavior is an open question.

In the same manner as cospectral vertices, we identify walk multiplets relative to a cospectral pair $\{u,v\}$ of a given graph by scanning through all vertex subsets of all possible sizes for those that fulfill \cref{eq:walkequivalent}.
For the graphs available in the above database, we have observed that, typically, the graphs have multiple walk multiplets (relative to the featured cospectral pair(s)) for each multiplet size---although there are e.\,g. cases where walk singlets are absent---with the number of multiplets typically increasing with their size.
Also, there is always at least one walk doublet, namely the cospectral pair itself.
The walk multiplet structure is further enriched by considering their generalized version (nonuniform and anti-multiplets, see \cref{sec:generalizedMultiplets} above), as described in detail in Ref.\,\cite{Morfonios2021_LAaiA_624_53_CospectralityPreservingGraphModifications}.
The relation of general walk multiplets to the structure of eigenvectors of graphs with cospectral vertices is an interesting topic to be pursued.

\section{Conclusions}
\label{sec:conclusion}

We have shown how flat bands can be induced by \emph{latent symmetry} between a pair of sites in the unit cells of discrete lattices.
This symmetry is revealed as an exchange permutation symmetry of the effective Hamiltonian upon reduction of the cell over the site pair subsystem, and imposes odd or even local parity of the original Hamiltonian eigenstates on those two sites.
Using recent concepts and tools from graph theory, where latent symmetry takes the form of \emph{cospectrality} between two vertices, we propose a framework for generating flat bands from the structural properties of graphs lacking permutation symmetries.
The key ingredient is the occurrence of \emph{walk equivalence} of cospectral vertices relative to vertex subsets called \emph{walk multiplets}.
This signifies a collective symmetry between possible walks along the edges of a graph from its cospectral vertices to a given walk multiplet, expressed in terms of corresponding \emph{walk matrices}.
Crucially, the amplitude sum on walk multiplets vanishes for any non-degenerate eigenvector with odd parity on cospectral vertices.

When connecting the graph as a unit cell into a lattice via its walk multiplets, those eigenvectors constitute compact localized states (CLSs) forming flat bands within an otherwise dispersive band structure.
We illustrate the scheme for 1D and 2D lattices using simple graphs with cospectral sites.
A generalization to more complex cell geometries, possibly with multiple latent symmetries, and to higher-dimensional lattices is straightforward.
As we demonstrate, the latent symmetry persists over flexible parametrizations of the lattice Hamiltonian elements, making the induced flat bands systematically tunable.
This should allow for a feasible generation of flat bands from latent symmetries in various realization platforms such as, e.\,g., photonic waveguide arrays or electric circuit networks, with tailored inter-site connections.
We thus offer a fundamental insight into a class of CLSs originating from hidden Hamiltonian symmetries, which may also provide a valuable tool in designing flat band setups.

\acknowledgments

We thank Jens Kwasniok for helpful discussions regarding $Q$-matrices and walk multiplets.
Funding by the Deutsche Forschungsgemeinschaft under grant DFG Schm 885/29-1 is gratefully acknowledged.
M.\,P. is thankful to the `Studienstiftung des deutschen Volkes' for financial support in the framework of a scholarship.

\appendix

\section{Cospectrality from walk matrices and orthogonal symmetry}
\label{app:Qmatrix}

We here give a brief account on the orthogonal symmetry matrix $Q$ describing vertex cospectrality.
The purpose is to provide an insightful connection between the latent symmetry of a graph, upon reduction over two cospectral vertices, and the underlying symmetry operation exchanging those vertices in the original graph.
The description is adapted from Ref.\,\cite{Godsil2012_AC_16_733_ControllableSubsetsGraphs} to a graph with $N$ vertices $\bH$ and symmetric weighted adjacency matrix $H$.

First, consider two arbitrary subsets $\bU, \bV \subseteq \bH$ with walk matrices 
\begin{equation} \label{eq:walkmatrices}
 W_\bX = [~\ket{e_\bX}, H \ket{e_\bX}, \dots, H^{N-1} \ket{e_\bX}~]
\end{equation}
for the indicator vectors $\ket{e_\bX}$, $\bX = \bU,\bV$.
If $W_\bV$ is invertible (that is, has full rank $N$), then the matrix
\begin{equation} \label{eq:genQMatrix}
 Q_{\bU\bV} =  W_\bU W_\bV^{-1}
\end{equation}
commutes with $H$, thus representing a general symmetry transformation.
To see this, recall that 
\begin{equation}
 H^N =  \sum_{\p=0}^{N-1} c_\p H^\p 
\end{equation}
by the Cayley-Hamilton theorem which states that $H$ fulfills its own characteristic equation $\chi(x) = \sum_{\p=0}^N a_\p x^\p = 0$, where $c_\p = -a_\p/a_N$.
Therefore, we have that
\begin{equation}
 H W_\bX = W_\bX C, \quad \bX = \bU,\bV
\end{equation}
where
\begin{equation}
C = 
\begin{bmatrix}
0        &     0    & \dots   &  0     &  c_0      \\
1        &     0    & \dots   &  0     &  c_1      \\
0        &     1    & \dots   &  0     &  c_2      \\
\vdots   &  \vdots  & \ddots  & \vdots & \vdots    \\
0        &     0    & \dots   &  1     &  c_{N-1} 
\end{bmatrix}
\end{equation}
is the companion matrix \cite{Meyer2000_MatrixAnalysisAppliedLinear} for $H$.
Thus,
\begin{align}
 H Q_{\bU\bV} &= W_\bU C W_\bV^{-1} \nonumber \\ &= W_\bU (W_\bV^{-1}H W_\bV) W_\bV^{-1} = Q_{\bU\bV} H.
\end{align}
Further, if both $W_\bU$ and $W_\bV$ are invertible and fulfill 
\begin{equation} \label{eq:cospectralityWalkmatrixSubsets}
 W_\bU^\top W_\bU= W_\bV^\top W_\bV,
\end{equation}
then $Q_{\bU\bV}$ is orthogonal:
\begin{align}
 Q_{\bU\bV}^\top 
 &= [W_\bV^{-1}]^\top W_\bU^\top = \nonumber \\   
 &= [W_\bV^{-1}]^\top W_\bV^\top W_\bV W_\bU^{-1} = 
 W_\bV W_\bU^{-1} =
 Q_{\bU\bV}^{-1}.
\end{align}
With invertible $W_\bX$ ($\bX =\bU$ or $\bV$), $H$ has simple eigenvalues \cite{Godsil2012_AC_16_733_ControllableSubsetsGraphs} $E_\nu$ (no degeneracies) and then, because $Q_{\bU\bV}$ commutes with $H$, it is a polynomial in $H$.
Thus, if $H$ is symmetric, so is $Q_{\bU\bV}$, and since it is also orthogonal, we obtain that
\begin{equation} \label{eq:symmetricQ}
 Q_{\bU\bV} = Q_{\bU\bV}^\top = Q_{\bU\bV}^{-1}.
\end{equation}

Now, if $\bU = \{u\}$ and $\bV = \{v\}$ constitute two cospectral vertices, we have \cite{Godsil2017_AM_StronglyCospectralVertices} % \,[Theorem 1.3 (f)] 
\begin{equation} \label{eq:cospectralityWalkmatrix}
 W_u^\top W_u = W_v^\top W_v.
\end{equation}
Then, if both $W_u$ and $W_v$ are invertible, \cref{eq:symmetricQ} holds for $Q_{\{u\},\{v\}} \equiv Q$ of the main text, that is, $Q^2 = Q^\top Q = I$.
In particular, since $Q W_v = W_u$ and $QH = HQ$, we have that $Q \ket{v} = \ket{u}$ and $Q \ket{u} = \ket{v}$.
Thus, being also orthogonal, $Q$ is block-diagonal with one block being the antidiagonal matrix 
\begin{equation} \label{eq:reflectionMatrix}
J_\bS = 
\begin{bmatrix}
0  & 1 \\
1  & 0 
\end{bmatrix}
\end{equation}
swapping $u$ and $v$ in $\bS=\{u,v\}$.

Further, for a walk multiplet $\bM$ the condition $[W_\bM]_{u*} = [W_\bM]_{v*}$, \cref{eq:walkequivalent},  yields $\braket{u|H^\p|e_\bM} = \braket{v|H^\p|e_\bM} = \braket{v|H^\p Q^\top Q|e_\bM} = \braket{u|H^\p Q|e_\bM}$ for $\p = 0, \dots, N-1$, where we used $H = H^\top$, $QH = HQ$, and $Q\ket{v} = \ket{u}$.
Since $W_u$ has full rank, the $N$ columns $H^\p \ket{u}$ span an $N$-dimensional column space, meaning that $Q\ket{e_\bM} = \ket{e_\bM}$ (both vectors have equal projections in all $N$ dimensions).
Thus, if $H$ has a vertex subset $\bF$ consisting of walk singlets, then $[W_s]_{u*} = [W_s]_{v*}$ $\forall s \in \bF$, so another block of $Q$ is the $|\bF| \times |\bF|$ unit matrix $I_\bF$ leaving the singlet vertices fixed (causing the odd-$\{u,v\}$-parity eigenvectors to vanish on them).

The remaining orthogonal block $Q_{\bO}$ operates on the remaining vertices within $\bO = \bH \setminus (\bS \cup \bF)$, transforming the corresponding rows of $W_v$ into those of $W_u$: 
$[W_u]_{\bO,*} = Q_{\bO}  [W_v]_{\bO,*}$.
With vertices labeled accordingly, $Q$ thus has the form
\begin{equation} 
Q = J_\bS ~\oplus~ I_\bF ~\oplus~ Q_{\bO}.
\end{equation}
As an example, for the graph of \cref{fig:graphA_reduction} the $|\bO| \times |\bO|$ block is given by
$Q_{\bO} = \frac{1}{2}
\scriptsize
\begin{bmatrix}
A  & B \\
B  & A 
\end{bmatrix}$,
with 
$A = 
\scriptsize
\begin{bmatrix}
1  & 1 \\
1  & 1 
\end{bmatrix}$
and
$B = 
\scriptsize
\begin{bmatrix}
-1  & 1 \\
1  & -1 
\end{bmatrix}$,
where $\bO = \{3,4,5,6\}$.

Notice here that, since $Q$ commutes with $H$, its eigenvector matrix block-diagonalizes $H$ accordingly under similarity transformation.
Such a transformation can be seen as reminiscent of the ``Fano detangling'' procedure of Ref.\,\cite{Flach2014_E_105_30001_DetanglingFlatBandsFano}, though here for a cospectral site pair $\{u,v\}$ (instead of a single site) and determined from the walk structure of $H$.

If the spectrum of $H$ is degenerate or has any eigenvector with vanishing amplitudes on $\{u,v\}$, then $W_u$ and $W_v$ do not have full rank \cite{Godsil2017_AM_StronglyCospectralVertices,Liu2019_AM_UnlockingWalkMatrixGraph} and are thus not invertible.
Hence, although a $Q$ matrix still exists, which is unique under the convention of treating eigenvectors vanishing on $\{u,v\}$ as $\{u,v\}$-even \cite{Godsil2017_AM_StronglyCospectralVertices}, it cannot be obtained directly from \cref{eq:genQMatrix}
\footnote{One may then attempt to apply the modification \ref{mod:multipletInterconnection} (see \cref{sec:modifications}) to lift the degeneracy and to remove the $\{u,v\}$-vanishing eigenstate(s), without altering the graph's multiplet structure (the simplest modification being an added loop of arbitrary weight to any walk singlet relative to $\{u,v\}$).
The $Q$-matrix can then be obtained from \cref{eq:genQMatrix}.
However, rigorously showing that this $Q$ is the same as for the unmodified graph requires a more general account on walk multiplets, to be given elsewhere.
}.

Alternatively, the following expression can be used for a $Q$-matrix (obeying, $Q^2 = Q^\top Q = I$ and $Q \ket{u} = \ket{v}$) \cite{Rontgen2021_PRL_126_180601_LatentSymmetryInducedDegeneracies}:
\begin{equation} \label{eq:QmatrixSpectral}
 Q = P_+ - P_- = I - 2P_-,
\end{equation}
where $P_\pm = \sum_\nu \ket{\varphi_{\nu}^\pm}\bra{\varphi_{\nu}^\pm}$ is the projector onto eigenvectors with $\pm$-parity on $\{u,v\}$ ($P^+$ also including eigenvectors vanishing on $\{u,v\}$), chosen in case of degeneracy such that there is at most one eigenvector of each parity non-vanishing on $\{u,v\}$ for any given eigenvalue.
This expression is not directly derived from the structure of the graph (specifically, its walk matrices $W_u, W_v$) but rather invokes the spectral properties of $H$---that is, one first needs to find its eigenvectors.

\section{General cospectrality-preserving graph extensions and intraconnections}
\label{app:modifications}

We here show that the cospectrality of a pair $\{u,v\}$ and the walk multiplets relative to it are preserved by the modifications \ref{mod:singletExtension}, \ref{mod:multipletExtension}, and \ref{mod:multipletInterconnection} listed in \cref{sec:modifications}.
Like in Ref.\,\cite{Morfonios2021_LAaiA_624_53_CospectralityPreservingGraphModifications}, the modifications are now stated in a more general form for nonuniform walk multiplets with weighted indicator vector $\ket{e^\gamma_\bM}$ (see \cref{sec:discussion}).

For an original weighted adjacency matrix of a graph $H$, the modified one $H'$ will have the form of a sum 
\begin{equation} \label{eq:graphModified}
 H' = A + B
\end{equation}
with 
\begin{equation} \label{eq:subgraphBlocks}
 A = H \oplus G
\end{equation}
generally being a block-diagonal matrix (including the case of absent or $0 \times 0$ block $G$) and 
\begin{equation} \label{eq:interconnectionMatrix}
 B = \ket{b_1}\bra{b_2} + \ket{b_2}\bra{b_1}
\end{equation}
being a symmetric sum of rank-one coupling matrices.
Setting the $\ket{b_{1,2}}$ to be site subset indicator vectors below, $B$ will express the interconnection of those subsets in the modified graph $H'$.

The powers of $H'$, appearing in the corresponding modified walk matrices $W'_\bM$, are given by
\begin{equation} \label{eq:matrixSumPower}
 [A + B]^\p = \sum_{p = 0}^\p \sum_{\pi(A,B)} \{A^{\p-p}B^p\},
\end{equation}
where $\sum_{\pi(A,B)} \{A^{\p-p}B^p\}$ denotes the sum of all distinct permutations of $A$'s and $B$'s in matrix products with $\p-p$ $A$'s and $p$ $B$'s;
for instance, $AAB + ABA + BAA$ for $\p=3$, $p=1$.
$H^{\prime\p}$ is thus generally a weighted sum of products of the matrices $H^{\p-p} \oplus G^{\p-p}$, $[\ket{b_1}\bra{b_1}]^{n_1}$, $[\ket{b_2}\bra{b_2}]^{n_2}$, $[\ket{b_1}\bra{b_2}]^{n_3}$, $[\ket{b_2}\bra{b_1}]^{n_4}$ with $p \in \{0,1,\dots,\p\}$ and $n_i \in \{0,1,\dots,p\}$.

In the following, we briefly prove preservation of cospectrality and walk multiplets under modifications \ref{mod:singletExtension}, \ref{mod:multipletExtension}, \ref{mod:multipletInterconnection}, which are depicted schematically in \cref{fig:cospectralGraphModifications}.

\subsection{Singlet extension \ref{mod:singletExtension}} \label{app:singletExtension}

For a singlet $c$ ($\neq u,v$) of $H$ connected symmetrically---i.\,e., so that $H'$ is symmetric---to an arbitrary graph $G$ with vertices $\bG$, we have $\ket{b_1} = \ket{c}$, which is the indicator vector of $c$ in $H'$, and $\ket{b_2} = \ket{e_\bG^\gamma}$, which is the arbitrarily weighted indicator vector of $\bG$, in \cref{eq:interconnectionMatrix}.

From \cref{eq:graphModified,eq:matrixSumPower}, elements $[H^{\prime\p}]_{uu} = \braket{u|H^{\prime\p}|u}$ thus only have contributions involving $\ket{u}$ in factors $[H^q]_{uu}$ and $[H^q]_{uc}$,$[H^q]_{cu}$ for different powers $q$.
For instance, with $\braket{c|e_\bG^\gamma} = 0$, we have 
$A^2 B^2 A = A^2 (\braket{e_\bG^\gamma |e_\bG^\gamma} \ket{c}\bra{c} + \ket{e_\bG^\gamma}\bra{e_\bG^\gamma})A$,
whose $uu$-element becomes
$[A^2 B^2 A]_{uu} = [H^2]_{uc} \braket{e_\bG^\gamma |e_\bG^\gamma} H_{cu}$.

Since $\{u,v\}$ are cospectral in $H$ and $c$ is a walk singlet, those factors $[H^q]_{uu}$,$[H^q]_{uc}$,$[H^q]_{cu}$ remain equal under the replacement $u \to v$, as do, trivially, factors not containing the index $u$.
This yields $[H^{\prime\p}]_{uu} = [H^{\prime\p}]_{vv}$, so $\{u,v\}$ remain cospectral in $H'$.

Similarly, walk matrix elements $[W'_\bM]_{u\p}$ for any walk multiplet $\bM$ of $H$ only have contributions involving $\ket{u}$ in factors $[W_\bM]_{uq}$ and $[H^q]_{uc}$.
Thus, since $[W_\bM]_{uq} = [W_\bM]_{vq}$ ($\bM$ walk multiplet in $H$) and $[H^q]_{uc} = [H^q]_{vc}$ ($c$ walk singlet in $H$), we have $[W'_\bM]_{u*} = [W'_\bM]_{v*}$, that is, $\bM$ is a walk multiplet also in $H'$.

\subsection{Multiplet extension \ref{mod:multipletExtension}}

For a walk multiplet $\bM$ of $H$ connected symmetrically to a single vertex $c'$ of an arbitrary graph $G$, we have $\ket{b_1} = \ket{c'}$, which is the indicator vector of $c'$ in $H'$, and $\ket{b_2} = \ket{e_\bM}$ in \cref{eq:interconnectionMatrix}.
With similar arguments as in \cref{app:singletExtension} above, again we get $[H^{\prime\p}]_{uu} = [H^{\prime\p}]_{vv}$ and $[W'_\bX]_{u*} = [W'_\bX]_{v*}$ for any walk multiplet $\bX$ of $H$.

\subsection{Multiplet interconnection \ref{mod:multipletInterconnection}}

If two disjoint walk multiplets $\bX$ and $\bY$ of $H$ are symmetrically and fully interconnected---that is, each vertex of one is connected to all of the other, with weights added to any already existing connection---we have $A = H$ in \cref{eq:subgraphBlocks}, with $G$ now being absent, and $\ket{b_1} = \ket{e_\bX}$, $\ket{b_2} = \ket{e_\bY}$ in \cref{eq:interconnectionMatrix}.
With similar arguments as in \cref{app:singletExtension}, cospectrality of the pair $\{u,v\}$ and any walk multiplet $\bM$ relative to it are preserved in $H'$.
Using the same form of the interconnection matrix $B$, this also holds if $\bX$ and $\bY$ overlap, that is, have common vertices.

\section{Symmetry versus Hermiticity} \label{app:hermiticity}

In this appendix we briefly comment on the relation between vertex cospectrality and latent symmetry when considering a complex Hermitian---as opposed to a real symmetric---Hamiltonian $H$.
Note that, for complex Hermitian $H$, the cospectrality condition for a pair $\{u,v\}$  in terms of walk matrices, \cref{eq:cospectralityWalkmatrix}, is replaced with
\begin{equation} \label{eq:cospectralityWalkmatrixSelfadjoint}
 W_u^\dagger W_u = W_v^\dagger W_v,
\end{equation}
with $(~)^\dagger = (~)^{\top *}$ denoting Hermitian conjugation.

It was recently shown \cite{Kempton2020_LAaiA_594_226_CharacterizingCospectralVerticesIsospectral} that cospectrality of a vertex pair $\{u,v\}$ of a graph $H$ is equivalent to latent symmetry between $u$ and $v$---that is, the $2 \times 2$ reduction $\tH_{\{u,v\}}$ of $H$ is bisymmetric---if $H$ is symmetric, that is, its graph is undirected.
Therefore, to relate vertex cospectrality to latent symmetry, we have assumed a symmetric unit cell Hamiltonian matrix $H$, which was also chosen real to generally possess a real eigenvalue spectrum.

Nevertheless, if $H$ is modified into a Bloch Hamiltonian $H_\bs{k}$ exclusively by interconnecting walk multiplets with self-adjoint complex weights (a special case of a directed graph; see $H_\bs{k}$ in \cref{fig:graphA_cellInterconnection} with each dotted line indicating complex conjugate weights $h e^{\pm i \bs{k}\cdot \lvec}$ in either direction), then vertex pair cospectrality does imply corresponding latent symmetry, and vice versa.
Indeed, the multiplet interconnection in \cref{app:modifications} above remains valid in the same form (with $\bra{x} = \ket{x}^\dagger$ and ``symmetric'' replaced by ``self-adjoint'') for walk multiplets with indicator vector $\ket{e_\bM^\gamma}$ weighted by a complex tuple $\gamma$ (see \cref{eq:weightedIndicator}).
For instance, in $H_\bs{k}$ in \cref{fig:graphA_cellInterconnection} the singlet $\{8\}$ is connected with complex weight $\gamma_8 = he^{ikL} = [H_\bs{k}]_{8m} = [H_\bs{k}]_{m8}^*$ ($m=1,2$) to the doublet $\{1,2\}$ and the singlet $\{7\}$ is connected with complex weight $\gamma_7 = he^{-ikL} = [H_\bs{k}]_{7m} = [H_\bs{k}]_{m7}^*$ ($m=3,4$) to the doublet $\{3,4\}$, for some real $h$.

Now, since such walk multiplet interconnections preserve $\{u,v\}$-cospectrality and relative multiplets (as shown in \cref{app:modifications}), in particular $\{u,v\}$ itself remains a walk doublet in $H_\bs{k}$: 
\begin{equation}
 [H_\bs{k}^\p]_{uu} + [H_\bs{k}^\p]_{uv} = [H_\bs{k}^\p]_{vv} + [H_\bs{k}^\p]_{vu}
\end{equation}
for all powers $\p \in \bN$.
As a consequence, 
\begin{equation}
 [H_\bs{k}^\p]_{uv} = [H_\bs{k}^\p]_{vu} \in \bR \quad \forall \, \p  \in \bN.
\end{equation}
Thus, the restriction of each power $H_\bs{k}^\p$ to the cospectral pair is bisymmetric, that is, commutes with the $2 \times 2$ exchange matrix $J_{\bS=\{u,v\}}$, \cref{eq:reflectionMatrix}.

As we showed very recently in Ref.\,\cite{Rontgen2021_PRL_126_180601_LatentSymmetryInducedDegeneracies}, a necessary and sufficient condition for a latent symmetry transformation $T$ upon reduction to a vertex subset $\bS$ is that all powers of $H$ restricted to $\bS$ have the same symmetry:
\begin{equation}
 T \tH_\bS = \tH_\bS T \quad \Longleftrightarrow \quad T [H^\p]_\bS = [H^\p]_\bS T \quad \forall \, \p\in \bN.
\end{equation}
In the present case $\bS = \{u,v\}$ is a cospectral pair and $T = J_\bS$, with $[H_\bs{k}^\p]_\bS J_\bS = J_\bS [H_\bs{k}^\p]_\bS$ implying $\tH_{\bs{k};\bS} J_\bS = J_\bS \tH_{\bs{k};\bS}$, meaning that $H_\bs{k}$ has a latent $J_\bS$-symmetry in its reduction over $\{u,v\}$.

To summarize: 
For a general directed graph $H$, $\{u,v\}$-cospectrality is necessary but in general not sufficient for corresponding latent symmetry \cite{Kempton2020_LAaiA_594_226_CharacterizingCospectralVerticesIsospectral}; 
but for a complex self-adjoint $H'$ (in our case the Bloch Hamiltonian $H_\bs{k}$) constructed from an undirected $H$ via Hermitian interconnection of walk multiplets relative to $\{u,v\}$, it is both necessary and sufficient.

\vfill

%%%%%%%%%%%%%%%%%%%%%%%%%%%%%%%%%%%%%%%%%%%%%%%%%%%%%

% \bibliographystyle{apsrev4-1}
% \bibliography{}

\begin{thebibliography}{65}%
\makeatletter
\providecommand \@ifxundefined [1]{%
 \@ifx{#1\undefined}
}%
\providecommand \@ifnum [1]{%
 \ifnum #1\expandafter \@firstoftwo
 \else \expandafter \@secondoftwo
 \fi
}%
\providecommand \@ifx [1]{%
 \ifx #1\expandafter \@firstoftwo
 \else \expandafter \@secondoftwo
 \fi
}%
\providecommand \natexlab [1]{#1}%
\providecommand \enquote  [1]{``#1''}%
\providecommand \bibnamefont  [1]{#1}%
\providecommand \bibfnamefont [1]{#1}%
\providecommand \citenamefont [1]{#1}%
\providecommand \href@noop [0]{\@secondoftwo}%
\providecommand \href [0]{\begingroup \@sanitize@url \@href}%
\providecommand \@href[1]{\@@startlink{#1}\@@href}%
\providecommand \@@href[1]{\endgroup#1\@@endlink}%
\providecommand \@sanitize@url [0]{\catcode `\\12\catcode `\$12\catcode
  `\&12\catcode `\#12\catcode `\^12\catcode `\_12\catcode `\%12\relax}%
\providecommand \@@startlink[1]{}%
\providecommand \@@endlink[0]{}%
\providecommand \url  [0]{\begingroup\@sanitize@url \@url }%
\providecommand \@url [1]{\endgroup\@href {#1}{\urlprefix }}%
\providecommand \urlprefix  [0]{URL }%
\providecommand \Eprint [0]{\href }%
\providecommand \doibase [0]{https://doi.org/}%
\providecommand \selectlanguage [0]{\@gobble}%
\providecommand \bibinfo  [0]{\@secondoftwo}%
\providecommand \bibfield  [0]{\@secondoftwo}%
\providecommand \translation [1]{[#1]}%
\providecommand \BibitemOpen [0]{}%
\providecommand \bibitemStop [0]{}%
\providecommand \bibitemNoStop [0]{.\EOS\space}%
\providecommand \EOS [0]{\spacefactor3000\relax}%
\providecommand \BibitemShut  [1]{\csname bibitem#1\endcsname}%
\let\auto@bib@innerbib\@empty
%</preamble>
\bibitem [{\citenamefont {Leykam}\ \emph {et~al.}(2018)\citenamefont {Leykam},
  \citenamefont {Andreanov},\ and\ \citenamefont
  {Flach}}]{Leykam2018_APX_3_1473052_ArtificialFlatBandSystems}%
  \BibitemOpen
  \bibfield  {author} {\bibinfo {author} {\bibfnamefont {D.}~\bibnamefont
  {Leykam}}, \bibinfo {author} {\bibfnamefont {A.}~\bibnamefont {Andreanov}},\
  and\ \bibinfo {author} {\bibfnamefont {S.}~\bibnamefont {Flach}},\ }\bibfield
   {title} {\emph {\bibinfo {title} {Artificial flat band systems: From lattice
  models to experiments}},\ }\href
  {https://doi.org/10.1080/23746149.2018.1473052} {\bibfield  {journal}
  {\bibinfo  {journal} {Adv. Phys. X}\ }\textbf {\bibinfo {volume} {3}},\
  \bibinfo {pages} {1473052} (\bibinfo {year} {2018})}\BibitemShut {NoStop}%
\bibitem [{\citenamefont {Leykam}\ and\ \citenamefont
  {Flach}(2018)}]{Leykam2018_AP_3_070901_PerspectivePhotonicFlatbands}%
  \BibitemOpen
  \bibfield  {author} {\bibinfo {author} {\bibfnamefont {D.}~\bibnamefont
  {Leykam}}\ and\ \bibinfo {author} {\bibfnamefont {S.}~\bibnamefont {Flach}},\
  }\bibfield  {title} {\emph {\bibinfo {title} {Perspective: {{Photonic}}
  flatbands}},\ }\href {https://doi.org/10.1063/1.5034365} {\bibfield
  {journal} {\bibinfo  {journal} {APL Photonics}\ }\textbf {\bibinfo {volume}
  {3}},\ \bibinfo {pages} {070901} (\bibinfo {year} {2018})}\BibitemShut
  {NoStop}%
\bibitem [{\citenamefont {Mukherjee}\ \emph {et~al.}(2015)\citenamefont
  {Mukherjee}, \citenamefont {Spracklen}, \citenamefont {Choudhury},
  \citenamefont {Goldman}, \citenamefont {{\"O}hberg}, \citenamefont
  {Andersson},\ and\ \citenamefont
  {Thomson}}]{Mukherjee2015_PRL_114_245504_ObservationLocalizedFlatBandState}%
  \BibitemOpen
  \bibfield  {author} {\bibinfo {author} {\bibfnamefont {S.}~\bibnamefont
  {Mukherjee}}, \bibinfo {author} {\bibfnamefont {A.}~\bibnamefont
  {Spracklen}}, \bibinfo {author} {\bibfnamefont {D.}~\bibnamefont
  {Choudhury}}, \bibinfo {author} {\bibfnamefont {N.}~\bibnamefont {Goldman}},
  \bibinfo {author} {\bibfnamefont {P.}~\bibnamefont {{\"O}hberg}}, \bibinfo
  {author} {\bibfnamefont {E.}~\bibnamefont {Andersson}},\ and\ \bibinfo
  {author} {\bibfnamefont {R.~R.}\ \bibnamefont {Thomson}},\ }\bibfield
  {title} {\emph {\bibinfo {title} {Observation of a {{Localized Flat}}-{{Band
  State}} in a {{Photonic Lieb Lattice}}}},\ }\href
  {https://doi.org/10.1103/PhysRevLett.114.245504} {\bibfield  {journal}
  {\bibinfo  {journal} {Phys. Rev. Lett.}\ }\textbf {\bibinfo {volume} {114}},\
  \bibinfo {pages} {245504} (\bibinfo {year} {2015})}\BibitemShut {NoStop}%
\bibitem [{\citenamefont {Vicencio}\ \emph {et~al.}(2015)\citenamefont
  {Vicencio}, \citenamefont {Cantillano}, \citenamefont {{Morales-Inostroza}},
  \citenamefont {Real}, \citenamefont {{Mej{\'i}a-Cort{\'e}s}}, \citenamefont
  {Weimann}, \citenamefont {Szameit},\ and\ \citenamefont
  {Molina}}]{Vicencio2015_PRL_114_245503_ObservationLocalizedStatesLieb}%
  \BibitemOpen
  \bibfield  {author} {\bibinfo {author} {\bibfnamefont {R.~A.}\ \bibnamefont
  {Vicencio}}, \bibinfo {author} {\bibfnamefont {C.}~\bibnamefont
  {Cantillano}}, \bibinfo {author} {\bibfnamefont {L.}~\bibnamefont
  {{Morales-Inostroza}}}, \bibinfo {author} {\bibfnamefont {B.}~\bibnamefont
  {Real}}, \bibinfo {author} {\bibfnamefont {C.}~\bibnamefont
  {{Mej{\'i}a-Cort{\'e}s}}}, \bibinfo {author} {\bibfnamefont {S.}~\bibnamefont
  {Weimann}}, \bibinfo {author} {\bibfnamefont {A.}~\bibnamefont {Szameit}},\
  and\ \bibinfo {author} {\bibfnamefont {M.~I.}\ \bibnamefont {Molina}},\
  }\bibfield  {title} {\emph {\bibinfo {title} {Observation of {{Localized
  States}} in {{Lieb Photonic Lattices}}}},\ }\href
  {https://doi.org/10.1103/PhysRevLett.114.245503} {\bibfield  {journal}
  {\bibinfo  {journal} {Phys. Rev. Lett.}\ }\textbf {\bibinfo {volume} {114}},\
  \bibinfo {pages} {245503} (\bibinfo {year} {2015})}\BibitemShut {NoStop}%
\bibitem [{\citenamefont {Taie}\ \emph {et~al.}(2015)\citenamefont {Taie},
  \citenamefont {Ozawa}, \citenamefont {Ichinose}, \citenamefont {Nishio},
  \citenamefont {Nakajima},\ and\ \citenamefont
  {Takahashi}}]{Taie2015_SA_1_e1500854_CoherentDrivingFreezingBosonic}%
  \BibitemOpen
  \bibfield  {author} {\bibinfo {author} {\bibfnamefont {S.}~\bibnamefont
  {Taie}}, \bibinfo {author} {\bibfnamefont {H.}~\bibnamefont {Ozawa}},
  \bibinfo {author} {\bibfnamefont {T.}~\bibnamefont {Ichinose}}, \bibinfo
  {author} {\bibfnamefont {T.}~\bibnamefont {Nishio}}, \bibinfo {author}
  {\bibfnamefont {S.}~\bibnamefont {Nakajima}},\ and\ \bibinfo {author}
  {\bibfnamefont {Y.}~\bibnamefont {Takahashi}},\ }\bibfield  {title} {\emph
  {{\selectlanguage {en}\bibinfo {title} {Coherent driving and freezing of
  bosonic matter wave in an optical {{Lieb}} lattice}}},\ }\href
  {https://doi.org/10.1126/sciadv.1500854} {\bibfield  {journal} {\bibinfo
  {journal} {Sci. Adv.}\ }\textbf {\bibinfo {volume} {1}},\ \bibinfo {pages}
  {e1500854} (\bibinfo {year} {2015})}\BibitemShut {NoStop}%
\bibitem [{\citenamefont {Apaja}\ \emph {et~al.}(2010)\citenamefont {Apaja},
  \citenamefont {Hyrk{\"a}s},\ and\ \citenamefont
  {Manninen}}]{Apaja2010_PRA_82_041402_FlatBandsDiracCones}%
  \BibitemOpen
  \bibfield  {author} {\bibinfo {author} {\bibfnamefont {V.}~\bibnamefont
  {Apaja}}, \bibinfo {author} {\bibfnamefont {M.}~\bibnamefont {Hyrk{\"a}s}},\
  and\ \bibinfo {author} {\bibfnamefont {M.}~\bibnamefont {Manninen}},\
  }\bibfield  {title} {\emph {\bibinfo {title} {Flat bands, {{Dirac}} cones,
  and atom dynamics in an optical lattice}},\ }\href
  {https://doi.org/10.1103/PhysRevA.82.041402} {\bibfield  {journal} {\bibinfo
  {journal} {Phys. Rev. A}\ }\textbf {\bibinfo {volume} {82}},\ \bibinfo
  {pages} {041402(R)} (\bibinfo {year} {2010})}\BibitemShut {NoStop}%
\bibitem [{\citenamefont {Abilio}\ \emph {et~al.}(1999)\citenamefont {Abilio},
  \citenamefont {Butaud}, \citenamefont {Fournier}, \citenamefont {Pannetier},
  \citenamefont {Vidal}, \citenamefont {Tedesco},\ and\ \citenamefont
  {Dalzotto}}]{Abilio1999_PRL_83_5102_MagneticFieldInducedLocalization}%
  \BibitemOpen
  \bibfield  {author} {\bibinfo {author} {\bibfnamefont {C.~C.}\ \bibnamefont
  {Abilio}}, \bibinfo {author} {\bibfnamefont {P.}~\bibnamefont {Butaud}},
  \bibinfo {author} {\bibfnamefont {T.}~\bibnamefont {Fournier}}, \bibinfo
  {author} {\bibfnamefont {B.}~\bibnamefont {Pannetier}}, \bibinfo {author}
  {\bibfnamefont {J.}~\bibnamefont {Vidal}}, \bibinfo {author} {\bibfnamefont
  {S.}~\bibnamefont {Tedesco}},\ and\ \bibinfo {author} {\bibfnamefont
  {B.}~\bibnamefont {Dalzotto}},\ }\bibfield  {title} {\emph {\bibinfo {title}
  {Magnetic {{Field Induced Localization}} in a {{Two}}-{{Dimensional
  Superconducting Wire Network}}}},\ }\href
  {https://doi.org/10.1103/PhysRevLett.83.5102} {\bibfield  {journal} {\bibinfo
   {journal} {Phys. Rev. Lett.}\ }\textbf {\bibinfo {volume} {83}},\ \bibinfo
  {pages} {5102} (\bibinfo {year} {1999})}\BibitemShut {NoStop}%
\bibitem [{\citenamefont {Drost}\ \emph {et~al.}(2017)\citenamefont {Drost},
  \citenamefont {Ojanen}, \citenamefont {Harju},\ and\ \citenamefont
  {Liljeroth}}]{Drost2017_NP_13_668_TopologicalStatesEngineeredAtomic}%
  \BibitemOpen
  \bibfield  {author} {\bibinfo {author} {\bibfnamefont {R.}~\bibnamefont
  {Drost}}, \bibinfo {author} {\bibfnamefont {T.}~\bibnamefont {Ojanen}},
  \bibinfo {author} {\bibfnamefont {A.}~\bibnamefont {Harju}},\ and\ \bibinfo
  {author} {\bibfnamefont {P.}~\bibnamefont {Liljeroth}},\ }\bibfield  {title}
  {\emph {{\selectlanguage {en}\bibinfo {title} {Topological states in
  engineered atomic lattices}}},\ }\href {https://doi.org/10.1038/nphys4080}
  {\bibfield  {journal} {\bibinfo  {journal} {Nat. Phys.}\ }\textbf {\bibinfo
  {volume} {13}},\ \bibinfo {pages} {668} (\bibinfo {year} {2017})}\BibitemShut
  {NoStop}%
\bibitem [{\citenamefont {Wan}\ \emph {et~al.}(2017)\citenamefont {Wan},
  \citenamefont {L{\"u}}, \citenamefont {Gao},\ and\ \citenamefont
  {Wu}}]{Wan2017_SR_7_15188_HybridInterferenceInducedFlat}%
  \BibitemOpen
  \bibfield  {author} {\bibinfo {author} {\bibfnamefont {L.-L.}\ \bibnamefont
  {Wan}}, \bibinfo {author} {\bibfnamefont {X.-Y.}\ \bibnamefont {L{\"u}}},
  \bibinfo {author} {\bibfnamefont {J.-H.}\ \bibnamefont {Gao}},\ and\ \bibinfo
  {author} {\bibfnamefont {Y.}~\bibnamefont {Wu}},\ }\bibfield  {title} {\emph
  {{\selectlanguage {En}\bibinfo {title} {Hybrid {{Interference Induced Flat
  Band Localization}} in {{Bipartite Optomechanical Lattices}}}}},\ }\href
  {https://doi.org/10.1038/s41598-017-15381-x} {\bibfield  {journal} {\bibinfo
  {journal} {Sci. Rep.}\ }\textbf {\bibinfo {volume} {7}},\ \bibinfo {pages}
  {15188} (\bibinfo {year} {2017})}\BibitemShut {NoStop}%
\bibitem [{\citenamefont {Helbig}\ \emph {et~al.}(2019)\citenamefont {Helbig},
  \citenamefont {Hofmann}, \citenamefont {Lee}, \citenamefont {Thomale},
  \citenamefont {Imhof}, \citenamefont {Molenkamp},\ and\ \citenamefont
  {Kiessling}}]{Helbig2019_PRB_99_161114_BandStructureEngineeringReconstruction}%
  \BibitemOpen
  \bibfield  {author} {\bibinfo {author} {\bibfnamefont {T.}~\bibnamefont
  {Helbig}}, \bibinfo {author} {\bibfnamefont {T.}~\bibnamefont {Hofmann}},
  \bibinfo {author} {\bibfnamefont {C.~H.}\ \bibnamefont {Lee}}, \bibinfo
  {author} {\bibfnamefont {R.}~\bibnamefont {Thomale}}, \bibinfo {author}
  {\bibfnamefont {S.}~\bibnamefont {Imhof}}, \bibinfo {author} {\bibfnamefont
  {L.~W.}\ \bibnamefont {Molenkamp}},\ and\ \bibinfo {author} {\bibfnamefont
  {T.}~\bibnamefont {Kiessling}},\ }\bibfield  {title} {\emph {\bibinfo {title}
  {Band structure engineering and reconstruction in electric circuit
  networks}},\ }\href {https://doi.org/10.1103/PhysRevB.99.161114} {\bibfield
  {journal} {\bibinfo  {journal} {Phys. Rev. B}\ }\textbf {\bibinfo {volume}
  {99}},\ \bibinfo {pages} {161114(R)} (\bibinfo {year} {2019})}\BibitemShut
  {NoStop}%
\bibitem [{\citenamefont {R{\"o}ntgen}\ \emph {et~al.}(2019)\citenamefont
  {R{\"o}ntgen}, \citenamefont {Morfonios}, \citenamefont {Brouzos},
  \citenamefont {Diakonos},\ and\ \citenamefont
  {Schmelcher}}]{Rontgen2019_PRL_123_080504_QuantumNetworkTransferStorage}%
  \BibitemOpen
  \bibfield  {author} {\bibinfo {author} {\bibfnamefont {M.}~\bibnamefont
  {R{\"o}ntgen}}, \bibinfo {author} {\bibfnamefont {C.~V.}\ \bibnamefont
  {Morfonios}}, \bibinfo {author} {\bibfnamefont {I.}~\bibnamefont {Brouzos}},
  \bibinfo {author} {\bibfnamefont {F.~K.}\ \bibnamefont {Diakonos}},\ and\
  \bibinfo {author} {\bibfnamefont {P.}~\bibnamefont {Schmelcher}},\ }\bibfield
   {title} {\emph {\bibinfo {title} {Quantum {{Network Transfer}} and
  {{Storage}} with {{Compact Localized States Induced}} by {{Local
  Symmetries}}}},\ }\href {https://doi.org/10.1103/PhysRevLett.123.080504}
  {\bibfield  {journal} {\bibinfo  {journal} {Phys. Rev. Lett.}\ }\textbf
  {\bibinfo {volume} {123}},\ \bibinfo {pages} {080504} (\bibinfo {year}
  {2019})}\BibitemShut {NoStop}%
\bibitem [{\citenamefont {Peotta}\ and\ \citenamefont
  {T{\"o}rm{\"a}}(2015)}]{Peotta2015_NC_6_8944_SuperfluidityTopologicallyNontrivialFlat}%
  \BibitemOpen
  \bibfield  {author} {\bibinfo {author} {\bibfnamefont {S.}~\bibnamefont
  {Peotta}}\ and\ \bibinfo {author} {\bibfnamefont {P.}~\bibnamefont
  {T{\"o}rm{\"a}}},\ }\bibfield  {title} {\emph {{\selectlanguage {en}\bibinfo
  {title} {Superfluidity in topologically nontrivial flat bands}}},\ }\href
  {https://doi.org/10.1038/ncomms9944} {\bibfield  {journal} {\bibinfo
  {journal} {Nat. Commun.}\ }\textbf {\bibinfo {volume} {6}},\ \bibinfo {pages}
  {8944} (\bibinfo {year} {2015})}\BibitemShut {NoStop}%
\bibitem [{\citenamefont {Julku}\ \emph {et~al.}(2016)\citenamefont {Julku},
  \citenamefont {Peotta}, \citenamefont {Vanhala}, \citenamefont {Kim},\ and\
  \citenamefont
  {T{\"o}rm{\"a}}}]{Julku2016_PRL_117_045303_GeometricOriginSuperfluidityLiebLattice}%
  \BibitemOpen
  \bibfield  {author} {\bibinfo {author} {\bibfnamefont {A.}~\bibnamefont
  {Julku}}, \bibinfo {author} {\bibfnamefont {S.}~\bibnamefont {Peotta}},
  \bibinfo {author} {\bibfnamefont {T.~I.}\ \bibnamefont {Vanhala}}, \bibinfo
  {author} {\bibfnamefont {D.-H.}\ \bibnamefont {Kim}},\ and\ \bibinfo {author}
  {\bibfnamefont {P.}~\bibnamefont {T{\"o}rm{\"a}}},\ }\bibfield  {title}
  {\emph {\bibinfo {title} {Geometric {{Origin}} of {{Superfluidity}} in the
  {{Lieb}}-{{Lattice Flat Band}}}},\ }\href
  {https://doi.org/10.1103/PhysRevLett.117.045303} {\bibfield  {journal}
  {\bibinfo  {journal} {Phys. Rev. Lett.}\ }\textbf {\bibinfo {volume} {117}},\
  \bibinfo {pages} {045303} (\bibinfo {year} {2016})}\BibitemShut {NoStop}%
\bibitem [{\citenamefont {Kopnin}\ \emph {et~al.}(2011)\citenamefont {Kopnin},
  \citenamefont {Heikkil{\"a}},\ and\ \citenamefont
  {Volovik}}]{Kopnin2011_PRB_83_220503_HightemperatureSurfaceSuperconductivityTopological}%
  \BibitemOpen
  \bibfield  {author} {\bibinfo {author} {\bibfnamefont {N.~B.}\ \bibnamefont
  {Kopnin}}, \bibinfo {author} {\bibfnamefont {T.~T.}\ \bibnamefont
  {Heikkil{\"a}}},\ and\ \bibinfo {author} {\bibfnamefont {G.~E.}\ \bibnamefont
  {Volovik}},\ }\bibfield  {title} {\emph {\bibinfo {title} {High-temperature
  surface superconductivity in topological flat-band systems}},\ }\href
  {https://doi.org/10.1103/PhysRevB.83.220503} {\bibfield  {journal} {\bibinfo
  {journal} {Phys. Rev. B}\ }\textbf {\bibinfo {volume} {83}},\ \bibinfo
  {pages} {220503(R)} (\bibinfo {year} {2011})}\BibitemShut {NoStop}%
\bibitem [{\citenamefont {Iglovikov}\ \emph {et~al.}(2014)\citenamefont
  {Iglovikov}, \citenamefont {H{\'e}bert}, \citenamefont {Gr{\'e}maud},
  \citenamefont {Batrouni},\ and\ \citenamefont
  {Scalettar}}]{Iglovikov2014_PRB_90_094506_SuperconductingTransitionsFlatbandSystems}%
  \BibitemOpen
  \bibfield  {author} {\bibinfo {author} {\bibfnamefont {V.~I.}\ \bibnamefont
  {Iglovikov}}, \bibinfo {author} {\bibfnamefont {F.}~\bibnamefont
  {H{\'e}bert}}, \bibinfo {author} {\bibfnamefont {B.}~\bibnamefont
  {Gr{\'e}maud}}, \bibinfo {author} {\bibfnamefont {G.~G.}\ \bibnamefont
  {Batrouni}},\ and\ \bibinfo {author} {\bibfnamefont {R.~T.}\ \bibnamefont
  {Scalettar}},\ }\bibfield  {title} {\emph {\bibinfo {title} {Superconducting
  transitions in flat-band systems}},\ }\href
  {https://doi.org/10.1103/PhysRevB.90.094506} {\bibfield  {journal} {\bibinfo
  {journal} {Phys. Rev. B}\ }\textbf {\bibinfo {volume} {90}},\ \bibinfo
  {pages} {094506} (\bibinfo {year} {2014})}\BibitemShut {NoStop}%
\bibitem [{\citenamefont {Kobayashi}\ \emph {et~al.}(2016)\citenamefont
  {Kobayashi}, \citenamefont {Okumura}, \citenamefont {Yamada}, \citenamefont
  {Machida},\ and\ \citenamefont
  {Aoki}}]{Kobayashi2016_PRB_94_214501_SuperconductivityRepulsivelyInteractingFermions}%
  \BibitemOpen
  \bibfield  {author} {\bibinfo {author} {\bibfnamefont {K.}~\bibnamefont
  {Kobayashi}}, \bibinfo {author} {\bibfnamefont {M.}~\bibnamefont {Okumura}},
  \bibinfo {author} {\bibfnamefont {S.}~\bibnamefont {Yamada}}, \bibinfo
  {author} {\bibfnamefont {M.}~\bibnamefont {Machida}},\ and\ \bibinfo {author}
  {\bibfnamefont {H.}~\bibnamefont {Aoki}},\ }\bibfield  {title} {\emph
  {\bibinfo {title} {Superconductivity in repulsively interacting fermions on a
  diamond chain: {{Flat}}-band-induced pairing}},\ }\href
  {https://doi.org/10.1103/PhysRevB.94.214501} {\bibfield  {journal} {\bibinfo
  {journal} {Phys. Rev. B}\ }\textbf {\bibinfo {volume} {94}},\ \bibinfo
  {pages} {214501} (\bibinfo {year} {2016})}\BibitemShut {NoStop}%
\bibitem [{\citenamefont {Tovmasyan}\ \emph {et~al.}(2016)\citenamefont
  {Tovmasyan}, \citenamefont {Peotta}, \citenamefont {T{\"o}rm{\"a}},\ and\
  \citenamefont
  {Huber}}]{Tovmasyan2016_PRB_94_245149_EffectiveTheoryEmergentSU2}%
  \BibitemOpen
  \bibfield  {author} {\bibinfo {author} {\bibfnamefont {M.}~\bibnamefont
  {Tovmasyan}}, \bibinfo {author} {\bibfnamefont {S.}~\bibnamefont {Peotta}},
  \bibinfo {author} {\bibfnamefont {P.}~\bibnamefont {T{\"o}rm{\"a}}},\ and\
  \bibinfo {author} {\bibfnamefont {S.~D.}\ \bibnamefont {Huber}},\ }\bibfield
  {title} {\emph {\bibinfo {title} {Effective theory and emergent {{SU}}(2)
  symmetry in the flat bands of attractive {{Hubbard}} models}},\ }\href
  {https://doi.org/10.1103/PhysRevB.94.245149} {\bibfield  {journal} {\bibinfo
  {journal} {Phys. Rev. B}\ }\textbf {\bibinfo {volume} {94}},\ \bibinfo
  {pages} {245149} (\bibinfo {year} {2016})}\BibitemShut {NoStop}%
\bibitem [{\citenamefont {Liang}\ \emph {et~al.}(2017)\citenamefont {Liang},
  \citenamefont {Vanhala}, \citenamefont {Peotta}, \citenamefont {Siro},
  \citenamefont {Harju},\ and\ \citenamefont
  {T{\"o}rm{\"a}}}]{Liang2017_PRB_95_024515_BandGeometryBerryCurvature}%
  \BibitemOpen
  \bibfield  {author} {\bibinfo {author} {\bibfnamefont {L.}~\bibnamefont
  {Liang}}, \bibinfo {author} {\bibfnamefont {T.~I.}\ \bibnamefont {Vanhala}},
  \bibinfo {author} {\bibfnamefont {S.}~\bibnamefont {Peotta}}, \bibinfo
  {author} {\bibfnamefont {T.}~\bibnamefont {Siro}}, \bibinfo {author}
  {\bibfnamefont {A.}~\bibnamefont {Harju}},\ and\ \bibinfo {author}
  {\bibfnamefont {P.}~\bibnamefont {T{\"o}rm{\"a}}},\ }\bibfield  {title}
  {\emph {\bibinfo {title} {Band geometry, {{Berry}} curvature, and superfluid
  weight}},\ }\href {https://doi.org/10.1103/PhysRevB.95.024515} {\bibfield
  {journal} {\bibinfo  {journal} {Phys. Rev. B}\ }\textbf {\bibinfo {volume}
  {95}},\ \bibinfo {pages} {024515} (\bibinfo {year} {2017})}\BibitemShut
  {NoStop}%
\bibitem [{\citenamefont {Tang}\ \emph {et~al.}(2011)\citenamefont {Tang},
  \citenamefont {Mei},\ and\ \citenamefont
  {Wen}}]{Tang2011_PRL_106_236802_HighTemperatureFractionalQuantumHall}%
  \BibitemOpen
  \bibfield  {author} {\bibinfo {author} {\bibfnamefont {E.}~\bibnamefont
  {Tang}}, \bibinfo {author} {\bibfnamefont {J.-W.}\ \bibnamefont {Mei}},\ and\
  \bibinfo {author} {\bibfnamefont {X.-G.}\ \bibnamefont {Wen}},\ }\bibfield
  {title} {\emph {\bibinfo {title} {High-{{Temperature Fractional Quantum Hall
  States}}}},\ }\href {https://doi.org/10.1103/PhysRevLett.106.236802}
  {\bibfield  {journal} {\bibinfo  {journal} {Phys. Rev. Lett.}\ }\textbf
  {\bibinfo {volume} {106}},\ \bibinfo {pages} {236802} (\bibinfo {year}
  {2011})}\BibitemShut {NoStop}%
\bibitem [{\citenamefont {Sun}\ \emph {et~al.}(2011)\citenamefont {Sun},
  \citenamefont {Gu}, \citenamefont {Katsura},\ and\ \citenamefont
  {Das~Sarma}}]{Sun2011_PRL_106_236803_NearlyFlatbandsNontrivialTopology}%
  \BibitemOpen
  \bibfield  {author} {\bibinfo {author} {\bibfnamefont {K.}~\bibnamefont
  {Sun}}, \bibinfo {author} {\bibfnamefont {Z.}~\bibnamefont {Gu}}, \bibinfo
  {author} {\bibfnamefont {H.}~\bibnamefont {Katsura}},\ and\ \bibinfo {author}
  {\bibfnamefont {S.}~\bibnamefont {Das~Sarma}},\ }\bibfield  {title} {\emph
  {\bibinfo {title} {Nearly {{Flatbands}} with {{Nontrivial Topology}}}},\
  }\href {https://doi.org/10.1103/PhysRevLett.106.236803} {\bibfield  {journal}
  {\bibinfo  {journal} {Phys. Rev. Lett.}\ }\textbf {\bibinfo {volume} {106}},\
  \bibinfo {pages} {236803} (\bibinfo {year} {2011})}\BibitemShut {NoStop}%
\bibitem [{\citenamefont {Neupert}\ \emph {et~al.}(2011)\citenamefont
  {Neupert}, \citenamefont {Santos}, \citenamefont {Chamon},\ and\
  \citenamefont
  {Mudry}}]{Neupert2011_PRL_106_236804_FractionalQuantumHallStates}%
  \BibitemOpen
  \bibfield  {author} {\bibinfo {author} {\bibfnamefont {T.}~\bibnamefont
  {Neupert}}, \bibinfo {author} {\bibfnamefont {L.}~\bibnamefont {Santos}},
  \bibinfo {author} {\bibfnamefont {C.}~\bibnamefont {Chamon}},\ and\ \bibinfo
  {author} {\bibfnamefont {C.}~\bibnamefont {Mudry}},\ }\bibfield  {title}
  {\emph {\bibinfo {title} {Fractional {{Quantum Hall States}} at {{Zero
  Magnetic Field}}}},\ }\href {https://doi.org/10.1103/PhysRevLett.106.236804}
  {\bibfield  {journal} {\bibinfo  {journal} {Phys. Rev. Lett.}\ }\textbf
  {\bibinfo {volume} {106}},\ \bibinfo {pages} {236804} (\bibinfo {year}
  {2011})}\BibitemShut {NoStop}%
\bibitem [{\citenamefont
  {Pal}(2018)}]{Pal2018_PRB_98_245116_NontrivialTopologicalFlatBands}%
  \BibitemOpen
  \bibfield  {author} {\bibinfo {author} {\bibfnamefont {B.}~\bibnamefont
  {Pal}},\ }\bibfield  {title} {\emph {\bibinfo {title} {Nontrivial topological
  flat bands in a diamond-octagon lattice geometry}},\ }\href
  {https://doi.org/10.1103/PhysRevB.98.245116} {\bibfield  {journal} {\bibinfo
  {journal} {Phys. Rev. B}\ }\textbf {\bibinfo {volume} {98}},\ \bibinfo
  {pages} {245116} (\bibinfo {year} {2018})}\BibitemShut {NoStop}%
\bibitem [{\citenamefont {Bhattacharya}\ and\ \citenamefont
  {Pal}(2019)}]{Bhattacharya2019_PRB_100_235145_FlatBandsNontrivialTopological}%
  \BibitemOpen
  \bibfield  {author} {\bibinfo {author} {\bibfnamefont {A.}~\bibnamefont
  {Bhattacharya}}\ and\ \bibinfo {author} {\bibfnamefont {B.}~\bibnamefont
  {Pal}},\ }\bibfield  {title} {\emph {\bibinfo {title} {Flat bands and
  nontrivial topological properties in an extended {{Lieb}} lattice}},\ }\href
  {https://doi.org/10.1103/PhysRevB.100.235145} {\bibfield  {journal} {\bibinfo
   {journal} {Phys. Rev. B}\ }\textbf {\bibinfo {volume} {100}},\ \bibinfo
  {pages} {235145} (\bibinfo {year} {2019})}\BibitemShut {NoStop}%
\bibitem [{\citenamefont {Danieli}\ \emph
  {et~al.}(2020{\natexlab{a}})\citenamefont {Danieli}, \citenamefont
  {Andreanov},\ and\ \citenamefont
  {Flach}}]{Danieli2020_PRB_102_041116_ManybodyFlatbandLocalization}%
  \BibitemOpen
  \bibfield  {author} {\bibinfo {author} {\bibfnamefont {C.}~\bibnamefont
  {Danieli}}, \bibinfo {author} {\bibfnamefont {A.}~\bibnamefont {Andreanov}},\
  and\ \bibinfo {author} {\bibfnamefont {S.}~\bibnamefont {Flach}},\ }\bibfield
   {title} {\emph {\bibinfo {title} {Many-body flatband localization}},\ }\href
  {https://doi.org/10.1103/PhysRevB.102.041116} {\bibfield  {journal} {\bibinfo
   {journal} {Phys. Rev. B}\ }\textbf {\bibinfo {volume} {102}},\ \bibinfo
  {pages} {041116(R)} (\bibinfo {year} {2020}{\natexlab{a}})}\BibitemShut
  {NoStop}%
\bibitem {Kuno2020_NJP_22_013032_FlatbandManybodyLocalizationErgodicity}%
  \BibitemOpen
  \bibfield  {author} {\bibinfo {author} {\bibfnamefont {Y.}~\bibnamefont
  {Kuno}}, \bibinfo {author} {\bibfnamefont {T.}~\bibnamefont {Orito}},\
  and\ \bibinfo {author} {\bibfnamefont {I.}~\bibnamefont {Ichinose}},\ }\bibfield
   {title} {\emph {\bibinfo {title} {Flat-Band Many-Body Localization and Ergodicity Breaking in the {{Creutz}} Ladder}},\ }\href
  {https://doi.org/10.1088/1367-2630/ab6352} {\bibfield  {journal} {\bibinfo
   {journal} {New J. Phys.}\ }\textbf {\bibinfo {volume} {22}},\ \bibinfo
  {pages} {013032} (\bibinfo {year} {2020}{\natexlab{a}})}\BibitemShut
  {NoStop}%
\bibitem {Orito2021_PRB_103_L060301_NonthermalizedDynamicsFlatbandManybody}%
  \BibitemOpen
  \bibfield  {author} {\bibinfo {author} {\bibfnamefont {T.}~\bibnamefont
  {Orito}}, \bibinfo {author} {\bibfnamefont {Y.}~\bibnamefont {Kuno}},\
  and\ \bibinfo {author} {\bibfnamefont {I.}~\bibnamefont {Ichinose}},\ }\bibfield
   {title} {\emph {\bibinfo {title} {Nonthermalized Dynamics of Flat-Band Many-Body Localization}},\ }\href
  {https://doi.org/10.1103/PhysRevB.103.L060301} {\bibfield  {journal} {\bibinfo
   {journal} {Phys. Rev. B}\ }\textbf {\bibinfo {volume} {103}},\ \bibinfo
  {pages} {L060301} (\bibinfo {year} {2021}{\natexlab{a}})}\BibitemShut
  {NoStop}%
\bibitem [{\citenamefont {Danieli}\ \emph
  {et~al.}(2020{\natexlab{b}})\citenamefont {Danieli}, \citenamefont
  {Andreanov}, \citenamefont {Mithun},\ and\ \citenamefont
  {Flach}}]{Danieli2020_ACP_NonlinearCagingAllBandsFlatLattices}%
  \BibitemOpen
  \bibfield  {author} {\bibinfo {author} {\bibfnamefont {C.}~\bibnamefont
  {Danieli}}, \bibinfo {author} {\bibfnamefont {A.}~\bibnamefont {Andreanov}},
  \bibinfo {author} {\bibfnamefont {T.}~\bibnamefont {Mithun}},\ and\ \bibinfo
  {author} {\bibfnamefont {S.}~\bibnamefont {Flach}},\ }\bibfield  {title}
  {\emph {\bibinfo {title} {Nonlinear caging in {{All}}-{{Bands}}-{{Flat
  Lattices}}}},\ }\href@noop {} \Eprint {https://arxiv.org/abs/2004.11871}
  {arXiv:2004.11871} {\bibfield  {journal} (\bibinfo {year}
  {2020}{\natexlab{b}})} \BibitemShut {NoStop}%
\bibitem [{\citenamefont {Danieli}\ \emph
  {et~al.}(2020{\natexlab{c}})\citenamefont {Danieli}, \citenamefont
  {Andreanov}, \citenamefont {Mithun},\ and\ \citenamefont
  {Flach}}]{Danieli2020_ACP_QuantumCagingInteractingManyBody}%
  \BibitemOpen
  \bibfield  {author} {\bibinfo {author} {\bibfnamefont {C.}~\bibnamefont
  {Danieli}}, \bibinfo {author} {\bibfnamefont {A.}~\bibnamefont {Andreanov}},
  \bibinfo {author} {\bibfnamefont {T.}~\bibnamefont {Mithun}},\ and\ \bibinfo
  {author} {\bibfnamefont {S.}~\bibnamefont {Flach}},\ }\bibfield  {title}
  {\emph {\bibinfo {title} {Quantum {{Caging}} in {{Interacting Many}}-{{Body
  All}}-{{Bands}}-{{Flat Lattices}}}},\ }\href@noop {} \Eprint {https://arxiv.org/abs/2004.11880}{arXiv:2004.11880} {\bibfield  {journal}
  (\bibinfo {year} {2020}{\natexlab{c}})} \BibitemShut {NoStop}%
\bibitem [{\citenamefont {He}\ \emph {et~al.}(2021)\citenamefont {He},
  \citenamefont {Mao}, \citenamefont {Cai}, \citenamefont {Zhang},
  \citenamefont {Li}, \citenamefont {Yuan}, \citenamefont {Zhu},\ and\
  \citenamefont
  {Wang}}]{He2021_PRL_126_103601_FlatBandLocalizationCreutzSuperradiance}%
  \BibitemOpen
  \bibfield  {author} {\bibinfo {author} {\bibfnamefont {Y.}~\bibnamefont
  {He}}, \bibinfo {author} {\bibfnamefont {R.}~\bibnamefont {Mao}}, \bibinfo
  {author} {\bibfnamefont {H.}~\bibnamefont {Cai}}, \bibinfo {author}
  {\bibfnamefont {J.-X.}\ \bibnamefont {Zhang}}, \bibinfo {author}
  {\bibfnamefont {Y.}~\bibnamefont {Li}}, \bibinfo {author} {\bibfnamefont
  {L.}~\bibnamefont {Yuan}}, \bibinfo {author} {\bibfnamefont {S.-Y.}\
  \bibnamefont {Zhu}},\ and\ \bibinfo {author} {\bibfnamefont {D.-W.}\
  \bibnamefont {Wang}},\ }\bibfield  {title} {\emph {\bibinfo {title}
  {Flat-{{Band Localization}} in {{Creutz Superradiance Lattices}}}},\ }\href
  {https://doi.org/10.1103/PhysRevLett.126.103601} {\bibfield  {journal}
  {\bibinfo  {journal} {Phys. Rev. Lett.}\ }\textbf {\bibinfo {volume}
  {126}},\ \bibinfo {pages} {103601} (\bibinfo {year} {2021})}\BibitemShut
  {NoStop}%
\bibitem [{\citenamefont {Rhim}\ and\ \citenamefont
  {Yang}(2019)}]{Rhim2019_PRB_99_045107_ClassificationFlatBandsAccording}%
  \BibitemOpen
  \bibfield  {author} {\bibinfo {author} {\bibfnamefont {J.-W.}\ \bibnamefont
  {Rhim}}\ and\ \bibinfo {author} {\bibfnamefont {B.-J.}\ \bibnamefont
  {Yang}},\ }\bibfield  {title} {\emph {\bibinfo {title} {Classification of
  flat bands according to the band-crossing singularity of {{Bloch}} wave
  functions}},\ }\href {https://doi.org/10.1103/PhysRevB.99.045107} {\bibfield
  {journal} {\bibinfo  {journal} {Phys. Rev. B}\ }\textbf {\bibinfo {volume}
  {99}},\ \bibinfo {pages} {045107} (\bibinfo {year} {2019})}\BibitemShut
  {NoStop}%
\bibitem [{\citenamefont {Maimaiti}\ \emph {et~al.}(2019)\citenamefont
  {Maimaiti}, \citenamefont {Flach},\ and\ \citenamefont
  {Andreanov}}]{Maimaiti2019_PRB_99_125129_Universald1FlatBand}%
  \BibitemOpen
  \bibfield  {author} {\bibinfo {author} {\bibfnamefont {W.}~\bibnamefont
  {Maimaiti}}, \bibinfo {author} {\bibfnamefont {S.}~\bibnamefont {Flach}},\
  and\ \bibinfo {author} {\bibfnamefont {A.}~\bibnamefont {Andreanov}},\
  }\bibfield  {title} {\emph {\bibinfo {title} {Universal d=1 flat band
  generator from compact localized states}},\ }\href
  {https://doi.org/10.1103/PhysRevB.99.125129} {\bibfield  {journal} {\bibinfo
  {journal} {Phys. Rev. B}\ }\textbf {\bibinfo {volume} {99}},\ \bibinfo
  {pages} {125129} (\bibinfo {year} {2019})}\BibitemShut {NoStop}%
\bibitem [{\citenamefont {Flach}\ \emph {et~al.}(2014)\citenamefont {Flach},
  \citenamefont {Leykam}, \citenamefont {Bodyfelt}, \citenamefont {Matthies},\
  and\ \citenamefont
  {Desyatnikov}}]{Flach2014_E_105_30001_DetanglingFlatBandsFano}%
  \BibitemOpen
  \bibfield  {author} {\bibinfo {author} {\bibfnamefont {S.}~\bibnamefont
  {Flach}}, \bibinfo {author} {\bibfnamefont {D.}~\bibnamefont {Leykam}},
  \bibinfo {author} {\bibfnamefont {J.~D.}\ \bibnamefont {Bodyfelt}}, \bibinfo
  {author} {\bibfnamefont {P.}~\bibnamefont {Matthies}},\ and\ \bibinfo
  {author} {\bibfnamefont {A.~S.}\ \bibnamefont {Desyatnikov}},\ }\bibfield
  {title} {\emph {{\selectlanguage {en}\bibinfo {title} {Detangling flat bands
  into {{Fano}} lattices}}},\ }\href
  {https://doi.org/10.1209/0295-5075/105/30001} {\bibfield  {journal} {\bibinfo
   {journal} {Europhys. Lett.}\ }\textbf {\bibinfo {volume} {105}},\ \bibinfo {pages}
  {30001} (\bibinfo {year} {2014})}\BibitemShut {NoStop}%
\bibitem [{\citenamefont {R{\"o}ntgen}\ \emph {et~al.}(2018)\citenamefont
  {R{\"o}ntgen}, \citenamefont {Morfonios},\ and\ \citenamefont
  {Schmelcher}}]{Rontgen2018_PRB_97_035161_CompactLocalizedStatesFlat}%
  \BibitemOpen
  \bibfield  {author} {\bibinfo {author} {\bibfnamefont {M.}~\bibnamefont
  {R{\"o}ntgen}}, \bibinfo {author} {\bibfnamefont {C.~V.}\ \bibnamefont
  {Morfonios}},\ and\ \bibinfo {author} {\bibfnamefont {P.}~\bibnamefont
  {Schmelcher}},\ }\bibfield  {title} {\emph {\bibinfo {title} {Compact
  localized states and flat bands from local symmetry partitioning}},\ }\href
  {https://doi.org/10.1103/PhysRevB.97.035161} {\bibfield  {journal} {\bibinfo
  {journal} {Phys. Rev. B}\ }\textbf {\bibinfo {volume} {97}},\ \bibinfo
  {pages} {035161} (\bibinfo {year} {2018})}\BibitemShut {NoStop}%
\bibitem [{\citenamefont {Ramachandran}\ \emph {et~al.}(2017)\citenamefont
  {Ramachandran}, \citenamefont {Andreanov},\ and\ \citenamefont
  {Flach}}]{Ramachandran2017_PRB_96_161104_ChiralFlatBandsExistence}%
  \BibitemOpen
  \bibfield  {author} {\bibinfo {author} {\bibfnamefont {A.}~\bibnamefont
  {Ramachandran}}, \bibinfo {author} {\bibfnamefont {A.}~\bibnamefont
  {Andreanov}},\ and\ \bibinfo {author} {\bibfnamefont {S.}~\bibnamefont
  {Flach}},\ }\bibfield  {title} {\emph {\bibinfo {title} {Chiral flat bands:
  {{Existence}}, engineering, and stability}},\ }\href
  {https://doi.org/10.1103/PhysRevB.96.161104} {\bibfield  {journal} {\bibinfo
  {journal} {Phys. Rev. B}\ }\textbf {\bibinfo {volume} {96}},\ \bibinfo
  {pages} {161104(R)} (\bibinfo {year} {2017})}\BibitemShut {NoStop}%
\bibitem [{\citenamefont {{Morales-Inostroza}}\ and\ \citenamefont
  {Vicencio}(2016)}]{MoralesInostroza2016_PRA_94_043831_SimpleMethodConstructFlatband}%
  \BibitemOpen
  \bibfield  {author} {\bibinfo {author} {\bibfnamefont {L.}~\bibnamefont
  {{Morales-Inostroza}}}\ and\ \bibinfo {author} {\bibfnamefont {R.~A.}\
  \bibnamefont {Vicencio}},\ }\bibfield  {title} {\emph {\bibinfo {title}
  {Simple method to construct flat-band lattices}},\ }\href
  {https://doi.org/10.1103/PhysRevA.94.043831} {\bibfield  {journal} {\bibinfo
  {journal} {Phys. Rev. A}\ }\textbf {\bibinfo {volume} {94}},\ \bibinfo
  {pages} {043831} (\bibinfo {year} {2016})}\BibitemShut {NoStop}%
\bibitem [{\citenamefont {Dias}\ and\ \citenamefont
  {Gouveia}(2015)}]{Dias2015_SR_5_16852_OrigamiRulesConstructionLocalized}%
  \BibitemOpen
  \bibfield  {author} {\bibinfo {author} {\bibfnamefont {R.~G.}\ \bibnamefont
  {Dias}}\ and\ \bibinfo {author} {\bibfnamefont {J.~D.}\ \bibnamefont
  {Gouveia}},\ }\bibfield  {title} {\emph {{\selectlanguage {en}\bibinfo
  {title} {Origami rules for the construction of localized eigenstates of the
  {{Hubbard}} model in decorated lattices}}},\ }\href
  {https://doi.org/10.1038/srep16852} {\bibfield  {journal} {\bibinfo
  {journal} {Sci. Rep.}\ }\textbf {\bibinfo {volume} {5}},\ \bibinfo {pages}
  {16852} (\bibinfo {year} {2015})}\BibitemShut {NoStop}%
\bibitem [{\citenamefont {Maimaiti}\ \emph {et~al.}(2017)\citenamefont
  {Maimaiti}, \citenamefont {Andreanov}, \citenamefont {Park}, \citenamefont
  {Gendelman},\ and\ \citenamefont
  {Flach}}]{Maimaiti2017_PRB_95_115135_CompactLocalizedStatesFlatband}%
  \BibitemOpen
  \bibfield  {author} {\bibinfo {author} {\bibfnamefont {W.}~\bibnamefont
  {Maimaiti}}, \bibinfo {author} {\bibfnamefont {A.}~\bibnamefont {Andreanov}},
  \bibinfo {author} {\bibfnamefont {H.~C.}\ \bibnamefont {Park}}, \bibinfo
  {author} {\bibfnamefont {O.}~\bibnamefont {Gendelman}},\ and\ \bibinfo
  {author} {\bibfnamefont {S.}~\bibnamefont {Flach}},\ }\bibfield  {title}
  {\emph {\bibinfo {title} {Compact localized states and flat-band generators
  in one dimension}},\ }\href {https://doi.org/10.1103/PhysRevB.95.115135}
  {\bibfield  {journal} {\bibinfo  {journal} {Phys. Rev. B}\ }\textbf {\bibinfo
  {volume} {95}},\ \bibinfo {pages} {115135} (\bibinfo {year}
  {2017})}\BibitemShut {NoStop}%
\bibitem [{\citenamefont {Xu}\ and\ \citenamefont
  {Pu}(2020)}]{Xu2020_PRA_102_053305_BuildingFlatbandLatticeModels}%
  \BibitemOpen
  \bibfield  {author} {\bibinfo {author} {\bibfnamefont {Y.}~\bibnamefont
  {Xu}}\ and\ \bibinfo {author} {\bibfnamefont {H.}~\bibnamefont {Pu}},\
  }\bibfield  {title} {\emph {\bibinfo {title} {Building flat-band lattice
  models from {{Gram}} matrices}},\ }\href
  {https://doi.org/10.1103/PhysRevA.102.053305} {\bibfield  {journal} {\bibinfo
   {journal} {Phys. Rev. A}\ }\textbf {\bibinfo {volume} {102}},\ \bibinfo
  {pages} {053305} (\bibinfo {year} {2020})}\BibitemShut {NoStop}%
\bibitem [{\citenamefont {Lee}\ \emph {et~al.}(2019)\citenamefont {Lee},
  \citenamefont {Fleurence}, \citenamefont {{Yamada-Takamura}},\ and\
  \citenamefont {Ozaki}}]{Lee2019_PRB_100_045150_HiddenMechanismEmbeddingFlat}%
  \BibitemOpen
  \bibfield  {author} {\bibinfo {author} {\bibfnamefont {C.-C.}\ \bibnamefont
  {Lee}}, \bibinfo {author} {\bibfnamefont {A.}~\bibnamefont {Fleurence}},
  \bibinfo {author} {\bibfnamefont {Y.}~\bibnamefont {{Yamada-Takamura}}},\
  and\ \bibinfo {author} {\bibfnamefont {T.}~\bibnamefont {Ozaki}},\ }\bibfield
   {title} {\emph {\bibinfo {title} {Hidden mechanism for embedding the flat
  bands of {{Lieb}}, kagome, and checkerboard lattices in other structures}},\
  }\href {https://doi.org/10.1103/PhysRevB.100.045150} {\bibfield  {journal}
  {\bibinfo  {journal} {Phys. Rev. B}\ }\textbf {\bibinfo {volume} {100}},\
  \bibinfo {pages} {045150} (\bibinfo {year} {2019})}\BibitemShut {NoStop}%
\bibitem [{\citenamefont {Smith}\ and\ \citenamefont
  {Webb}(2019)}]{Smith2019_PASMaiA_514_855_HiddenSymmetriesRealTheoretical}%
  \BibitemOpen
  \bibfield  {author} {\bibinfo {author} {\bibfnamefont {D.}~\bibnamefont
  {Smith}}\ and\ \bibinfo {author} {\bibfnamefont {B.}~\bibnamefont {Webb}},\
  }\bibfield  {title} {\emph {\bibinfo {title} {Hidden symmetries in real and
  theoretical networks}},\ }\href {https://doi.org/10.1016/j.physa.2018.09.131}
  {\bibfield  {journal} {\bibinfo  {journal} {Physica A}\ }\textbf {\bibinfo {volume} {514}},\ \bibinfo {pages}
  {855} (\bibinfo {year} {2019})}\BibitemShut {NoStop}%
\bibitem [{\citenamefont {R{\"o}ntgen}\ \emph {et~al.}(2021)\citenamefont
  {R{\"o}ntgen}, \citenamefont {Pyzh}, \citenamefont {Morfonios}, \citenamefont
  {Palaiodimopoulos}, \citenamefont {Diakonos},\ and\ \citenamefont
  {Schmelcher}}]{Rontgen2021_PRL_126_180601_LatentSymmetryInducedDegeneracies}%
  \BibitemOpen
  \bibfield  {author} {\bibinfo {author} {\bibfnamefont {M.}~\bibnamefont
  {R{\"o}ntgen}}, \bibinfo {author} {\bibfnamefont {M.}~\bibnamefont {Pyzh}},
  \bibinfo {author} {\bibfnamefont {C.~V.}\ \bibnamefont {Morfonios}}, \bibinfo
  {author} {\bibfnamefont {N.~E.}\ \bibnamefont {Palaiodimopoulos}}, \bibinfo
  {author} {\bibfnamefont {F.~K.}\ \bibnamefont {Diakonos}},\ and\ \bibinfo
  {author} {\bibfnamefont {P.}~\bibnamefont {Schmelcher}},\ }\bibfield  {title}
  {\emph {\bibinfo {title} {Latent {{Symmetry Induced Degeneracies}}}},\ }\href
  {https://doi.org/10.1103/PhysRevLett.126.180601} {\bibfield  {journal}
  {\bibinfo  {journal} {Phys. Rev. Lett.}\ }\textbf {\bibinfo {volume}
  {126}},\ \bibinfo {pages} {180601} (\bibinfo {year} {2021})}\BibitemShut
  {NoStop}%
\bibitem [{\citenamefont {Kempton}\ \emph {et~al.}(2020)\citenamefont
  {Kempton}, \citenamefont {Sinkovic}, \citenamefont {Smith},\ and\
  \citenamefont
  {Webb}}]{Kempton2020_LAaiA_594_226_CharacterizingCospectralVerticesIsospectral}%
  \BibitemOpen
  \bibfield  {author} {\bibinfo {author} {\bibfnamefont {M.}~\bibnamefont
  {Kempton}}, \bibinfo {author} {\bibfnamefont {J.}~\bibnamefont {Sinkovic}},
  \bibinfo {author} {\bibfnamefont {D.}~\bibnamefont {Smith}},\ and\ \bibinfo
  {author} {\bibfnamefont {B.}~\bibnamefont {Webb}},\ }\bibfield  {title}
  {\emph {{\selectlanguage {en}\bibinfo {title} {Characterizing cospectral
  vertices via isospectral reduction}}},\ }\href
  {https://doi.org/10.1016/j.laa.2020.02.020} {\bibfield  {journal} {\bibinfo
  {journal} {Linear Algebra Appl.}\ }\textbf {\bibinfo {volume}
  {594}},\ \bibinfo {pages} {226} (\bibinfo {year} {2020})}\BibitemShut
  {NoStop}%
\bibitem [{\citenamefont {Priyadarshy}\ \emph {et~al.}(1996)\citenamefont
  {Priyadarshy}, \citenamefont {Skourtis}, \citenamefont {Risser},\ and\
  \citenamefont
  {Beratan}}]{Priyadarshy1996_JCP_104_9473_BridgemediatedElectronicInteractionsDifferences}%
  \BibitemOpen
  \bibfield  {author} {\bibinfo {author} {\bibfnamefont {S.}~\bibnamefont
  {Priyadarshy}}, \bibinfo {author} {\bibfnamefont {S.~S.}\ \bibnamefont
  {Skourtis}}, \bibinfo {author} {\bibfnamefont {S.~M.}\ \bibnamefont
  {Risser}},\ and\ \bibinfo {author} {\bibfnamefont {D.~N.}\ \bibnamefont
  {Beratan}},\ }\bibfield  {title} {\emph {\bibinfo {title} {Bridge-mediated
  electronic interactions: {{Differences}} between {{Hamiltonian}} and
  {{Green}} function partitioning in a non-orthogonal basis}},\ }\href
  {https://doi.org/10.1063/1.471690} {\bibfield  {journal} {\bibinfo  {journal}
  {J. Chem. Phys.}\ }\textbf {\bibinfo {volume} {104}},\ \bibinfo {pages}
  {9473} (\bibinfo {year} {1996})}\BibitemShut {NoStop}%
\bibitem [{\citenamefont {Jin}\ and\ \citenamefont
  {Song}(2011)}]{Jin2011_PRA_83_062118_PartitioningTechniqueDiscreteQuantum}%
  \BibitemOpen
  \bibfield  {author} {\bibinfo {author} {\bibfnamefont {L.}~\bibnamefont
  {Jin}}\ and\ \bibinfo {author} {\bibfnamefont {Z.}~\bibnamefont {Song}},\
  }\bibfield  {title} {\emph {\bibinfo {title} {Partitioning technique for
  discrete quantum systems}},\ }\href
  {https://doi.org/10.1103/PhysRevA.83.062118} {\bibfield  {journal} {\bibinfo
  {journal} {Phys. Rev. A}\ }\textbf {\bibinfo {volume} {83}},\ \bibinfo
  {pages} {062118} (\bibinfo {year} {2011})}\BibitemShut {NoStop}%
\bibitem [{\citenamefont {Grosso}\ and\ \citenamefont
  {Parravicini}(2013)}]{Grosso2013_SolidStatePhysics}%
  \BibitemOpen
  \bibfield  {author} {\bibinfo {author} {\bibfnamefont {G.}~\bibnamefont
  {Grosso}}\ and\ \bibinfo {author} {\bibfnamefont {G.~P.}\ \bibnamefont
  {Parravicini}},\ }\href@noop {} {{\selectlanguage {en}\emph {\bibinfo {title}
  {Solid {{State Physics}}}}}}\ (\bibinfo  {publisher} {{Academic Press}},\
  \bibinfo {year} {2013})\BibitemShut {NoStop}%
\bibitem [{\citenamefont
  {Feshbach}(1962)}]{Feshbach1962_AoP_19_287_UnifiedTheoryNuclearReactions}%
  \BibitemOpen
  \bibfield  {author} {\bibinfo {author} {\bibfnamefont {H.}~\bibnamefont
  {Feshbach}},\ }\bibfield  {title} {\emph {{\selectlanguage {en}\bibinfo
  {title} {A unified theory of nuclear reactions. {{II}}}}},\ }\href
  {https://doi.org/10.1016/0003-4916(62)90221-X} {\bibfield  {journal}
  {\bibinfo  {journal} {Ann. Phys. (NY)}\ }\textbf {\bibinfo {volume} {19}},\
  \bibinfo {pages} {287} (\bibinfo {year} {1962})}\BibitemShut {NoStop}%
\bibitem [{\citenamefont
  {Datta}(1995)}]{Datta1995_ElectronicTransportMesoscopicSystems}%
  \BibitemOpen
  \bibfield  {author} {\bibinfo {author} {\bibfnamefont {S.}~\bibnamefont
  {Datta}},\ }\href@noop {} {\emph {\bibinfo {title} {Electronic {{Transport}}
  in {{Mesoscopic Systems}}}}}\ (\bibinfo  {publisher} {{Cambridge University
  Press}},\ \bibinfo {address} {{Cambridge, UK}},\ \bibinfo {year}
  {1995})\BibitemShut {NoStop}%
\bibitem [{\citenamefont {Pal}\ \emph {et~al.}(2013)\citenamefont {Pal},
  \citenamefont {K.Maiti},\ and\ \citenamefont
  {Chakrabarti}}]{Pal2013_E_102_17004_CompleteAbsenceLocalizationFamily}%
  \BibitemOpen
  \bibfield  {author} {\bibinfo {author} {\bibfnamefont {B.}~\bibnamefont
  {Pal}}, \bibinfo {author} {\bibfnamefont {S.}~\bibnamefont {K.Maiti}},\ and\
  \bibinfo {author} {\bibfnamefont {A.}~\bibnamefont {Chakrabarti}},\
  }\bibfield  {title} {\emph {{\selectlanguage {en}\bibinfo {title} {Complete
  absence of localization in a family of disordered lattices}}},\ }\href
  {https://doi.org/10.1209/0295-5075/102/17004} {\bibfield  {journal} {\bibinfo
   {journal} {Europhys. Lett.}\ }\textbf {\bibinfo {volume} {102}},\
  \bibinfo {pages} {17004} (\bibinfo {year} {2013})}\BibitemShut {NoStop}%
\bibitem [{\citenamefont {Pal}\ and\ \citenamefont
  {Chakrabarti}(2014{\natexlab{a}})}]{Pal2014_PELSaN_60_188_AbsolutelyContinuousEnergyBands}%
  \BibitemOpen
  \bibfield  {author} {\bibinfo {author} {\bibfnamefont {B.}~\bibnamefont
  {Pal}}\ and\ \bibinfo {author} {\bibfnamefont {A.}~\bibnamefont
  {Chakrabarti}},\ }\bibfield  {title} {\emph {{\selectlanguage {en}\bibinfo
  {title} {Absolutely continuous energy bands in the electronic spectrum of
  quasiperiodic ladder networks}}},\ }\href
  {https://doi.org/10.1016/j.physe.2014.02.022} {\bibfield  {journal} {\bibinfo
   {journal} {Physica E}\ }\textbf
  {\bibinfo {volume} {60}},\ \bibinfo {pages} {188} (\bibinfo {year}
  {2014}{\natexlab{a}})}\BibitemShut {NoStop}%
\bibitem [{\citenamefont {Pal}\ and\ \citenamefont
  {Chakrabarti}(2014{\natexlab{b}})}]{Pal2014_PLA_378_2782_EngineeringBandsExtendedElectronic}%
  \BibitemOpen
  \bibfield  {author} {\bibinfo {author} {\bibfnamefont {B.}~\bibnamefont
  {Pal}}\ and\ \bibinfo {author} {\bibfnamefont {A.}~\bibnamefont
  {Chakrabarti}},\ }\bibfield  {title} {\emph {{\selectlanguage {en}\bibinfo
  {title} {Engineering bands of extended electronic states in a class of
  topologically disordered and quasiperiodic lattices}}},\ }\href
  {https://doi.org/10.1016/j.physleta.2014.07.034} {\bibfield  {journal}
  {\bibinfo  {journal} {Phys. Lett. A}\ }\textbf {\bibinfo {volume}
  {378}},\ \bibinfo {pages} {2782} (\bibinfo {year}
  {2014}{\natexlab{b}})}\BibitemShut {NoStop}%
\bibitem [{\citenamefont {Duarte}\ and\ \citenamefont
  {Torres}(2015)}]{Duarte2015_LAaiA_474_110_EigenvectorsIsospectralGraphTransformations}%
  \BibitemOpen
  \bibfield  {author} {\bibinfo {author} {\bibfnamefont {P.}~\bibnamefont
  {Duarte}}\ and\ \bibinfo {author} {\bibfnamefont {M.~J.}\ \bibnamefont
  {Torres}},\ }\bibfield  {title} {\emph {{\selectlanguage {en}\bibinfo {title}
  {Eigenvectors of isospectral graph transformations}}},\ }\href
  {https://doi.org/10.1016/j.laa.2015.01.038} {\bibfield  {journal} {\bibinfo
  {journal} {Linear Algebra Appl.}\ }\textbf {\bibinfo {volume}
  {474}},\ \bibinfo {pages} {110} (\bibinfo {year} {2015})}\BibitemShut
  {NoStop}%
\bibitem [{\citenamefont {Bunimovich}\ \emph {et~al.}(2019)\citenamefont
  {Bunimovich}, \citenamefont {Smith},\ and\ \citenamefont
  {Webb}}]{Bunimovich2019_AMNS_4_231_FindingHiddenStructuresHierarchies}%
  \BibitemOpen
  \bibfield  {author} {\bibinfo {author} {\bibfnamefont {L.}~\bibnamefont
  {Bunimovich}}, \bibinfo {author} {\bibfnamefont {D.}~\bibnamefont {Smith}},\
  and\ \bibinfo {author} {\bibfnamefont {B.}~\bibnamefont {Webb}},\ }\bibfield
  {title} {\emph {{\selectlanguage {en}\bibinfo {title} {Finding {{Hidden
  Structures}}, {{Hierarchies}}, and {{Cores}} in {{Networks}} via
  {{Isospectral Reduction}}}}},\ }\href
  {https://doi.org/10.2478/AMNS.2019.1.00021} {\bibfield  {journal} {\bibinfo
  {journal} {Appl. Math. Nonlinear Sci.}\ }\textbf {\bibinfo {volume} {4}},\
  \bibinfo {pages} {231} (\bibinfo {year} {2019})}\BibitemShut {NoStop}%
\bibitem [{\citenamefont {Bunimovich}\ and\ \citenamefont
  {Webb}(2014)}]{Bunimovich2014_IsospectralTransformationsNewApproach}%
  \BibitemOpen
  \bibfield  {author} {\bibinfo {author} {\bibfnamefont {L.}~\bibnamefont
  {Bunimovich}}\ and\ \bibinfo {author} {\bibfnamefont {B.}~\bibnamefont
  {Webb}},\ }\href {https://doi.org/10.1007/978-1-4939-1375-6}
  {{\selectlanguage {en}\emph {\bibinfo {title} {Isospectral
  {{Transformations}}: {{A New Approach}} to {{Analyzing Multidimensional
  Systems}} and {{Networks}}}}}},\ Springer {{Monographs}} in {{Mathematics}}\
  (\bibinfo  {publisher} {{Springer-Verlag}},\ \bibinfo {address} {{New
  York}},\ \bibinfo {year} {2014})\BibitemShut {NoStop}%
\bibitem [{\citenamefont {Godsil}\ and\ \citenamefont
  {Smith}(2017)}]{Godsil2017_AM_StronglyCospectralVertices}%
  \BibitemOpen
  \bibfield  {author} {\bibinfo {author} {\bibfnamefont {C.}~\bibnamefont
  {Godsil}}\ and\ \bibinfo {author} {\bibfnamefont {J.}~\bibnamefont {Smith}},\
  }\bibfield  {title} {\emph {\bibinfo {title} {Strongly {{Cospectral
  Vertices}}}},\ }\href@noop {} \Eprint
  {https://arxiv.org/abs/1709.07975} {arXiv:1709.07975} {\bibfield  {journal} (\bibinfo {year} {2017})} \BibitemShut
  {NoStop}%
\bibitem [{\citenamefont {Estrada}\ and\ \citenamefont
  {Knight}(2015)}]{Estrada2015_FirstCourseNetworkTheory}%
  \BibitemOpen
  \bibfield  {author} {\bibinfo {author} {\bibfnamefont {E.}~\bibnamefont
  {Estrada}}\ and\ \bibinfo {author} {\bibfnamefont {P.~A.}\ \bibnamefont
  {Knight}},\ }\href@noop {} {\emph {\bibinfo {title} {A {{First Course}} in
  {{Network Theory}}}}}\ (\bibinfo  {publisher} {{Oxford University Press}},\
  \bibinfo {address} {{Oxford, New York}},\ \bibinfo {year} {2015})\BibitemShut
  {NoStop}%
\bibitem [{\citenamefont {Tsuji}\ and\ \citenamefont
  {Estrada}(2019)}]{Tsuji2019_JCP_150_204123_InfluenceLongrangeInteractionsQuantum}%
  \BibitemOpen
  \bibfield  {author} {\bibinfo {author} {\bibfnamefont {Y.}~\bibnamefont
  {Tsuji}}\ and\ \bibinfo {author} {\bibfnamefont {E.}~\bibnamefont
  {Estrada}},\ }\bibfield  {title} {\emph {\bibinfo {title} {Influence of
  long-range interactions on quantum interference in molecular conduction.
  {{A}} tight-binding ({{H\"uckel}}) approach}},\ }\href
  {https://doi.org/10.1063/1.5097330} {\bibfield  {journal} {\bibinfo
  {journal} {J. Chem. Phys.}\ }\textbf {\bibinfo {volume} {150}},\ \bibinfo
  {pages} {204123} (\bibinfo {year} {2019})}\BibitemShut {NoStop}%
\bibitem [{\citenamefont {Brualdi}\ and\ \citenamefont
  {Cvetkovic}(2008)}]{Brualdi2008_CombinatorialApproachMatrixTheory}%
  \BibitemOpen
  \bibfield  {author} {\bibinfo {author} {\bibfnamefont {R.~A.}\ \bibnamefont
  {Brualdi}}\ and\ \bibinfo {author} {\bibfnamefont {D.}~\bibnamefont
  {Cvetkovic}},\ }\href@noop {} {{\selectlanguage {en}\emph {\bibinfo {title}
  {A {{Combinatorial Approach}} to {{Matrix Theory}} and {{Its
  Applications}}}}}}\ (\bibinfo  {publisher} {{Chapman and Hall/CRC Press}},\ \bibinfo {year}
  {2008})\BibitemShut {NoStop}%
\bibitem [{\citenamefont
  {Godsil}(2012)}]{Godsil2012_AC_16_733_ControllableSubsetsGraphs}%
  \BibitemOpen
  \bibfield  {author} {\bibinfo {author} {\bibfnamefont {C.}~\bibnamefont
  {Godsil}},\ }\bibfield  {title} {\emph {{\selectlanguage {en}\bibinfo {title}
  {Controllable {{Subsets}} in {{Graphs}}}}},\ }\href
  {https://doi.org/10.1007/s00026-012-0156-3} {\bibfield  {journal} {\bibinfo
  {journal} {Ann. Comb.}\ }\textbf {\bibinfo {volume} {16}},\ \bibinfo {pages}
  {733} (\bibinfo {year} {2012})}\BibitemShut {NoStop}%
\bibitem [{\citenamefont {Liu}\ and\ \citenamefont
  {Siemons}(2019)}]{Liu2019_AM_UnlockingWalkMatrixGraph}%
  \BibitemOpen
  \bibfield  {author} {\bibinfo {author} {\bibfnamefont {F.}~\bibnamefont
  {Liu}}\ and\ \bibinfo {author} {\bibfnamefont {J.}~\bibnamefont {Siemons}},\
  }\bibfield  {title} {\emph {\bibinfo {title} {Unlocking the walk matrix of a
  graph}},\ }\href@noop {} \Eprint {https://arxiv.org/abs/1911.00062} {arXiv:1911.00062} {\bibfield  {journal} (\bibinfo {year} {2019})} \BibitemShut
  {NoStop}%
\bibitem [{\citenamefont
  {Meyer}(2000)}]{Meyer2000_MatrixAnalysisAppliedLinear}%
  \BibitemOpen
  \bibfield  {author} {\bibinfo {author} {\bibfnamefont {C.~D.}\ \bibnamefont
  {Meyer}},\ }\href@noop {} {\emph {\bibinfo {title} {Matrix Analysis and
  Applied Linear Algebra}}}\ (\bibinfo  {publisher} {{Society for Industrial
  and Applied Mathematics}},\ \bibinfo {address} {{USA}},\ \bibinfo {year}
  {2000})\BibitemShut {NoStop}%
\bibitem [{\citenamefont
  {Xu}(2011)}]{Xu2011_LAaiA_434_185_FunctionsMatrixKrylovMatrices}%
  \BibitemOpen
  \bibfield  {author} {\bibinfo {author} {\bibfnamefont {H.}~\bibnamefont
  {Xu}},\ }\bibfield  {title} {\emph {{\selectlanguage {en}\bibinfo {title}
  {Functions of a matrix and {{Krylov}} matrices}}},\ }\href
  {https://doi.org/10.1016/j.laa.2010.08.044} {\bibfield  {journal} {\bibinfo
  {journal} {Linear Algebra Appl.}\ }\textbf {\bibinfo {volume}
  {434}},\ \bibinfo {pages} {185} (\bibinfo {year} {2011})}\BibitemShut
  {NoStop}%
\bibitem [{\citenamefont {Morfonios}\ \emph {et~al.}(2021)\citenamefont
  {Morfonios}, \citenamefont {Pyzh}, \citenamefont {R{\"o}ntgen},\ and\
  \citenamefont
  {Schmelcher}}]{Morfonios2021_LAaiA_624_53_CospectralityPreservingGraphModifications}%
  \BibitemOpen
  \bibfield  {author} {\bibinfo {author} {\bibfnamefont {C.~V.}\ \bibnamefont
  {Morfonios}}, \bibinfo {author} {\bibfnamefont {M.}~\bibnamefont {Pyzh}},
  \bibinfo {author} {\bibfnamefont {M.}~\bibnamefont {R{\"o}ntgen}},\ and\
  \bibinfo {author} {\bibfnamefont {P.}~\bibnamefont {Schmelcher}},\ }\bibfield
   {title} {\emph {{\selectlanguage {en}\bibinfo {title} {Cospectrality
  preserving graph modifications and eigenvector properties via walk
  equivalence of vertices}}},\ }\href
  {https://doi.org/10.1016/j.laa.2021.04.004} {\bibfield  {journal} {\bibinfo
  {journal} {Linear Algebra Appl.}\ }\textbf {\bibinfo {volume}
  {624}},\ \bibinfo {pages} {53} (\bibinfo {year} {2021})}\BibitemShut
  {NoStop}%
\bibitem [{Note1()}]{Note1}%
  \BibitemOpen
  \bibinfo {note} {We note that the occurrence of a walk equivalent pair
  relative to some ${\protect \mathbb {M}}$ necessarily renders $W_{\protect
  \mathbb {M}}$ non-invertible by reducing its rank. Incidentally, this lifts
  the so-called ``controllability'' \cite
  {Godsil2012_AC_16_733_ControllableSubsetsGraphs,Farrugia2014_LAaiA_454_138_ControllabilityUndirectedGraphs,Aguilar2015_ITAC_60_1611_GraphControllabilityClassesLaplacian}
  of ${\protect \mathbb {M}}$ relative to $H$, and as a consequence allows for
  local permutation symmetries in $H$ mapping ${\protect \mathbb {M}}$ to
  itself \cite {Godsil2012_AC_16_733_ControllableSubsetsGraphs}.}\BibitemShut
  {Stop}%
\bibitem [{Note2()}]{Note2}%
  \BibitemOpen
  \bibinfo {note} {We find such parametrizations numerically by starting with
  the unweighted graph and then (i) setting one weight to a random value
  (representing an independent parameter) and scanning through the graph for
  edges which can be set to that same value while preserving cospectrality,
  (ii) repeating successively for unaltered edges until all edges are
  parametrized.}\BibitemShut {Stop}%
\bibitem [{\citenamefont {Eisenberg}\ \emph {et~al.}(2019)\citenamefont
  {Eisenberg}, \citenamefont {Kempton},\ and\ \citenamefont
  {Lippner}}]{Eisenberg2019_DM_342_2821_PrettyGoodQuantumState}%
  \BibitemOpen
  \bibfield  {author} {\bibinfo {author} {\bibfnamefont {O.}~\bibnamefont
  {Eisenberg}}, \bibinfo {author} {\bibfnamefont {M.}~\bibnamefont {Kempton}},\
  and\ \bibinfo {author} {\bibfnamefont {G.}~\bibnamefont {Lippner}},\
  }\bibfield  {title} {\emph {{\selectlanguage {en}\bibinfo {title} {Pretty
  good quantum state transfer in asymmetric graphs via potential}}},\ }\href
  {https://doi.org/10.1016/j.disc.2018.10.037} {\bibfield  {journal} {\bibinfo
  {journal} {Discrete Math.}\ } \textbf {\bibinfo {volume} {342}},\ \bibinfo
  {pages} {2821} (\bibinfo {year} {2019})}\BibitemShut {NoStop}%
\bibitem [{\citenamefont
  {Vanderbilt}(2018)}]{Vanderbilt2018_BerryPhasesElectronicStructure}%
  \BibitemOpen
  \bibfield  {author} {\bibinfo {author} {\bibfnamefont {D.}~\bibnamefont
  {Vanderbilt}},\ }\href {https://doi.org/10.1017/9781316662205} {\emph
  {\bibinfo {title} {Berry {{Phases}} in {{Electronic Structure Theory}}:
  {{Electric Polarization}}, {{Orbital Magnetization}} and {{Topological
  Insulators}}}}}\ (\bibinfo  {publisher} {{Cambridge University Press}},\
  \bibinfo {address} {{Cambridge}},\ \bibinfo {year} {2018})\BibitemShut
  {NoStop}%
\bibitem [{Note3()}]{Note3}%
  \BibitemOpen
  \bibinfo {note} {For clarity, we note that the bands were computed by standard numerical matrix diagonalization of the Bloch Hamiltonian $H_\bs{k}$ (in varying $\bs{k}$) and not from the corresponding nonlinear eigenvalue problem $\tH_{\bS;\bs{k}}(E)\ket{\varphi}=E\ket{\varphi}$ of the reduced $H_\bs{k}$ by finding the roots of $\det(E - \tH_{\bS;\bs{k}}(E))=0$ (which, as mentioned in \cref{sec:latSymmetry}, would generally yield a subset of the full eigenvalue spectrum).}\BibitemShut {Stop}%
\bibitem [{Note4()}]{Note4}%
  \BibitemOpen
  \bibinfo {note} {Note that two sites $u,v$ related by an involutory
  permutation symmetry $\varPi $ are automatically cospectral, since
  $\mathinner {\langle {u|H^{r}|u}\rangle } = \mathinner {\langle
  {u|H^{r}\varPi ^2|u}\rangle } = \mathinner {\langle {u|\varPi H^{r}\varPi
  |u}\rangle } = \mathinner {\langle {v|H^{r}|v}\rangle }$ $\forall \protect
  \,{r}$ (see \protect \cref {eq:cospectrality}), and any site $c$ fixed by
  $\varPi $ is a walk singlet relative to $\{u,v\}$, since $\mathinner {\langle
  {u|H^{r}|c}\rangle } = \mathinner {\langle {u|H^{r}\varPi |c}\rangle } =
  \mathinner {\langle {u|\varPi H^{r}|c}\rangle } = \mathinner {\langle
  {v|H^{r}|c}\rangle }$ $\forall \protect \,{r}$.}\BibitemShut {Stop}%
\bibitem [{\citenamefont {Maimaiti}\ \emph {et~al.}(2021)\citenamefont
  {Maimaiti}, \citenamefont {Andreanov},\ and\ \citenamefont
  {Flach}}]{Maimaiti2021_PRB_103_165116_FlatbandGeneratorTwoDimensions}%
  \BibitemOpen
  \bibfield  {author} {\bibinfo {author} {\bibfnamefont {W.}~\bibnamefont
  {Maimaiti}}, \bibinfo {author} {\bibfnamefont {A.}~\bibnamefont
  {Andreanov}},\ and\ \bibinfo {author} {\bibfnamefont {S.}~\bibnamefont
  {Flach}},\ }\bibfield  {title} {\emph {\bibinfo {title} {Flat-band generator
  in two dimensions}},\ }\href {https://doi.org/10.1103/PhysRevB.103.165116}
  {\bibfield  {journal} {\bibinfo  {journal} {Phys. Rev. B}\ }\textbf
  {\bibinfo {volume} {103}},\ \bibinfo {pages} {165116} (\bibinfo {year}
  {2021})}\BibitemShut {NoStop}%
\bibitem [{\citenamefont {R{\"o}ntgen}\ \emph
  {et~al.}(2020{\natexlab{c}})\citenamefont {R{\"o}ntgen}, \citenamefont
  {Palaiodimopoulos}, \citenamefont {Morfonios}, \citenamefont {Brouzos},
  \citenamefont {Pyzh}, \citenamefont {Diakonos},\ and\ \citenamefont
  {Schmelcher}}]{Rontgen2020_PRA_101_042304_DesigningPrettyGoodState}%
  \BibitemOpen
  \bibfield  {author} {\bibinfo {author} {\bibfnamefont {M.}~\bibnamefont
  {R{\"o}ntgen}}, \bibinfo {author} {\bibfnamefont {N.~E.}\ \bibnamefont
  {Palaiodimopoulos}}, \bibinfo {author} {\bibfnamefont {C.~V.}\ \bibnamefont
  {Morfonios}}, \bibinfo {author} {\bibfnamefont {I.}~\bibnamefont {Brouzos}},
  \bibinfo {author} {\bibfnamefont {M.}~\bibnamefont {Pyzh}}, \bibinfo {author}
  {\bibfnamefont {F.~K.}\ \bibnamefont {Diakonos}},\ and\ \bibinfo {author}
  {\bibfnamefont {P.}~\bibnamefont {Schmelcher}},\ }\bibfield  {title} {\emph
  {\bibinfo {title} {Designing pretty good state transfer via isospectral
  reductions}},\ }\href {https://doi.org/10.1103/PhysRevA.101.042304}
  {\bibfield  {journal} {\bibinfo  {journal} {Phys. Rev. A}\ }\textbf {\bibinfo
  {volume} {101}},\ \bibinfo {pages} {042304} (\bibinfo {year}
  {2020}{\natexlab{c}})}\BibitemShut {NoStop}%
\bibitem [{Note5()}]{Note5}%
  \BibitemOpen
  \bibinfo {note} {One may then attempt to apply the modification \ref
  {mod:multipletInterconnection} (see \protect \cref {sec:modifications}) to
  lift the degeneracy and to remove the $\{u,v\}$-vanishing eigenstate(s),
  without altering the graph's multiplet structure (the simplest modification
  being an added loop of arbitrary weight to any walk singlet relative to
  $\{u,v\}$). The $Q$-matrix can then be obtained from \protect \cref
  {eq:genQMatrix}. However, rigorously showing that this $Q$ is the same as for
  the unmodified graph requires a more general account on walk multiplets, to
  be given elsewhere.}\BibitemShut {Stop}%
\bibitem [{\citenamefont {Farrugia}\ and\ \citenamefont
  {Sciriha}(2014)}]{Farrugia2014_LAaiA_454_138_ControllabilityUndirectedGraphs}%
  \BibitemOpen
  \bibfield  {author} {\bibinfo {author} {\bibfnamefont {A.}~\bibnamefont
  {Farrugia}}\ and\ \bibinfo {author} {\bibfnamefont {I.}~\bibnamefont
  {Sciriha}},\ }\bibfield  {title} {\emph {{\selectlanguage {en}\bibinfo
  {title} {Controllability of undirected graphs}}},\ }\href
  {https://doi.org/10.1016/j.laa.2014.04.022} {\bibfield  {journal} {\bibinfo
  {journal} {Linear Algebra Appl.}\ }\textbf {\bibinfo {volume}
  {454}},\ \bibinfo {pages} {138} (\bibinfo {year} {2014})}\BibitemShut
  {NoStop}%
\bibitem [{\citenamefont {Aguilar}\ and\ \citenamefont
  {Gharesifard}(2015)}]{Aguilar2015_ITAC_60_1611_GraphControllabilityClassesLaplacian}%
  \BibitemOpen
  \bibfield  {author} {\bibinfo {author} {\bibfnamefont {C.~O.}\ \bibnamefont
  {Aguilar}}\ and\ \bibinfo {author} {\bibfnamefont {B.}~\bibnamefont
  {Gharesifard}},\ }\bibfield  {title} {\emph {\bibinfo {title} {Graph
  {{Controllability Classes}} for the {{Laplacian Leader}}-{{Follower
  Dynamics}}}},\ }\href {https://doi.org/10.1109/TAC.2014.2381435} {\bibfield
  {journal} {\bibinfo  {journal} {IEEE Trans. Autom. Contr.}\ }\textbf
  {\bibinfo {volume} {60}},\ \bibinfo {pages} {1611} (\bibinfo {year}
  {2015})}\BibitemShut {NoStop}%
\end{thebibliography}

%%%%%%%%%%%%%%%%%%%%%%%%%%%%%%%%%%%%%%%%%%%%%%%%%%%%%%%%%%%%%%%%%%%%%%%%%%%%%%


%%%%%%%%%%%%%%%%%%%%%%%%%%%%%%%%%%%%%%%%%%%%%%%%%%%%%%%%%%%%%%%%%%%%%%%%%%%%%%%

\end{document}